\documentclass[useAMS,usenatbib]{mn2e}
\usepackage{rotating, graphicx}
\usepackage{subfig}
\usepackage{pdflscape}
\usepackage{natbib}
\usepackage{amsmath}
\usepackage{courier}
\usepackage{aecompl}
\usepackage{color}
\usepackage[T1]{fontenc}

\title[The Pillars of Creation revisited with MUSE]{The Pillars of Creation revisited with MUSE: gas kinematics and high-mass stellar feedback traced by optical spectroscopy}
\author[A. F. Mc Leod]{A. F. Mc Leod$^{1}$\thanks{E-mail:amcleod@eso.org}, J. E. Dale$^{2}$$^{,}$$^{3}$, A. Ginsburg$^{1}$, B. Ercolano$^{2}$$^{,}$$^{3}$,  \newauthor M. Gritschneder$^{2}$, S. Ramsay$^{1}$ and L. Testi$^{1}$$^{,}$$^{4}$ \\
$^{1}$European Southern Observatory, Karl-Schwarzschild-Str. 2, D-85748 Garching bei M{\"u}nchen, Germany\\
$^{2}$Universit{\"a}ts-Sternwarte M{\"u}nchen, Scheinerstr. 1, D-81679 M{\"u}nchen, Germany\\
$^{3}$Excellence Cluster 'Universe', Boltzmannstr. 2, D-85748 Garching bei M{\"u}nchen, Germany\\
$^{4}$INAF/Osservatorio Astrofisico of Arcetri, Largo E. Fermi, 5, 50125 Firenze, Italy}

\begin{document}

\date{Accepted yyyy month dd. Received yyyy month dd; in original form yyyy month dd}

\pagerange{\pageref{firstpage}--\pageref{lastpage}} \pubyear{2015}

\maketitle

\label{firstpage}

\begin{abstract}
Integral field unit (IFU) data of the iconic Pillars of Creation in M16 are presented. The ionisation structure of the pillars was studied in great detail over almost the entire visible wavelength range, and maps of the relevant physical parameters, e.g. extinction, electron density, electron temperature, line-of-sight velocity of the ionised and neutral gas are shown. In agreement with previous authors, we find that the pillar tips are being ionised and photo-evaporated by the massive members of the nearby cluster NGC 6611. They display a stratified ionisation structure where the emission lines peak in a descending order according to their ionisation energies. The IFU data allowed us to analyse the kinematics of the photo-evaporative flow in terms of the stratified ionisation structure, and we find that, in agreement with simulations, the photo-evaporative flow is traced by a blueshift in the position-velocity profile. The gas kinematics and ionisation structure have allowed us to produce a sketch of the 3D geometry of the Pillars, positioning the pillars with respect to the ionising cluster stars. We use a novel method to detect a previously unknown bipolar outflow at the tip of the middle pillar and suggest that it has an embedded protostar as its driving source. Furthermore we identify a candidate outflow in the leftmost pillar. With the derived physical parameters and ionic abundances, we estimate a mass loss rate due to the photo-evaporative flow of 70  M$_{\odot}$ Myr$^{-1}$ which yields an expected lifetime of approximately 3 Myr.
\end{abstract}


\section{Introduction}

Throughout their entire lifetime, massive stars influence their immediate surroundings via strong stellar winds, potent ionising radiation and powerful supernovae events. Simulations show that their feedback is responsible for clearing vast bubble-shaped gas voids, inflating expanding HII regions, exposing pillar-like dust and gas lanes and regulating the formation of new stellar populations in their surrounding natal molecular clouds (\citealt{2010ApJ...723..971G}, \citealt{2011ApJ...729...72P}, \citealt{2011ApJ...736..142B}, \citealt{2012A&A...546A..33T}, \citealt{2013MNRAS.435..917W}, \citealt{2013MNRAS.431.1062D}, \citealt{2013MNRAS.430..234D}, \citealt{2015ApJ...798...32N}). The simplified and idealised conditions of the simulations can yield important insight on feedback mechanisms, but even the state of the art simulations fail to represent the complexity of reality, and the details of this feedback have yet to be fully understood. Furthermore, massive star formation dominates the energetics and feedback in star forming galaxies, and properly accounting for the star formation feedback is a critical ingredient of galaxy evolution models. This is the first publication from our FuSIOn (Feedback in massive star forming regions: from Simulations to Observations) study in which we seek to test the predictions of the above mentioned numerical simulations, to put better constraints on their initial conditions and to ultimately understand feedback mechanisms by comparing them with observations of high mass stellar feedback.

Pillar-like structures are a common morphological phenomenon in star forming regions where massive O- and B-type stars shape the surrounding material via ionisation and stellar winds, and they arise naturally in simulations of cluster formation and evolution that include radiative feedback (\citealt{2010ApJ...723..971G}, \citealt{2012MNRAS.427.2852D}, \citealt{2012A&A...546A..33T}, \citealt{2013MNRAS.435..917W}). Such structures are found to host young stellar objects and are observed to be  between the hot and hostile environment created by massive stars and the material of the parent molecular cloud that has not yet been reached by the stellar feedback. The nearby massive stars compress, ionise and photo-evaporate the top layers of gas that face directly toward them \citep{1991PASP..103..853H}, and therefore ultimately cause the pillars to be destroyed over time. Whether indeed the massive stars are able to trigger star formation at the pillar-ambient medium interface, and how efficient their photo-evaporating effect really is are still open questions that need to be addressed.

One of the prime examples of pillars in a massive star forming region is the Pillars of Creation in the Eagle Nebula (M16), which can be seen as protruding from a molecular cloud into a vast HII region inflated by the 2-3 Myr old and $\sim$ 2 kpc distant \citep{1993AJ....106.1906H} massive cluster NGC 6611, whose brightest and most massive object (the O3-O5 V star W205) lies - in projection - just short of 2 pc North-East of the Pillars. The Pillars can be divided into three main structures, all of them pointing towards the cluster stars of NGC 6611. In this work we will refer to them as P1 (the easternmost and biggest pillar, subdivided into P1a and P1b), P2 (middle pillar) and P3 (the westernmost pillar), as is shown in Fig. \ref{museptg}. This pillar-like shape seems to have its origin in the fact that the low-density material in the bodies of the pillars could be easily removed, but it is capped by much denser cores which shield the pillars from the intense ionising sources in NGC 6611 \citep{2008hsf2.book..599O}.

These Pillars have been studied by many authors in the past two decades since they were first captured in spectacular detail with the Hubble Space Telescope (HST) by \cite{1996AJ....111.2349H} (hereafter referred to as H96), and as it is too great a task to collect and cite all publications regarding these famous structures, we can here only give a brief historical research overview. H96 first described the morphology in great detail and analysed the stratified ionisation structure in terms of a photo-evaporative flow and identified star forming cometary globules currently emerging from the Pillars with the HST images obtained in the H$\alpha$, [SII]$\lambda$6717,6731 and [OIII]$\lambda$5007 filters. By comparing the optical observations with near-IR images, these authors suggest the presence of a population of YSOs in the pillars and so called evaporating gaseous globules (EGGs), indicating active star formation in the region. \cite{2002A&A...389..513M} however find that only about 15 \% of the EGGs show signs of active star formation, while the rest appear not to host young stellar objects. For completeness we show the spectrum of an EGG not covered by the HST data of H96 which is located to the left of P1 (the spectrum is shown in Fig. \ref{lineid}c and its location is shown in Fig. \ref{ratiomaps}a). A molecular line and continuum emission study, combined with mid-IR observations, was presented by \cite{1999A&A...342..233W}, who reported that the pillars are capped by dense and relatively cold ($\sim$ 10-60 M$_{\odot}$) cores in a pre-protostellar evolutionary stage. These authors detected no embedded IR sources or molecular outflows at the pillar tips, found very small ($\sim$ 1.7 km/s) velocity gradients along the pillar bodies and estimated the total mass of the Pillars to be $\sim$ 200 M$_{\odot}$. \cite{2002ApJ...568L.127F} used millimetre data to analyse the age sequence of CO cores, finding a linear age gradient with more evolved objects closer to NGC 6611 and less evolved ones closer to the Pillars, suggesting the propagation of star formation. Whether this corresponds to triggered star formation is debated, as suggested by \cite{2007ApJ...666..321I}, but the presence of young stellar objects (YSOs) and water masers near or in the pillar tips is discussed by several different authors (\citealt{2002ApJ...570..749T}, \citealt{2004ApJ...610..835H}, \citealt{2007ApJ...654..347L}). In the more recent years, much effort has been invested in the simulation of such objects to better understand the extent of the stellar feedback, the physical effects occurring at the pillar-ambient interface and their kinematic structure (\citealt{2006MNRAS.369..143M}, \citealt{2006ApJ...647.1151M}, \citealt{2012MNRAS.420..141E}). To celebrate Hubble's 25th anniversary, the NASA/ESA/Hubble Heritage Team recently released a new, revisited HST image of the Pillars with a much higher spatial resolution and a bigger field of view than H96, bringing new media attention to these structures. 

As part of the science verification of the optical integral field spectrograph MUSE at the Very Large Telescope (VLT), we obtained a 3x3 arcmin mosaic of the Pillars. In this work we  analyse the ionisation structure of these objects in greater detail than ever before, and tie their evaporation and active star formation directly back to the members of NGC 6611. Following the theme of our FuSIOn project, we seek to compare our observations with the results obtained by \citealt{2012MNRAS.420..141E}, who created synthetic observations of pillars under the influence of ionising O- and B-type stars in their immediate neighbourhood. They computed line profiles as a function of position within the pillars for H$\alpha$, [OIII]$\lambda$5007, [NII]$\lambda$6584 and [SII]$\lambda$6717, as well as line ratios for the just mentioned emission lines. 

This paper is organised as follows. We introduce both the observations and simulations in Section 2; we compute the main physical parameters such as electron density and temperature, emission line and ratio maps, as well as analyse the ionisation structure in Section 3. In Section 4 we describe the velocity structure obtained via emission line fitting and compute the expected lifetime of the Pillars. Conclusions are presented in Section 5. The appendix is available as online supplementary material.

\section[]{Observations and simulations}

\subsection{IFU data}
\begin{figure*}
\includegraphics[scale=0.5]{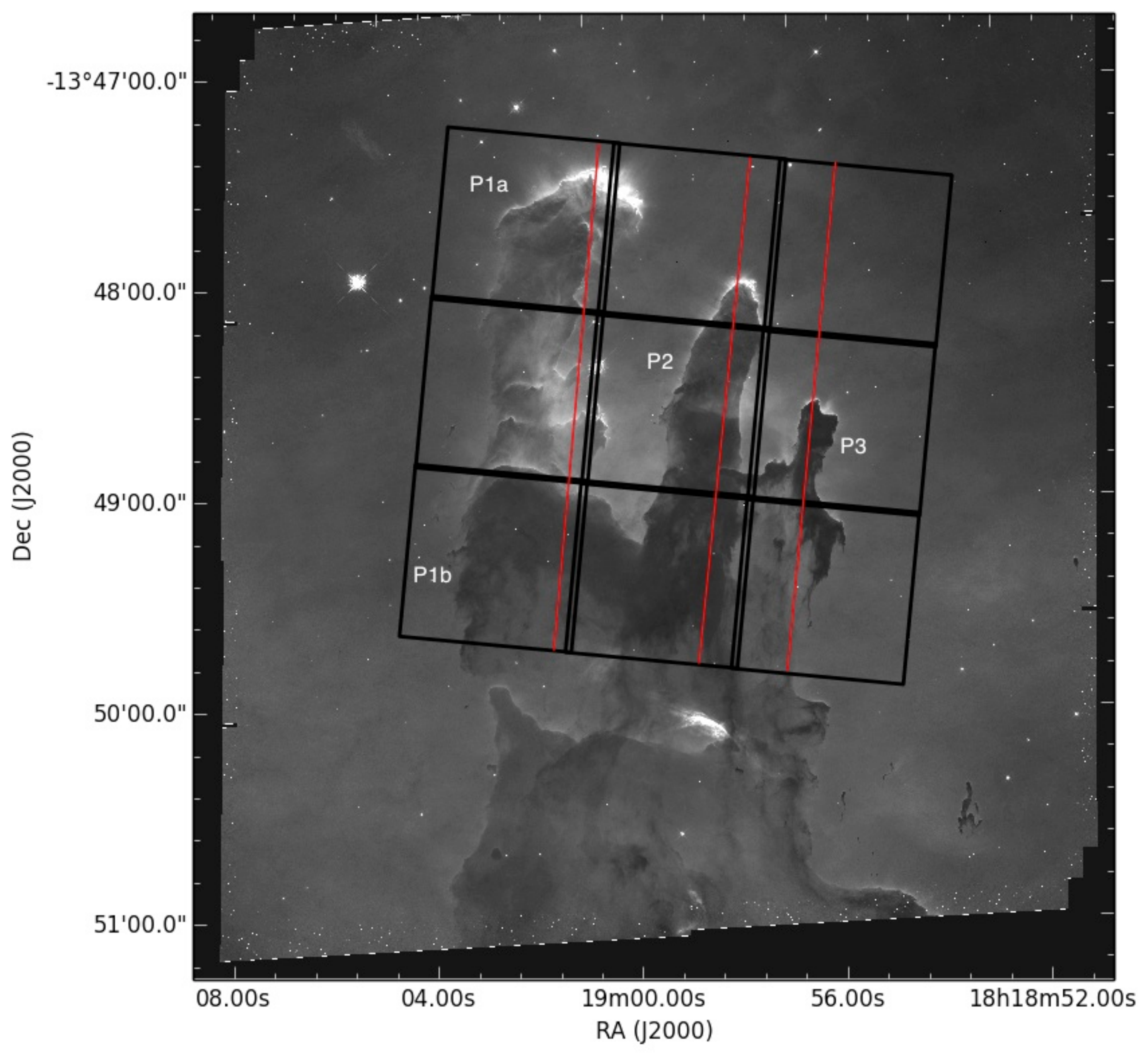}
\caption{The nine MUSE pointings overlaid on the Hubble Space Telescope H$\alpha$+[NII] image (credit: STScI). Each pointing spans approximately 1$\times$1 arcmin$^{2}$. The red lines represent the cuts along which the line intensity profiles were computed (Section 3.2).}
  \label{museptg}
\end{figure*}

The optical IFU observations of the Pillars of Creation in M16 were obtained with the Multi Unit Spectroscopy Explorer MUSE instrument \citep{2010SPIE.7735E..08B} mounted on the VLT. The data was obtained during the instrument's science verification run in service mode on June 21 and June 22 (2014) for the proposal 60.A-9309(A) (PI Mc Leod), using the nominal wavelength range  spanning from 4750 \AA\ to 9350 \AA\ (and a wavelength-dependent spectral resolution of $\sim$ 75 and 150 km s$^{-1}$ respectively). The Wide Field Mode of MUSE has a field of view of 1x1 arcmin$^{2}$, and we obtained two 3x3 arcmin$^{2}$ mosaics with an exposure time of 30 and 130 seconds each, thus requiring 9 pointings per mosaic. The nine fields of the two mosaics are shown in Fig. \ref{museptg} overlaid on the HST H$\alpha$+[NII] image, and their central coordinates and number of spaxels (spectral pixels) are listed in Table \ref{coord}. Each field was observed three times via a dither pattern by rotating the instrument position angle by 90 degrees at each exposure. The full dataset thus consists of two subsets of 27 exposures each, where the subsets differ in exposure time only. The effective spatial resolution of the presented data is about 0.2", which corresponds to $\sim$ 1.9$\times$10$^{-3}$ pc (or $\sim$ 390 AU).

\begin{table}
\begin{center}
\caption{Central coordinates of the MUSE mosaic pointings shown in Fig. \ref{museptg}. The number of spaxels (spectral pixels) corresponds to the number of valid spaxels in the H$\alpha$ slice of the data cube.}
\begin{tabular}{lccc}
\hline
\hline
Field no. & RA (J2000) & Dec (J2000) & No. of spaxels \\
\hline
1 & 18:18:51.918 & -13:48:45.90 & 94586 \\
2 & 18:18:48.892 & -13:49:23.52 & 93936 \\
3 & 18:18:45.866 & -13:50:01.18 & 94129 \\
4 & 18:18:48.493 & -13:50:45.96 & 94746 \\
5 & 18:18:51.519 & -13:50:08.34 & 94723 \\
6 & 18:18:54.545 & -13:49:30.58 & 94034 \\
7 & 18:18:57.172 & -13:50:15.50 & 94207 \\
8 & 18:18:54.146 & -13:50:53.12 & 94690 \\
9 & 18:18:51.120 & -13:51:30.78 & 94610 \\
\hline
\end{tabular}
\label{coord}
\end{center}
\end{table}

\begin{figure}
\mbox{
\subfloat[]{\includegraphics[scale=0.25]{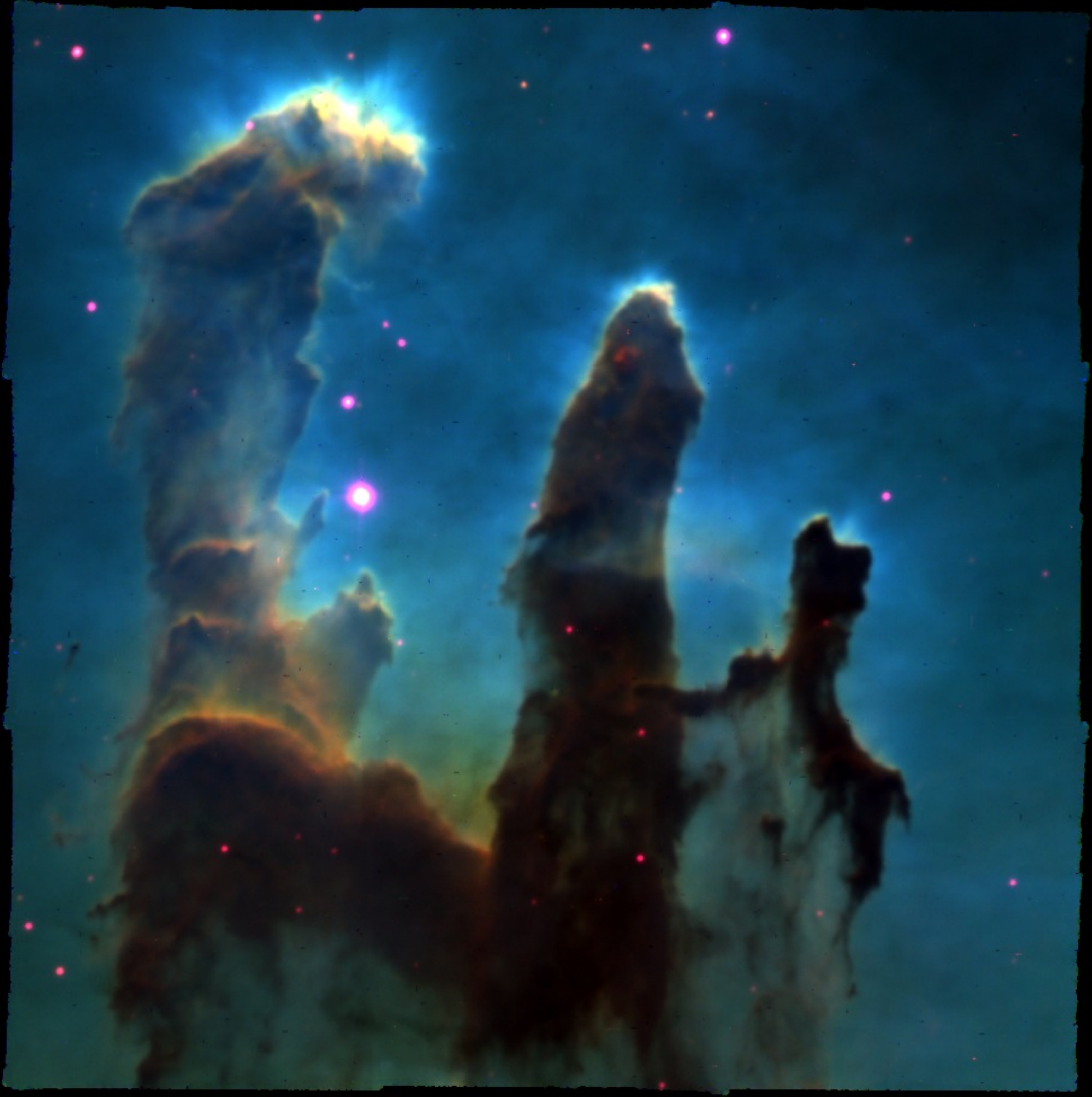}}}
\mbox{
\subfloat[]{\includegraphics[scale=0.25]{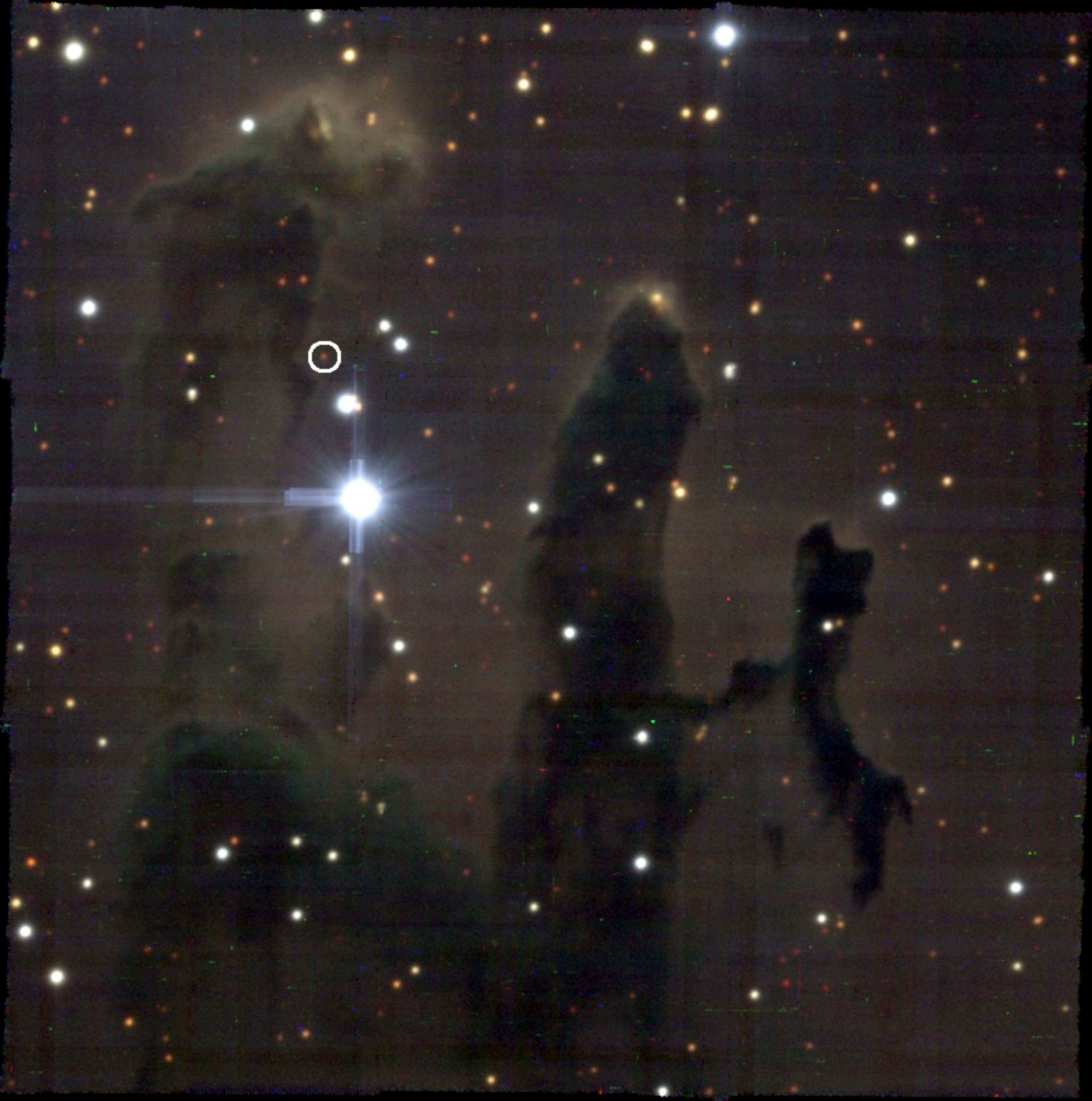}}}
  \caption{RGB composite (panel a): red is [SII]$\lambda$6717, green is H$\alpha$, and blue is [OIII]$\lambda$5007. Continuum RGB composite (panel b, red 8050 - 8250 \AA, green 6850 - 7000 \AA, blue 5200 - 5400 \AA ). The tips of P1 and P2 are to be seen as reflection nebulae illuminated by the nearby stars. The white circle in panel b indicates the continuum source discussed in Section 4.1.}
  \label{30sRGB}
\end{figure}

The 54 exposures were reduced using the standard recipes of the MUSE pipeline (v0.18.2) and the required calibrations. To make the integrated line maps of the entire mosaic discussed in Section 3.1, the integrated line maps of the single pointings were combined using Montage. After this step of making the mosaic a fault in the world coordinate system (WCS) of the single fields, resulting from the rotating dither pattern, manifested itself. This lead to the single fields being shifted and rotated with respect to one another, and an updating of the WCS of each field by comparing observed stellar coordinates with a catalog was required. This was done with the \textsc{iraf} tasks \texttt{MSCTPEAK} and \texttt{MSCSETWCS} of the \texttt{MSCRED} package.  

As a counterpart to the iconic and very well known HST 3-color composite by \cite{1996AJ....111.2349H} and the recently released 25th HST anniversary image, the RGB composite (where red is [SII]$\lambda$6717, green is H$\alpha$, and blue is [OIII]$\lambda$5007) of the 30 second exposure is shown in Fig. \ref{30sRGB}a. This RGB composite clearly shows the structure and evaporation of the Pillars and their tips which are being illuminated and ionised by the nearby cluster NGC 6611, and also clearly to be seen are the dust lanes at the lower end of the Pillars that point towards their natal molecular cloud, as well as several detached globules to the left of P1 (better appreciable in the digital version if this article).

Figure \ref{30sRGB}b shows a continuum 3-color composite of the region: the pillar tips (P1 and P2) are to be seen as reflection nebulae illuminated by the nearby stars. The star at the tip of P2 corresponds to a known T Tauri star (\citealt{2002ApJ...565L..25S}, \citealt{2007ApJ...666..321I}).

\subsection{Hydrodynamical simulations}
We compare our observations with three-dimensional radiation-hydrodynamical simulations performed with the smoothed-particle hydrodynamics (SPH) code \textsc{divine}. The code is able to model the plane-parallel irradiation by ionising photons of an arbitrary gas distribution, as described by \cite{2009MNRAS.393...21G}, and also includes the effects of the diffuse ionising field resulting from recombinations \citep{2011MNRAS.413..401E}.\\
\indent \cite{2011MNRAS.413..401E} simulated the photoionisation of a (4pc)$^{3}$ turbulent box initially containing $474$M$_{\odot}$ of neutral gas at 10K and with a turbulent RMS Mach number of 5. The box is illuminated with a Lyman continuum flux of 5$\times10^{9}$ photons cm$^{-2}$ s$^{-1}$ and allowed to evolve for $\approx$ 500 kyr, corresponding to $\approx$ 0.17 freefall times. The turbulent density field is sculpted by the photoevaporation flows and the pressure in the HII region into a complex morphology which includes three prominent pillar structures pointing roughly in the direction from which the photons are propagating. It is these structures which form the basis for our comparative work.\\

The SPH simulations and subsequent \textsc{mocassin} (described in Section 2.3) calculations do not specifically reproduce the pillars in M16, as they have smaller mass and the radiation field is not as strong as the one produced by NGC 6611 (see Section 3.3). The comparison we seek to do is more of a qualitative one which helps us to see whether the characteristic ionisation structure of the pillar tips is recovered in the simulations after running them through the radiative transfer calculation, ultimately testing our numerical computations.  

\subsection{Monte Carlo radiation-transport calculations}
Since the SPH simulations described above are only intended to reproduce the dynamical effects of photoevaporation and ionised gas pressure, they do not lend themselves to detailed comparison with observations. For this purpose, we post-processed the SPH simulation output with the 3D Monte Carlo photoionisation code \textsc{mocassin} \citep{2003MNRAS.340.1136E,2005MNRAS.362.1038E,2008ApJS..175..534E}. \textsc{mocassin} uses a Monte Carlo formalism, enabling the selfÑconsistent treatment of the direct and diffuse radiation field.\\
\indent We mapped a 1$\times$3$\times$1 pc cutout of the hydrodynamical simulation snapshot onto a uniform 128$\times$384$\times$128 pixel grid. We assumed typical HII region gas-phase abundances as follows: He/H = 0.1, C/H = 2.2$\times10^{-4}$, N/H = 4.0$\times10^{-5}$, O/H = 3.3$\times10^{-4}$, Ne/H = 5.0$\times10^{-5}$, S/H = 9.0$\times10^{-6}$. The gas is heated mainly by photoionisation of hydrogen and other abundant elements. Gas cooling is mainly via collisionally excited lines of heavy elements, although contributions from recombination lines, free-bound, free-free and two-photon continuum emission are also included. To compute the emission line flux from each cell, \textsc{mocassin} solves the statistical equilibrium problem for all atoms and ions, given the local temperature and ionisation state.\\
\indent These \textsc{mocassin} calculations were already used by \cite{2012MNRAS.420..141E} to create synthetic observations of the simulations of \cite{2011MNRAS.413..401E}. Here, for the first time, we make use of the very rich new dataset provided by MUSE to compare the hydrodynamic simulations in detail to the Pillars of Creation.\\

\section{Ionisation structure}
\subsection{Extinction correction}

\begin{figure*} 
\center
\hspace{-2cm}
\includegraphics[angle=90,height=22cm,width=17cm]{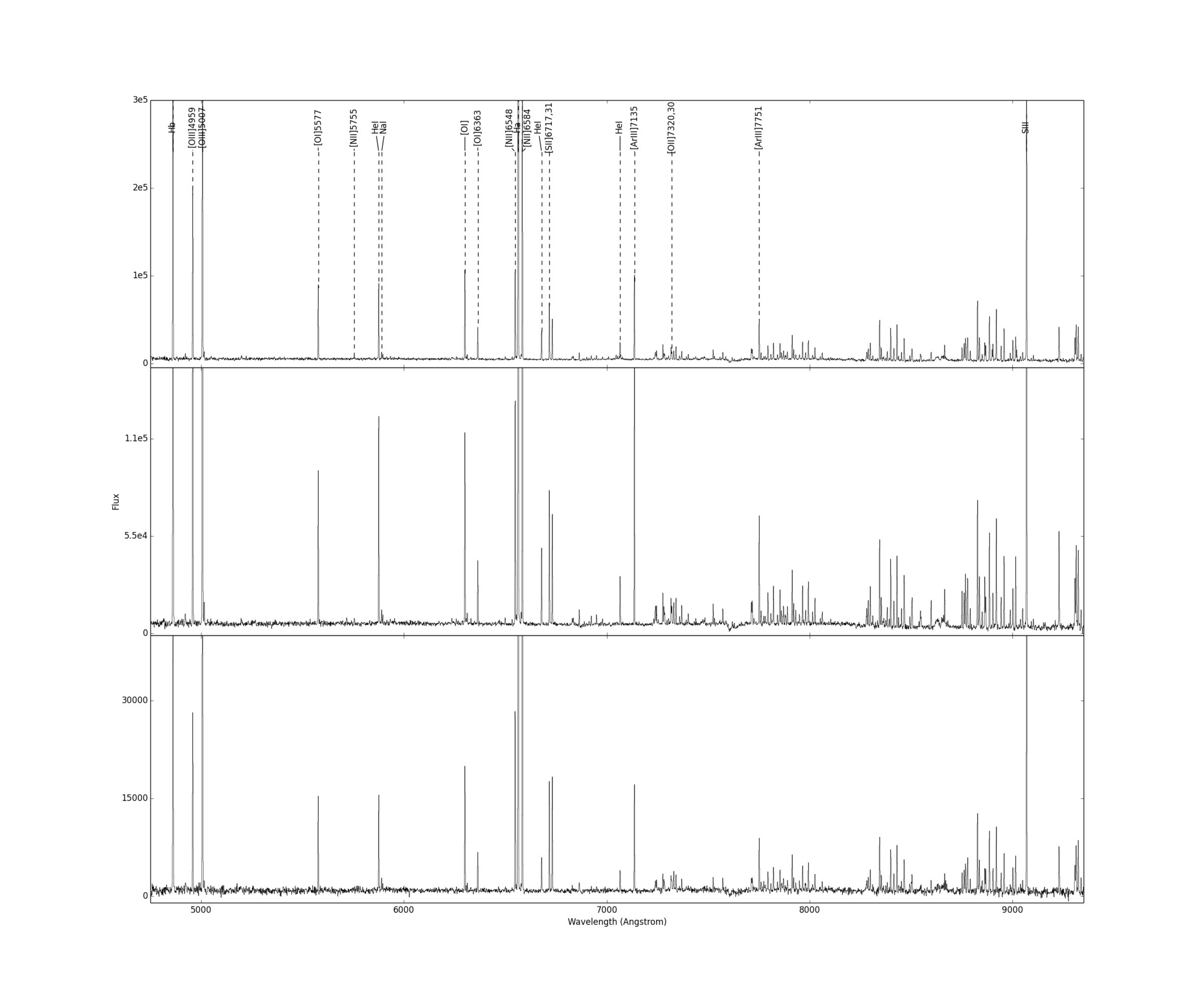}
\caption{Example spectra of the HII region (top panel), the pillar-ambient matter interface (middle panel) and the EGG (bottom panel) extracted from the positions marked in Fig. \ref{ratiomaps}a (see text Section 3.1). The flux is measured in 10$^{-20}$ ergs s$^{-1}$ cm$^{-2}$ \AA$^{-1}$ pixel$^{-1}$.}
\label{lineid}
\end{figure*}

The extinction toward M16 has been studied by many authors in the past and has been found to vary significantly over the members of NGC 6611. \cite{1990A&A...227..213C} found that for $\lambda <$ 5500 \AA\  the extinction law deviates from the standard galactic extinction law (R$_{V}$ = 3.1, \citealt{1979MNRAS.187P..73S}), this effect is caused by the dust grains being larger than those found in the standard interstellar medium \citep{2006AJ....132.1783O}. Indeed, \cite{1993AJ....106.1906H} find a typical value of R$_{V}\sim$ 3.75 for NGC 6611. Therefore, to correct for extinction along the line of sight toward M16 we computed the reddening coefficient C via the Balmer decrement method:

$$C=3.1\times \bigg(log\frac{F(H\alpha)}{F(H\beta)}-log\frac{I(H\alpha)}{I(H\beta)}\bigg)$$

assuming a reddening curve parametrised by \cite{1989ApJ...345..245C} with R$_{V}$ = 3.1 for $\lambda >$ 5500 \AA, and R$_{V}$ = 3.75 for shorter wavelengths. An intrinsic flux ratio of $I(H\alpha)/I(H\beta)=2.86$ (corresponding to a case B Balmer recombination decrement, \citealt{2006agna.book.....O}) was adopted, and the measured $F(H\alpha)/F(H\beta)$ flux ratio used to compute C was taken to be the mean value obtained from the map of that particular line ratio, thus obtaining $\langle F(H\alpha)/F(H\beta) \rangle \cong 5.93$ and therefore C $\cong$ 0.98. We find extinction values between $\sim$ 1.5 and $\sim$ 3.2 mag, which agree with literature values \citep{2010A&A...521A..61G}. Fig. \ref{lineid} shows co-added spectra of the HII region (top panel), the pillar-ambient interface (the pillar tips, middle panel) and an EGG (bottom panel). The prominent features in the 8000 \AA\ to 9000 \AA\ range are OH sky lines that have not been removed from the spectra). Table \ref{lines} lists the detected emission lines together with their observed wavelength (column 2), the adopted reddening curve (column 4) and the flux ratio prior and after correcting for extinction (columns 3 and 5) for the example spectrum of the HII region shown in Fig. \ref{lineid}. The integrated line intensity maps were de-reddened according to the appropriate reddening curve value for each wavelength. The [OII]$\lambda$7320,30 as well as the [NII]$\lambda$5755 have intensities (relative to H$\beta$) $<$ 0.07 and were not included in the fit used to derive the parameters in Table \ref{lines}.

\begin{figure*}
\mbox{
\subfloat[]{\includegraphics[scale=0.3]{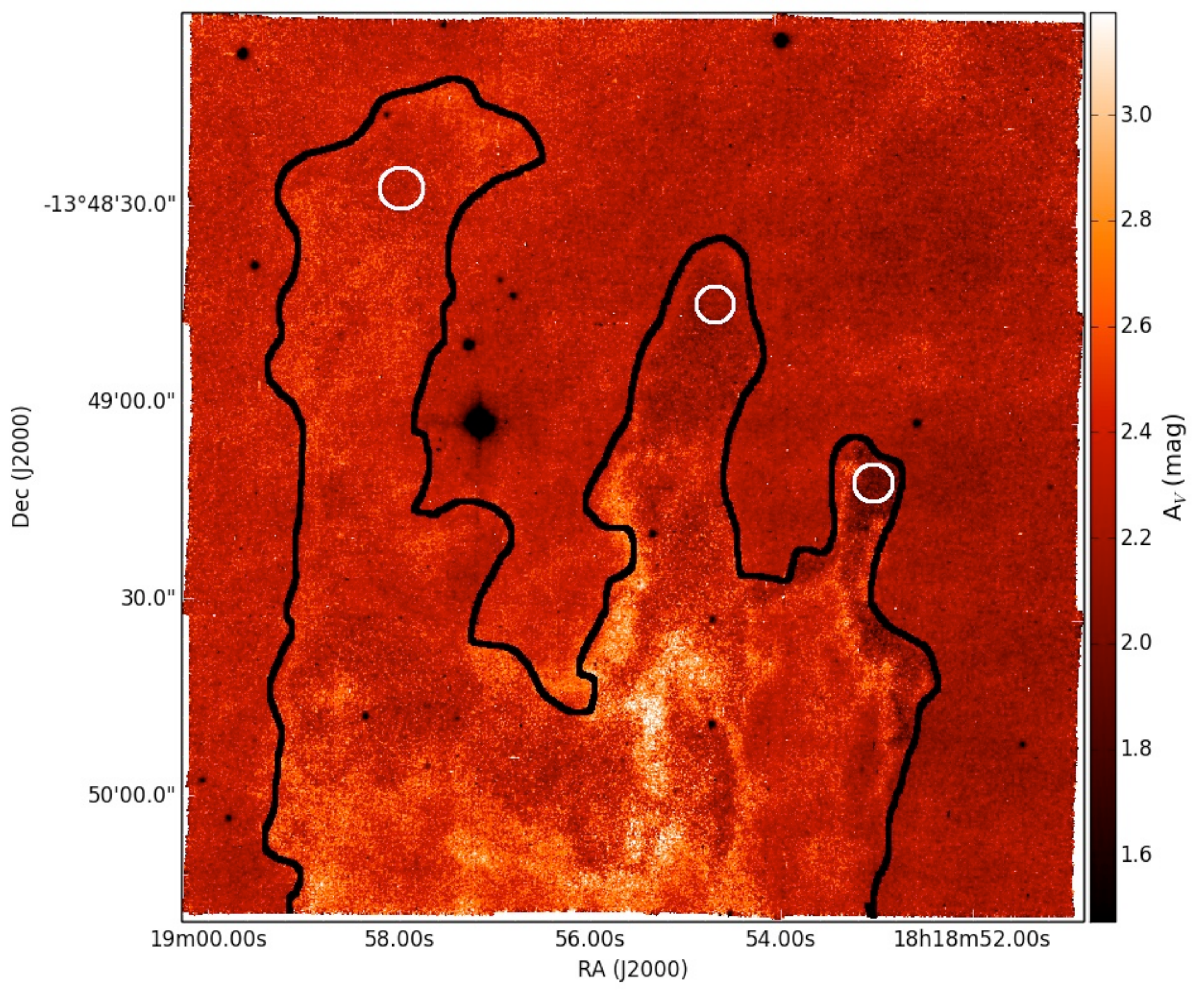}}
\subfloat[]{\includegraphics[scale=0.48]{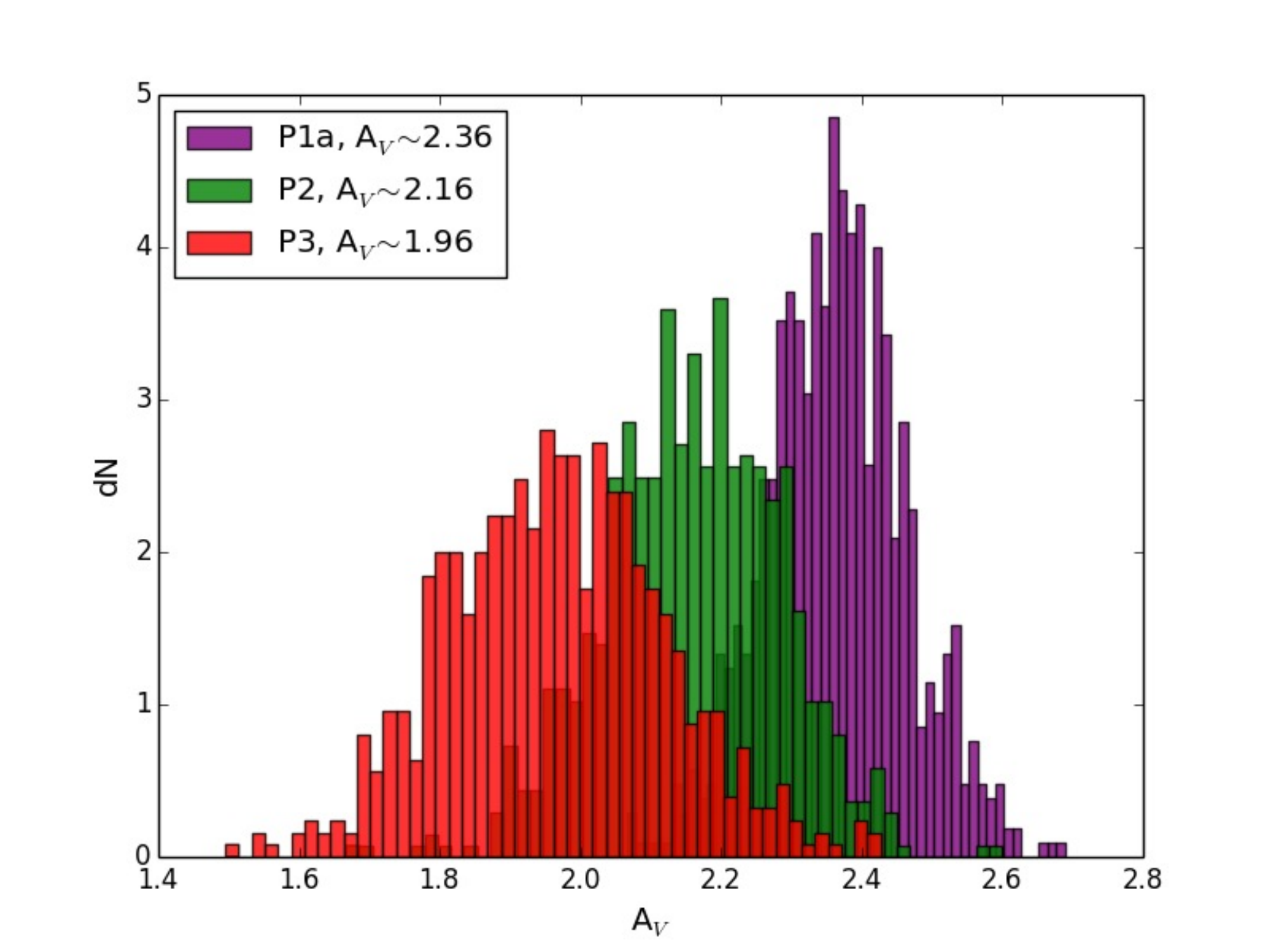}}}
  \caption{Left: extinction map in terms of A$_{V}$ (linear scale to maximum/minimum). To help the reader identify the location of the Pillars has been marked with a schematic black contour. The white circles in panel (a) correspond to the regions used to quantify the extinction for the three Pillars (panel b and text Section 3.1). Right: normalised extinction histograms of circular regions (r $\sim$ 3 arcsec) at the tip of the three pillars.}
  \label{hahb}
\end{figure*}

A clear gradient can be seen in the extinction map shown in Fig. \ref{hahb}a where P1a shows a higher overall extinction, P3 the lowest, and P2 values in between. This is quantified in Fig. \ref{hahb}b, where the normalised histograms of circular regions (r $\sim$ 3 arcsec, white circles in panel a) in the three pillar tips are shown together with the best fit values to each histogram. We therefore believe that P3 is the pillar closest along the line of sight, behind that lies P2, and behind P2 lies P1. A description of the 3D geometry of the Pillars is given in Section 4.

\begin{table*}
\begin{center}
\caption{Detected emission lines (column 1) and line centroids obtained via gaussian line fitting (column 2), line fluxes relative to the observed H$\beta$ flux prior (column 3) and after (column 5) correcting for extinction, and the reddening curve for each line (column 4).}
\begin{tabular}{lcccc}
\hline
\hline
Line & $\lambda_{obs}$ & F($\lambda$)/F(H$\beta$) & f($\lambda$) & F($\lambda$)/F(H$\beta$)$_{corr}$ \\
\hline
4861 H$\beta$ & 4861.42 & 1.0$\pm$0.019 & 0.0 & 1.000\\
4959 [OIII] & 4959.01 & 0.433$\pm$0.015 & -0.025 & 0.419\\
5007 [OIII] & 5006.93 & 1.36$\pm$0.023 & -0.036 & 1.297\\
5577 [OI] &  5577.42 & 0.298$\pm$0.014 & -0.153 & 0.244\\
5876 HeI & 5875.76 & 0.233$\pm$0.015 & -0.203 & 0.179\\
6300 [OI] & 6300.39 & 0.382$\pm$0.015 & -0.267 & 0.269\\
6363 [OI] & 6363.86 & 0.121$\pm$0.014 & -0.276 & 0.084\\
6548 [NII] & 6548.12 & 0.312$\pm$0.015 & -0.303 & 0.21\\
6563 H$\alpha$ & 6562.89 & 6.476$\pm$0.088 & -0.305 & 4.343\\
6584 [NII] & 6583.5 & 0.975$\pm$0.019 & -0.308 & 0.552\\
6678 HeI & 6678.24 & 0.083$\pm$0.015 & -0.322 & 0.055\\
6717 [SII] & 6716.56 & 0.197$\pm$0.015 & -0.327 & 0.129\\
6731 [SII] & 6730.94 & 0.144$\pm$0.015 & -0.329 & 0.093\\
7065 HeI & 7065.37 & 0.047$\pm$0.015 & -0.377 & 0.029\\
7135 [ArIII] & 7135.89 & 0.231$\pm$0.015 & -0.387 & 0.139\\
7751 [ArIII] & 7751.19 & 0.129$\pm$0.014 & -0.475 & 0.069\\
9068 [SIII] & 9068.69 & 0.521$\pm$0.015 & -0.531 & 0.272\\
\hline
\label{lines}
\end{tabular}
\end{center}
\end{table*}

\subsection{Integrated line maps}
The extinction-corrected integrated line maps were created using the moment map subroutine of the \textsc{python} package \textsc{spectral$\_$cube}\footnote{spectral-cube.readthedocs.org}. All line and line ratio maps shown in this paper (comprised the appendix) are linearly scaled to minimum/maximum, please refer to the online version for Appendix figures. The H$\alpha$, [NII]$\lambda$6548, [SII]$\lambda$6717,31 and [OIII]$\lambda$5007 maps are shown in Fig. \ref{intensitymap}, the maps of the other detected lines are shown in Appendix A, available as supplementary online material. H96 discussed the normal-oriented striations coming from the pillar-ambient interface in the context of photo-evaporation: as the strong ionising radiation from NGC 6611 hits the dense molecular material of the pillars, the matter is ionised and the pressure at the pillar-ambient matter interface is increased causing matter to flow away from the latter, creating a photo-ionised flow. The fact that the evaporation is normal to the pillar-ambient interface is an indication that neither the interaction between the flow and the ambient matter, nor the interaction of the impinging photons with the outflowing material are able to influence the direction of the photo evaporative flow (see H96). The photo-evaporative striations can be clearly seen in the H$\alpha$ maps (and slightly in the [SIII]$\lambda$9068 and [ArIII]$\lambda$7135 maps). The emission from the other species is either very diffuse ([OIII], H$\beta$) or very localised (rest).

The strongest emission is the H$\alpha$ emission coming from the tip of P1. In general we identify two types of emission: a more or less localised line emission that traces the tips of the pillars as well as the protrusions on the inner side of P1 can be seen in the H$\alpha$, [ArIII], HeI, [NII], [OI], [OII], [SII], [SIII] maps, and a more diffuse emission that surrounds the pillars in a halo-like manner ([OIII] and H$\beta$). In the first type the emission peak of some lines is sharper and more localised than others, e.g. [SII] compared to H$\alpha$, and we furthermore see a spatial shift in the peak emission of different lines corresponding to a stratified ionisation structure which correlates with the ionisation energy of the identified species. This stratification was already discussed in H96, who modelled the scenario with CLOUDY84 (see \citealt{2013RMxAA..49..137F} for the latest release of the code) and found a very good agreement between the model and the data. While these authors were limited to three emission lines (H$\alpha$, [SII] and [OIII]), MUSE now offers the possibility of investigating this stratified ionisation structure in great detail across almost the entire optical regime and thus covering almost all of the optical ionisation-tracing emission lines. Fig. \ref{intprof} shows the intensity profiles of the emitting line species listed in Table \ref{lines} for the three pillars along the cuts marked in Fig. \ref{museptg} as a function of distance from the top of the image (wherever two lines of the same ionised state and atom were present these were averaged): the stratified ionisation structure is clearly seen from the fact that the lines from the ions/atoms with higher ionisation energy peak first, followed by the ones with lower and lower ionisation energy. The locations of the peaks with respect to the [OIII] peak are listed in Table \ref{peaks}: it is clear that MUSE is not able to spatially resolve the ionisation front, as for some lines (especially in P2 and P3 where the emission is weaker compared to P1) it is impossible to spatially separate the maxima.

\begin{table*}
\begin{center}
\caption{Location of the peaks (in 10$^{-2}$ pc) of the lines seen in Fig. \ref{intprof}a, \ref{intprof}b and \ref{intprof}c with respect to the position of the [OIII] peak (E$_{ion}$ = 35.12 eV) and their ionisation energy (see text Section 3.2).}
\begin{tabular}{lcccccccc}
\hline
\hline
 & [ArIII] & HeI & [SIII] & H$\alpha,\beta$ & [NII] & [SII] & [OII] & [OI] \\
\hline
P1  & 0.296 & 0.739 & 1.775 & 1.923 & 2.218 & 2.218 & 2.218 & 2.514 \\
\hline
P2 & 0.296 & 0.296 & 0.444 & 0.444 & 0.592 & 0.739 & 0.739 & 1.627 \\
\hline
P3 & 0.296 & 0.296 & 0.296 & 0.296 & 0.592 & 0.592 & 0.592 & 0.887 \\
\hline
E$_{ion}$ (eV) & 27.63 & 24.59 & 23.23 & 13.59 & 14.53 & 10.36 & 13.62 & - \\
\hline
\label{peaks}
\end{tabular}
\end{center}
\end{table*}

Fig. \ref{intprof_sim} shows the line intensity profiles obtained from our simulations, and although we cannot resolve the separation between the H$\alpha$, [SII] and [NII] peaks, the shape and trend of the profiles is in good agreement. The distance between the [OIII] maximum and the maxima of the other three lines is about d $\approx$ 0.015 pc, which agrees with what we find from the observations of P1 where d$_{H\alpha,[NII]}\approx$ 0.019 and d$_{[SII]}\approx$ 0.022. Our models therefore quantitatively predict the ionisation structure of pillar-like objects.

\begin{figure*}
\mbox{
  \subfloat[]{\includegraphics[scale=0.39]{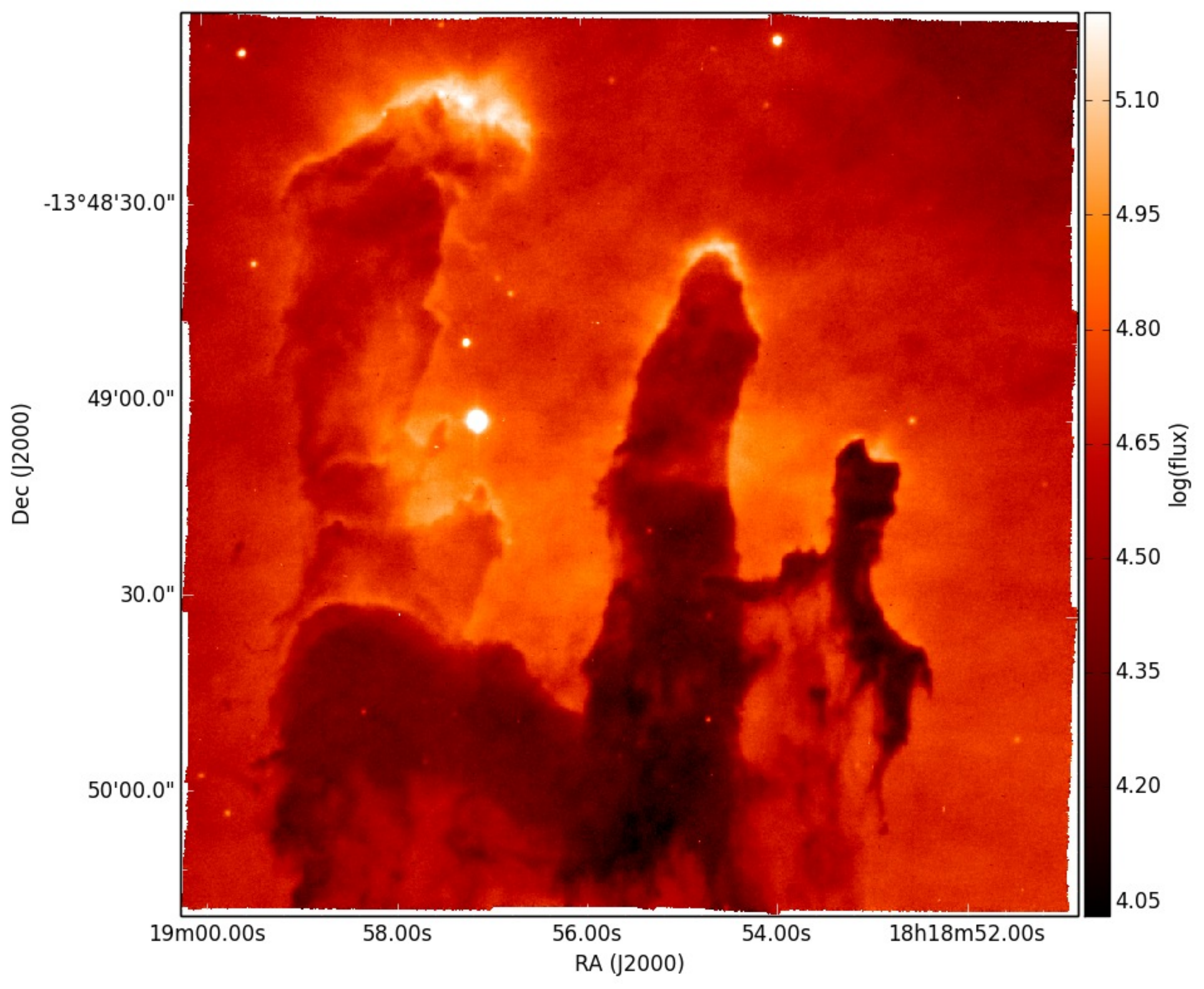}}
  \subfloat[]{\includegraphics[scale=0.39]{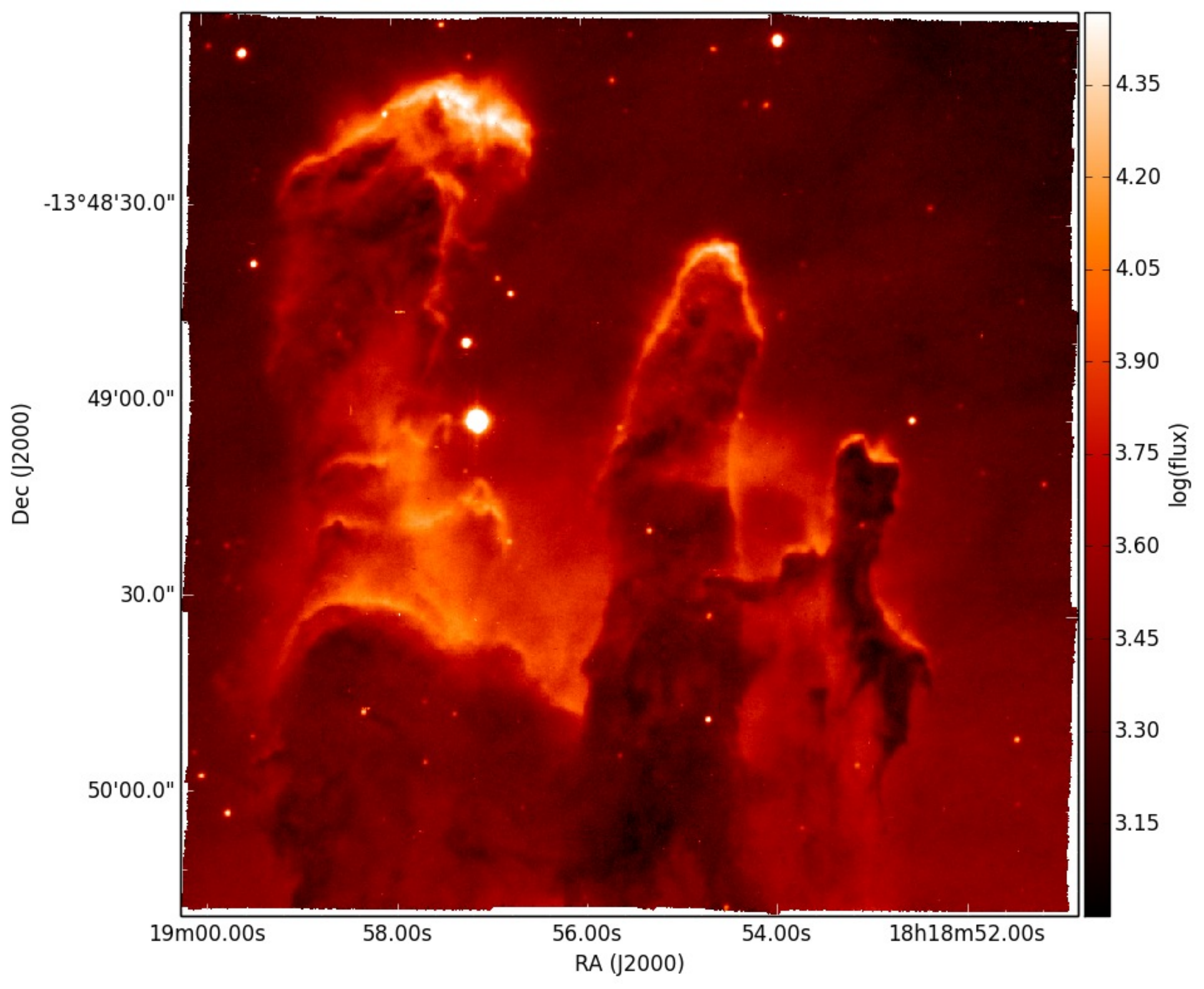}}}
  \mbox{
  \subfloat[]{\includegraphics[scale=0.39]{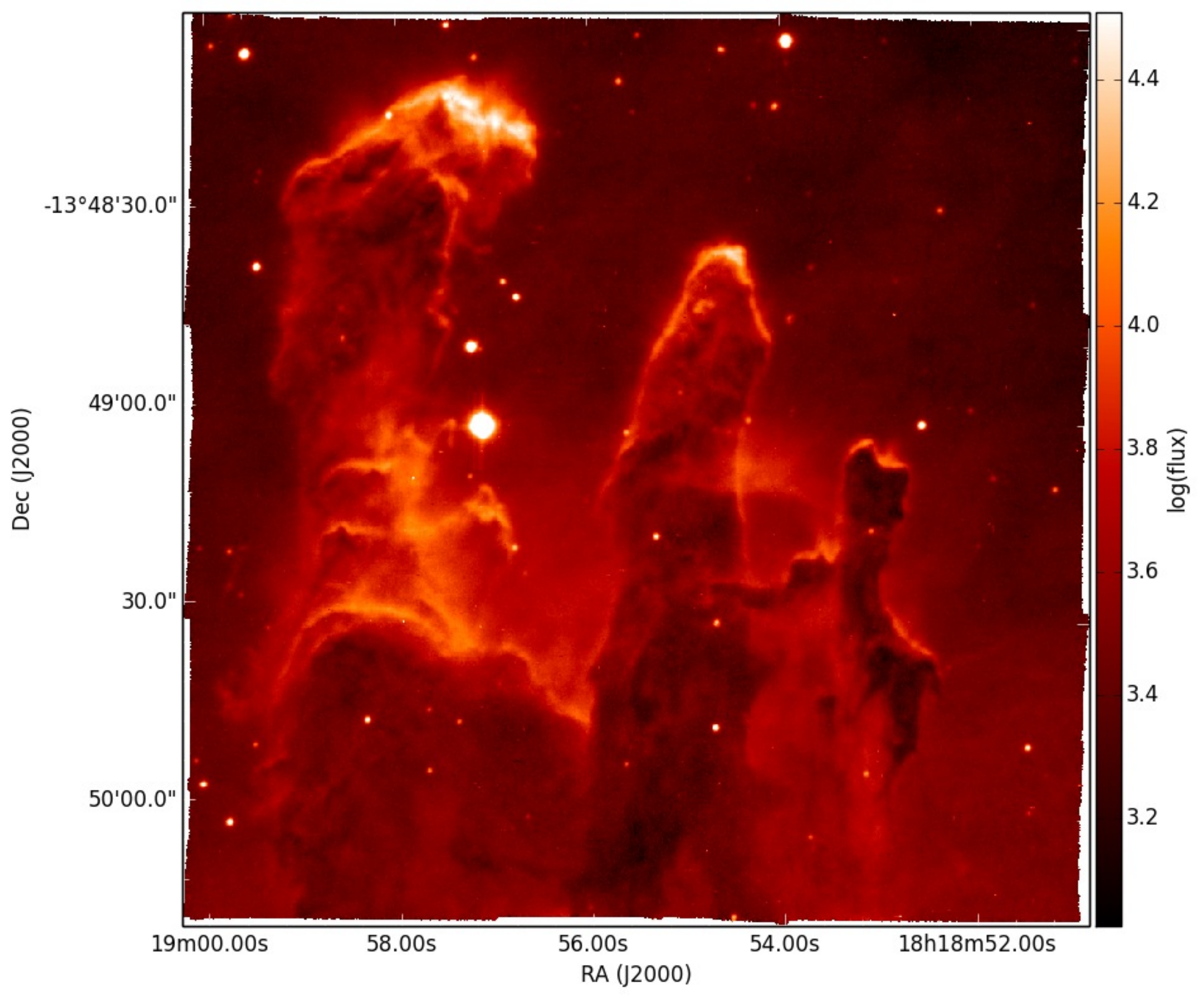}}
  \subfloat[]{\includegraphics[scale=0.39]{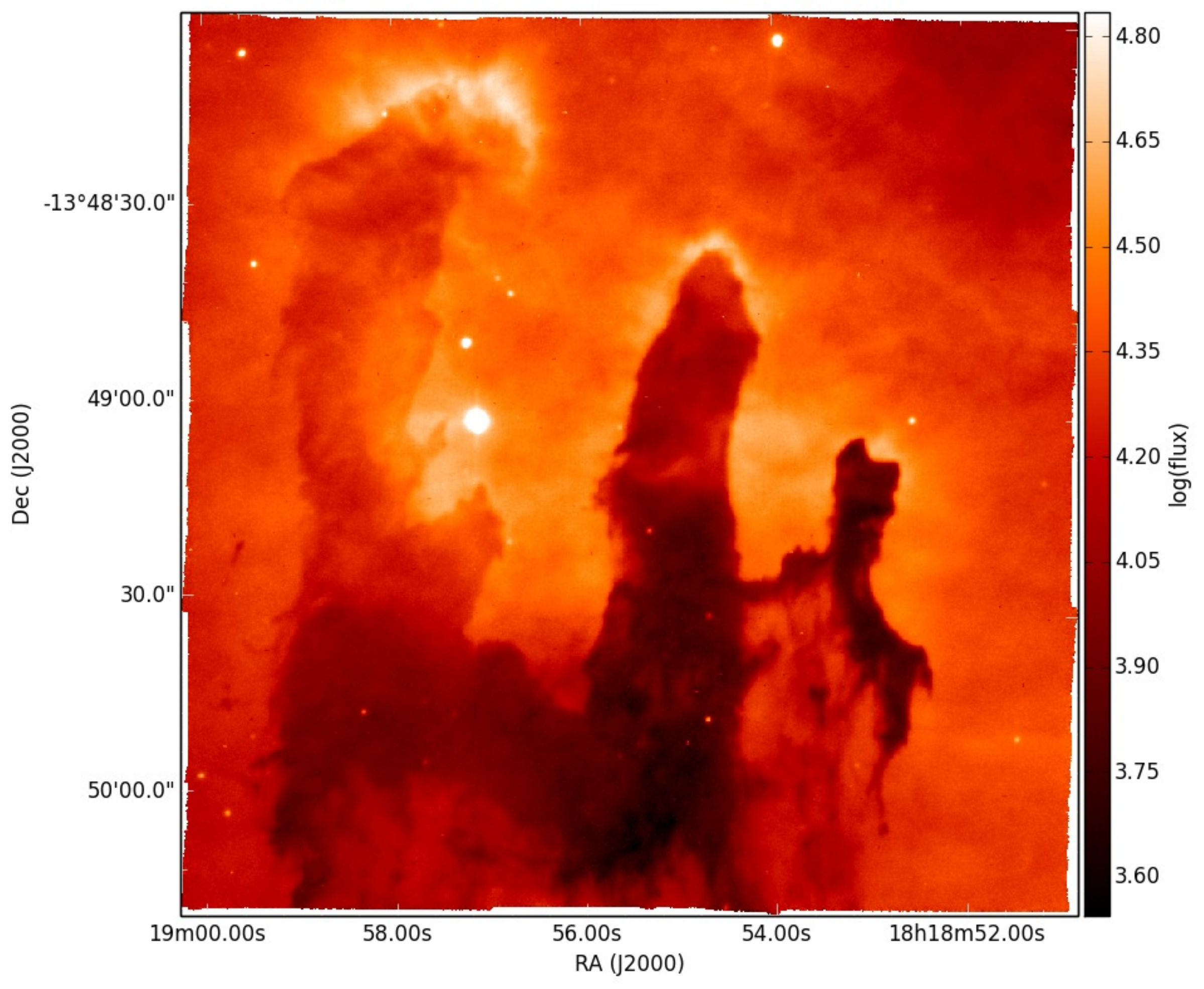}}}
  \caption{Integrated line maps derived from the MUSE data: H$\alpha$ (a), [NII]$\lambda$6548 (b), [SII]$\lambda$6717,31 (c) and [OIII]$\lambda$5007 (d). The flux scale bar is in $10^{-20}$ erg s$^{-1}$ cm$^{-2}$ pixel$^{-1}$, all maps are are linearly scaled to minimum/maximum. See Section 3.2 for discussion.}
  \label{intensitymap}
\end{figure*}

\begin{figure*}
\mbox{
\subfloat[]{\includegraphics[scale=0.48]{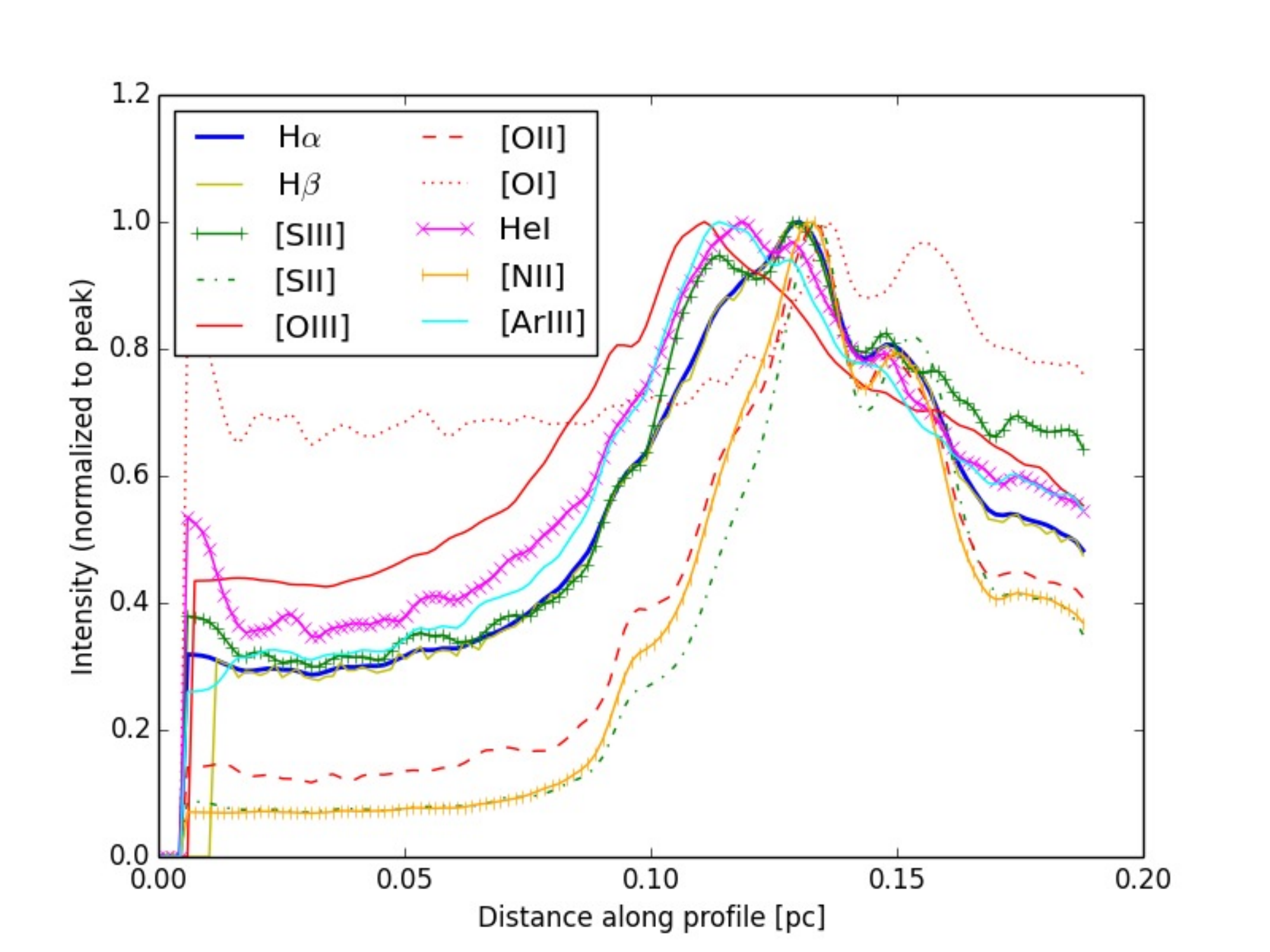}}
\subfloat[]{\includegraphics[scale=0.48]{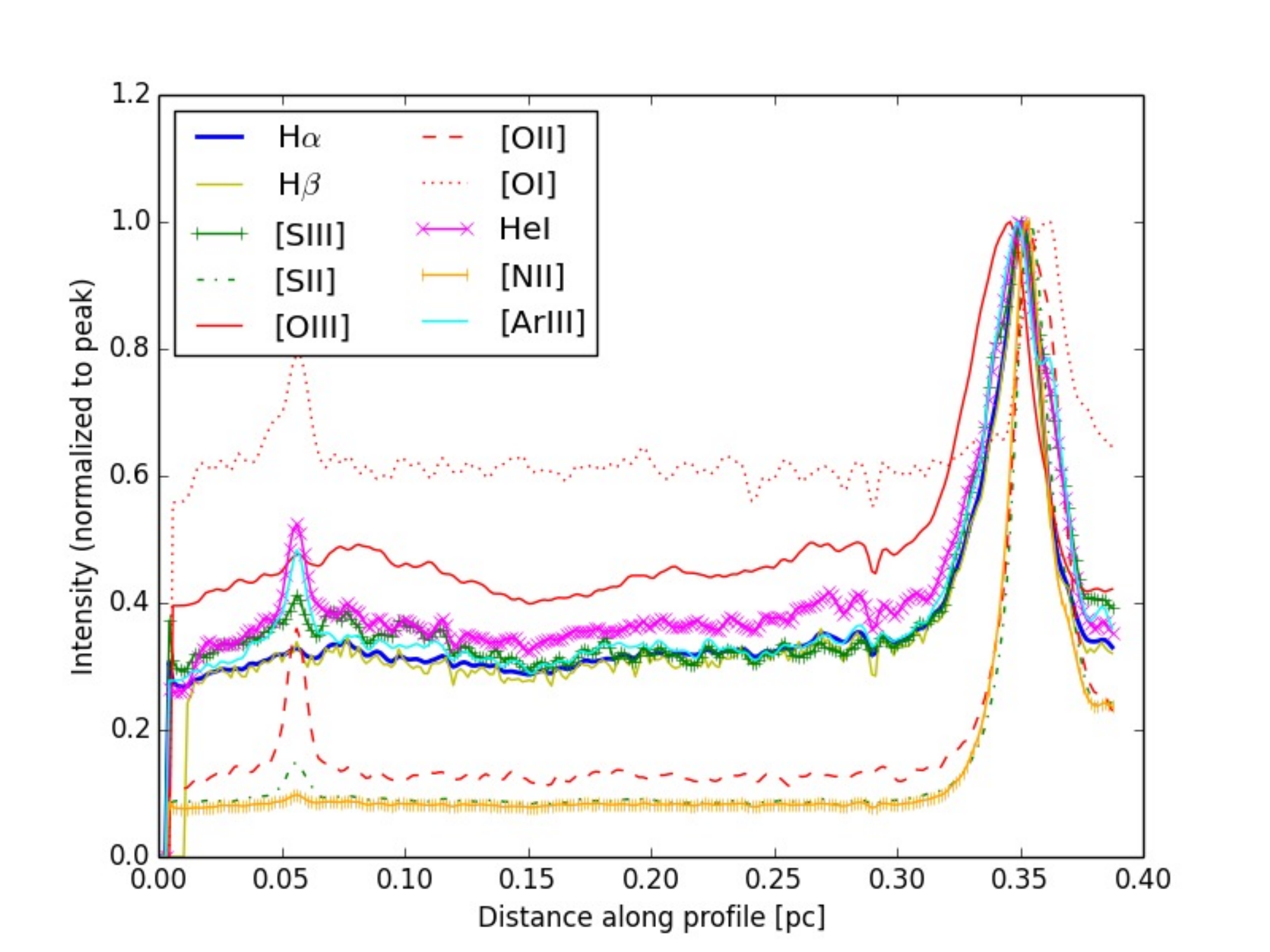}}}
\mbox{
\subfloat[]{\includegraphics[scale=0.48]{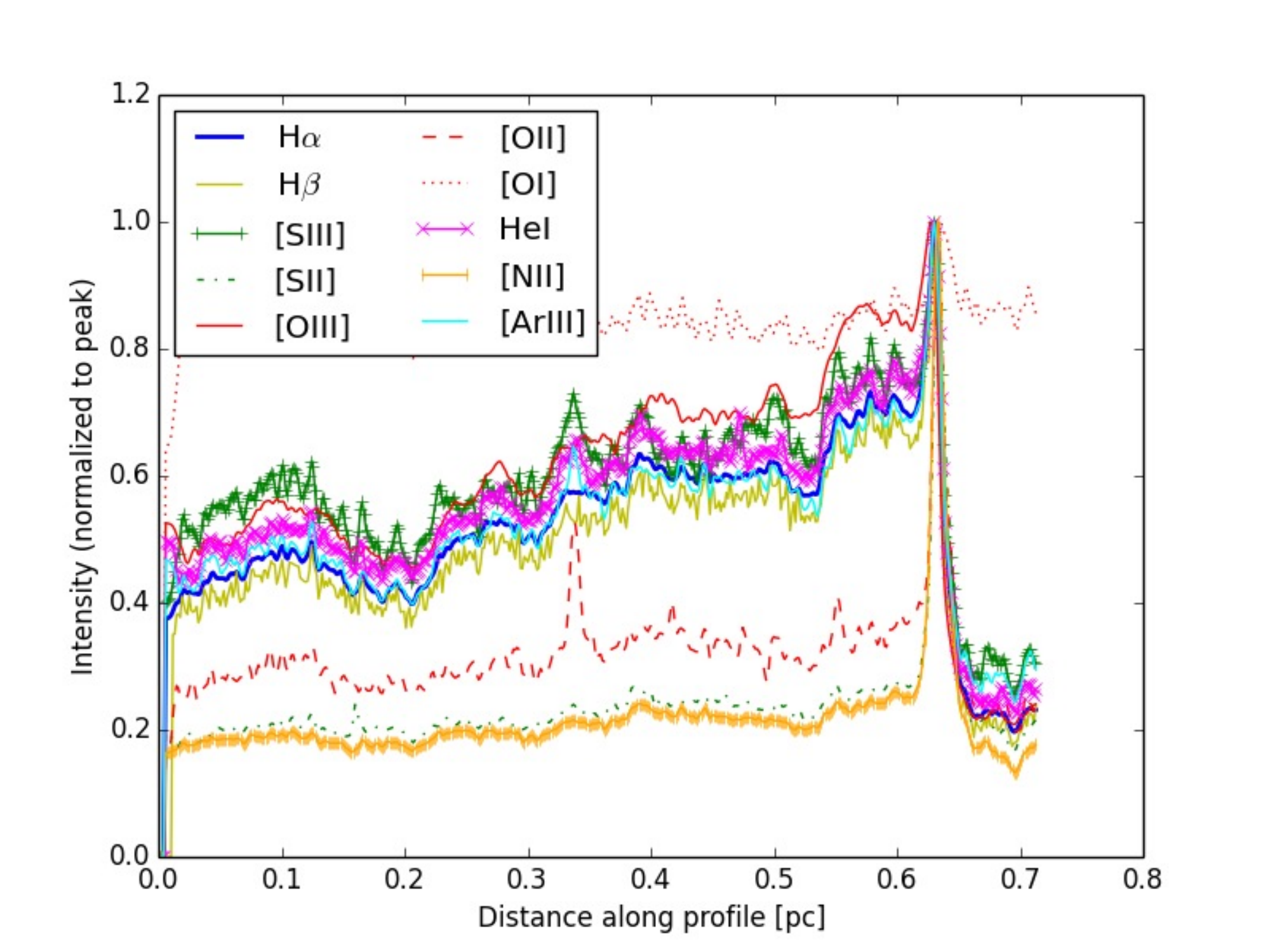}}
\subfloat[]{\includegraphics[scale=0.48]{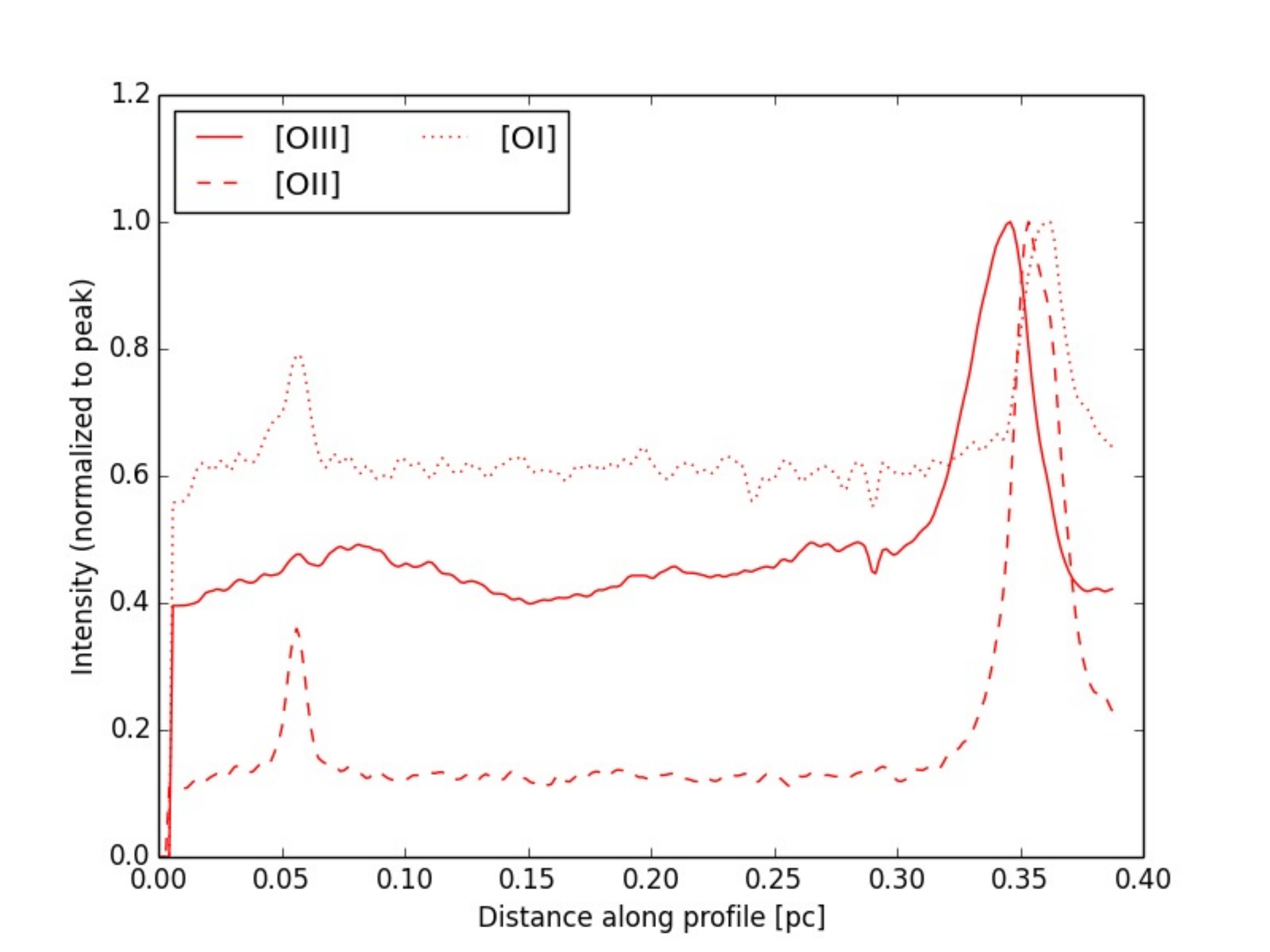}}}
  \caption{Line intensity profiles for the emission lines listed in Table \ref{lines} for P1, P2 and P3 (panels a, b, and c respectively, see text Section 3.2 for discussion). The lines are the same in the first three panels: H$\alpha$ (blue), H$\beta$ (yellow), [SIII]$\lambda$9068 (solid green), [SII]$\lambda$6717,31 (dashed green), [OIII]$\lambda$4969,5007 (solid red), [OII]$\lambda$7320,30 (dashed red), [OI]$\lambda$5577,6300 (dotted red), HeI $\lambda$5876,6678 (magenta), [ArIII]$\lambda$7135,7751 (cyan), [NII]$\lambda$6548,6584 (orange). Panel d shows the oxygen lines only for P2: [OIII] (solid line), [OII] (dashed line) and [OI] (dotted line). The profiles in these figures only cover the pillar tips, the do not follow the whole length of the slits shown in Fig. \ref{museptg}.}
 \label{intprof}
\end{figure*}

\begin{figure*}
\mbox{
\subfloat[]{\includegraphics[scale=0.5]{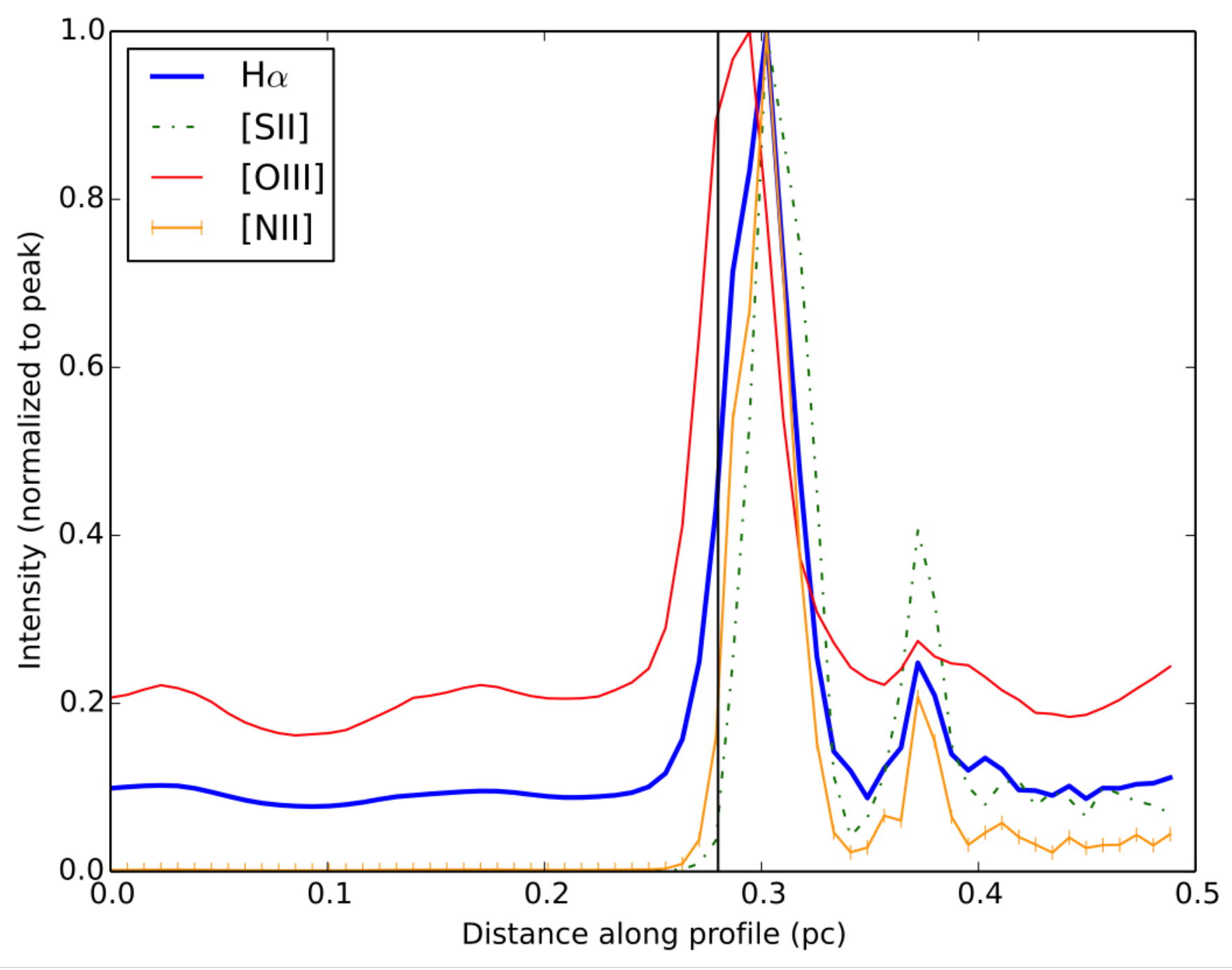}}
\subfloat[]{\includegraphics[scale=0.5]{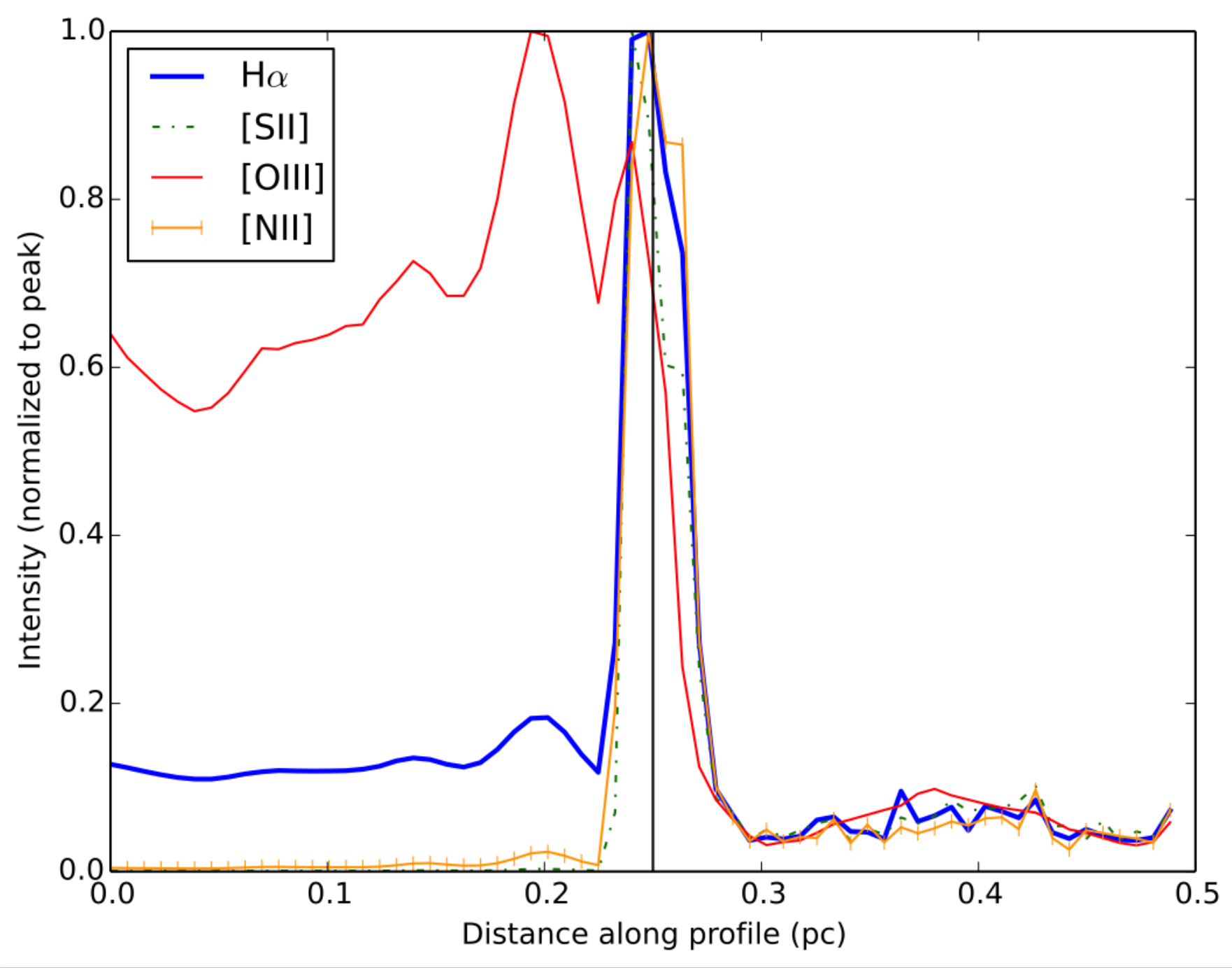}}}
\mbox{
\subfloat[]{\includegraphics[scale=0.5]{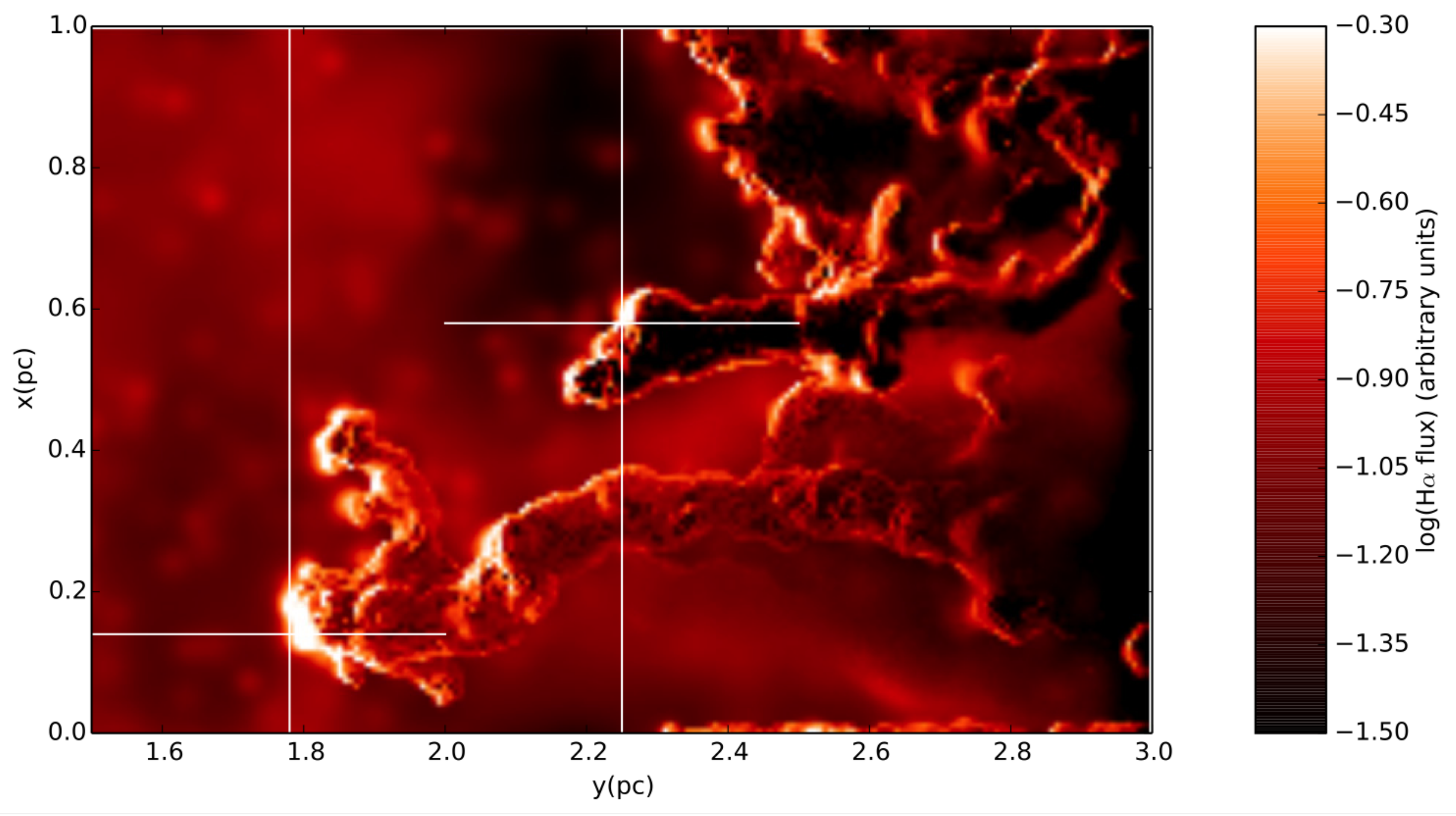}}}
  \caption{Line intensity profiles for the emission lines obtained from our simulations (panels a and b) along the (horizontal) slits shown in the H$\alpha$ intensity map (panel c). The colour-coding is the same as in Fig. \ref{intprof}. See text Section 3.2 for discussion.}
 \label{intprof_sim}
\end{figure*}

\subsection{Electron temperature and density maps}
The electron density was estimated via the ratio of the [SII]$\lambda$6717 and [SII]$\lambda$6731 lines. The sulfur ion has a single ground level, and two first excited levels are energetically very close and their ratio is thus sensitive to density but not temperature. We used the analytical solution given in \cite{1984MNRAS.208..253M} for a three-level atom:

\begin{equation}
\frac{[SII]\lambda6717}{[SII]\lambda6731} \equiv R_{[SII]}\simeq 1.49\frac{1+3.77x}{1+12.8x}
\label{ratio}
\end{equation}
 
 where $x=0.01N_{e}/T^{1/2}$ and assuming that $T=10^{4}$ K, so that the electron density $N_{e}$ in cm$^{-3}$ is given by

\begin{equation}
N_{e} = \frac{R_{[SII]}-1.49}{5.6713-12.8R_{[SII]}}\times10^{4}
\label{ne}
\end{equation}

\begin{figure*}
\mbox{
  \subfloat[]{\includegraphics[scale=0.38]{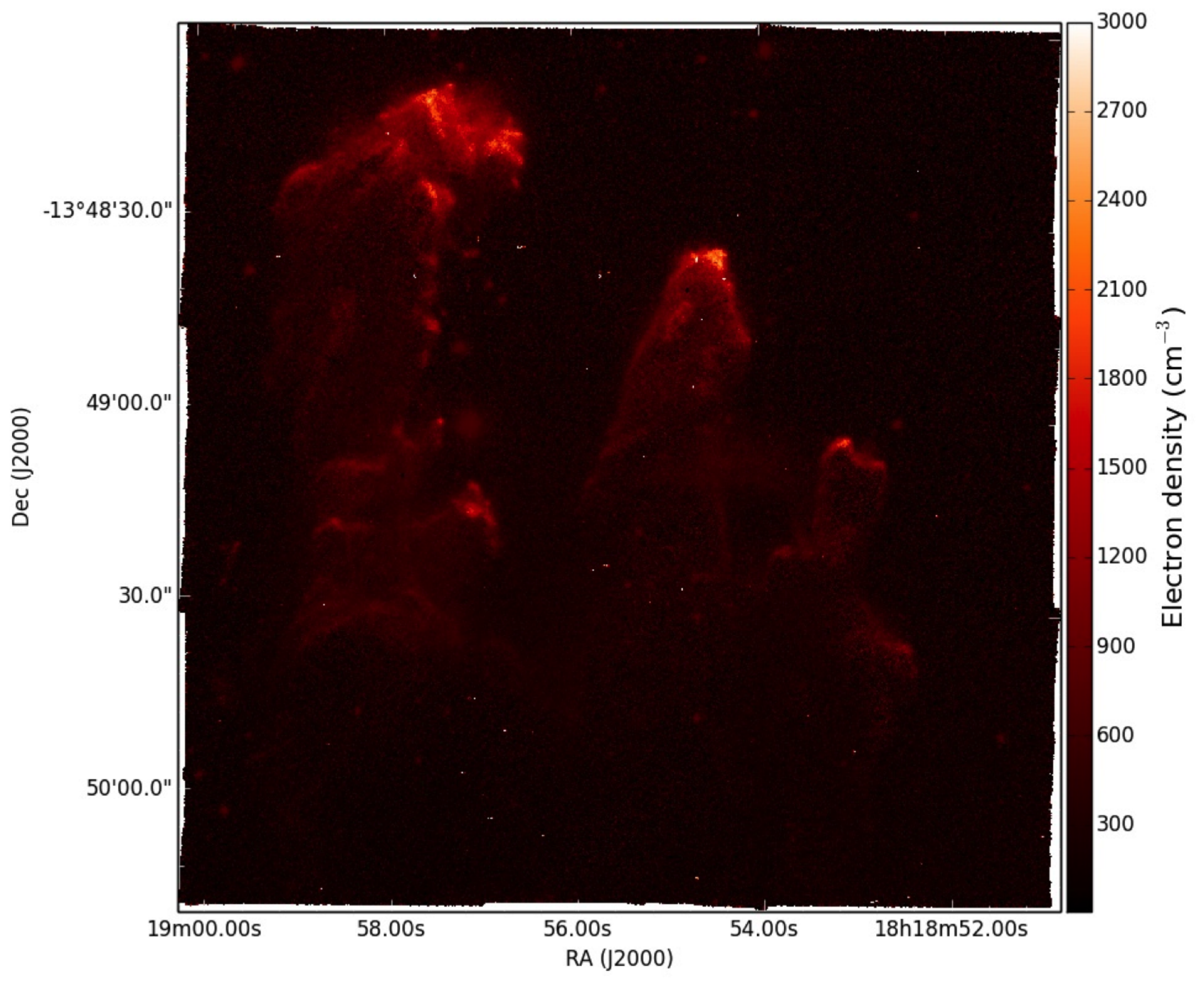}}
  \subfloat[]{\includegraphics[scale=0.38]{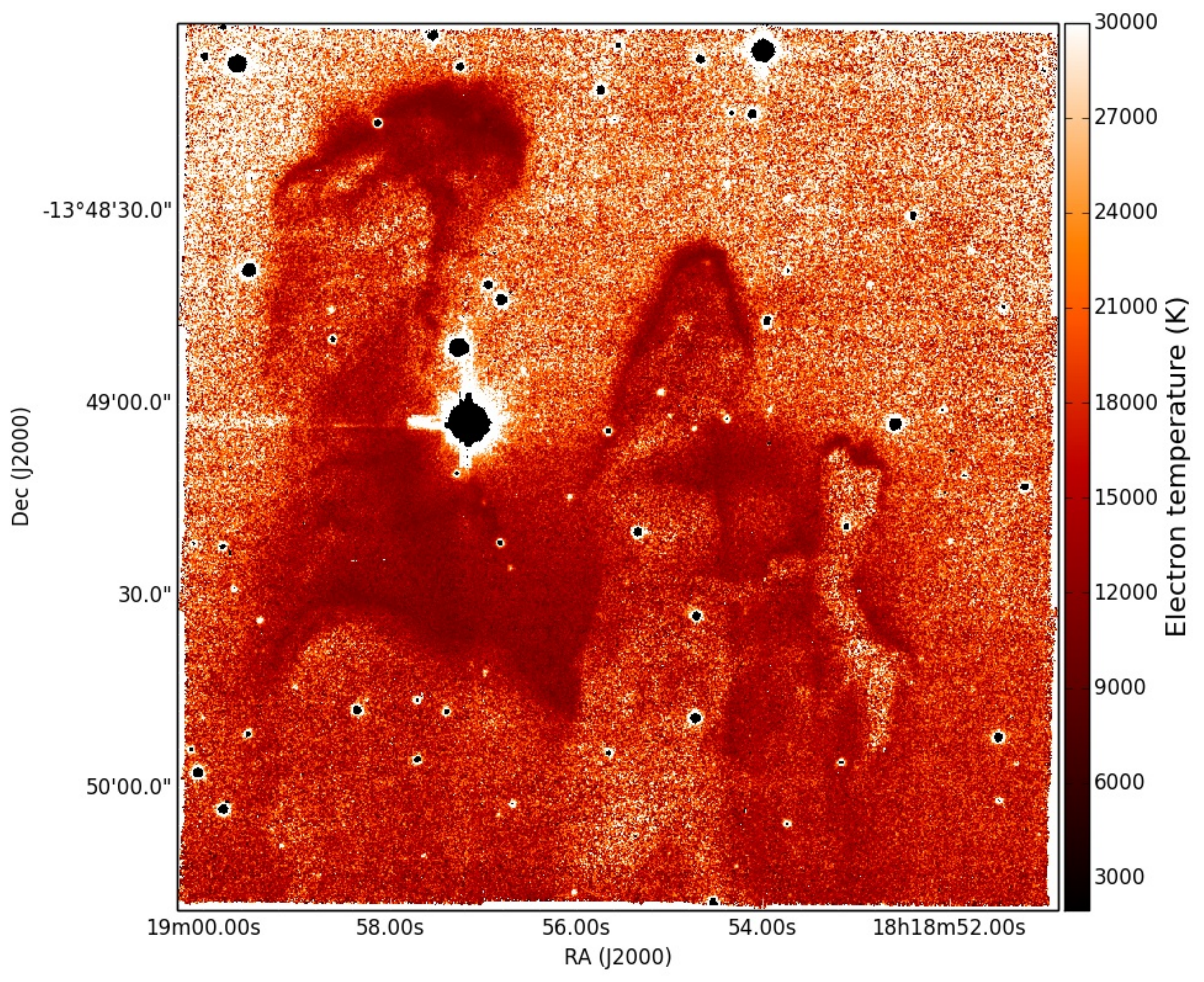}}}
  \mbox{
  \subfloat[]{\includegraphics[scale=0.55]{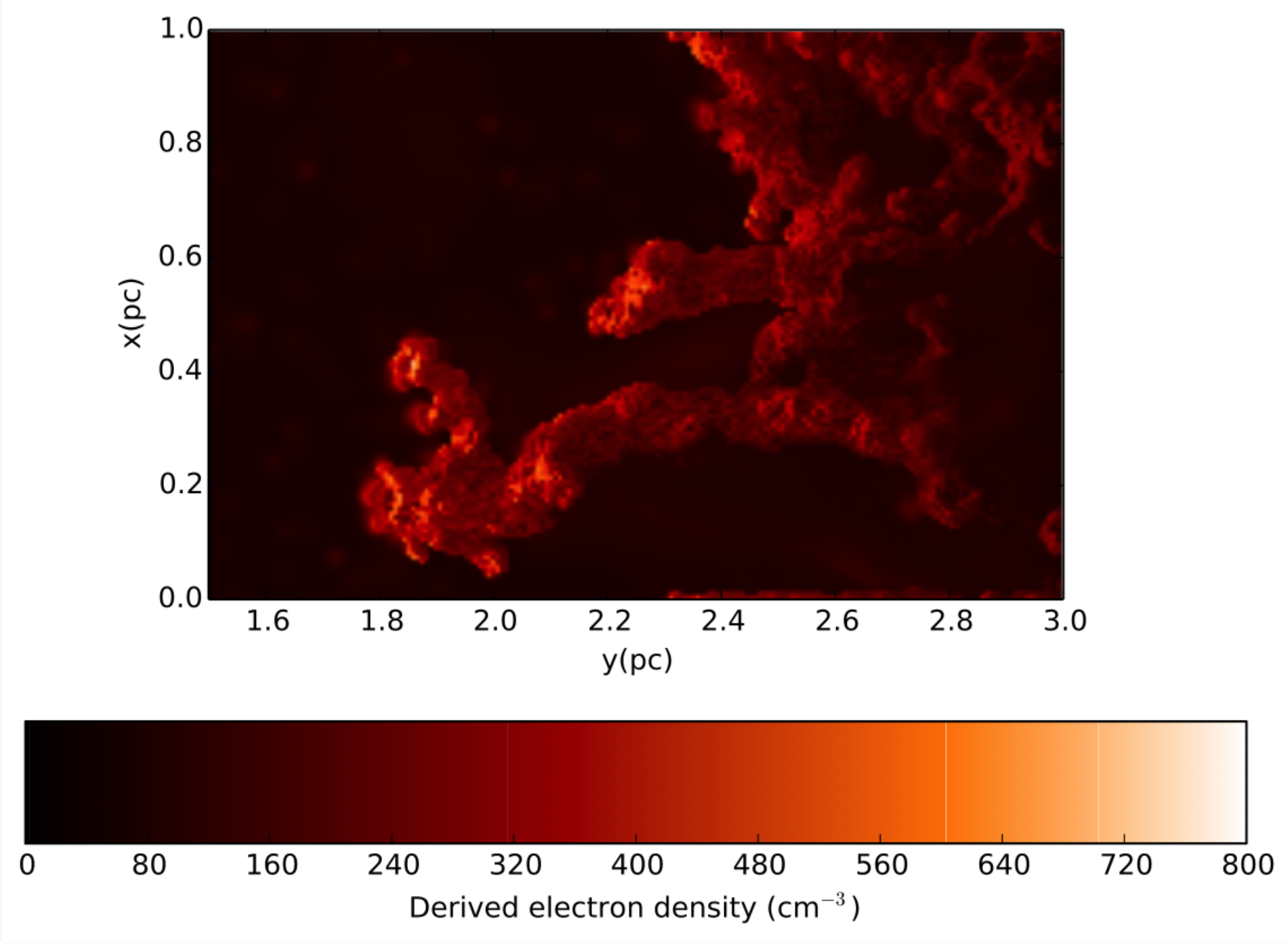}}}
  \caption{The electron density map derived via Eq. \ref{ne} (a), the electron temperature derived via Eq. \ref{te} (b), and the electron density of the simulation computed with the sulfur line ratio (c) (see text Section 3.3).}
  \label{NT}
\end{figure*}

\begin{figure*}
\mbox{
\subfloat[]{\includegraphics[scale=0.55]{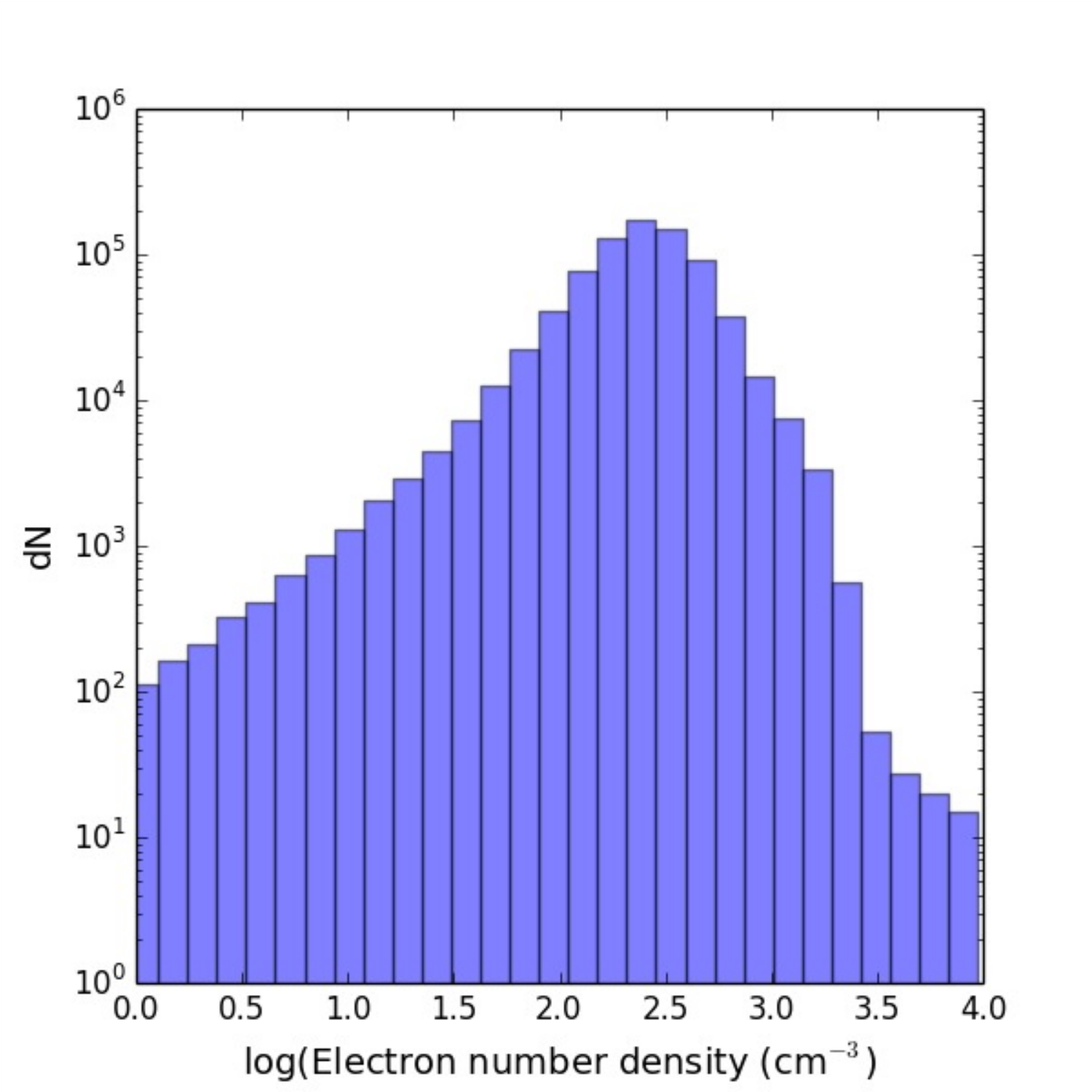}}
\subfloat[]{\includegraphics[scale=0.48]{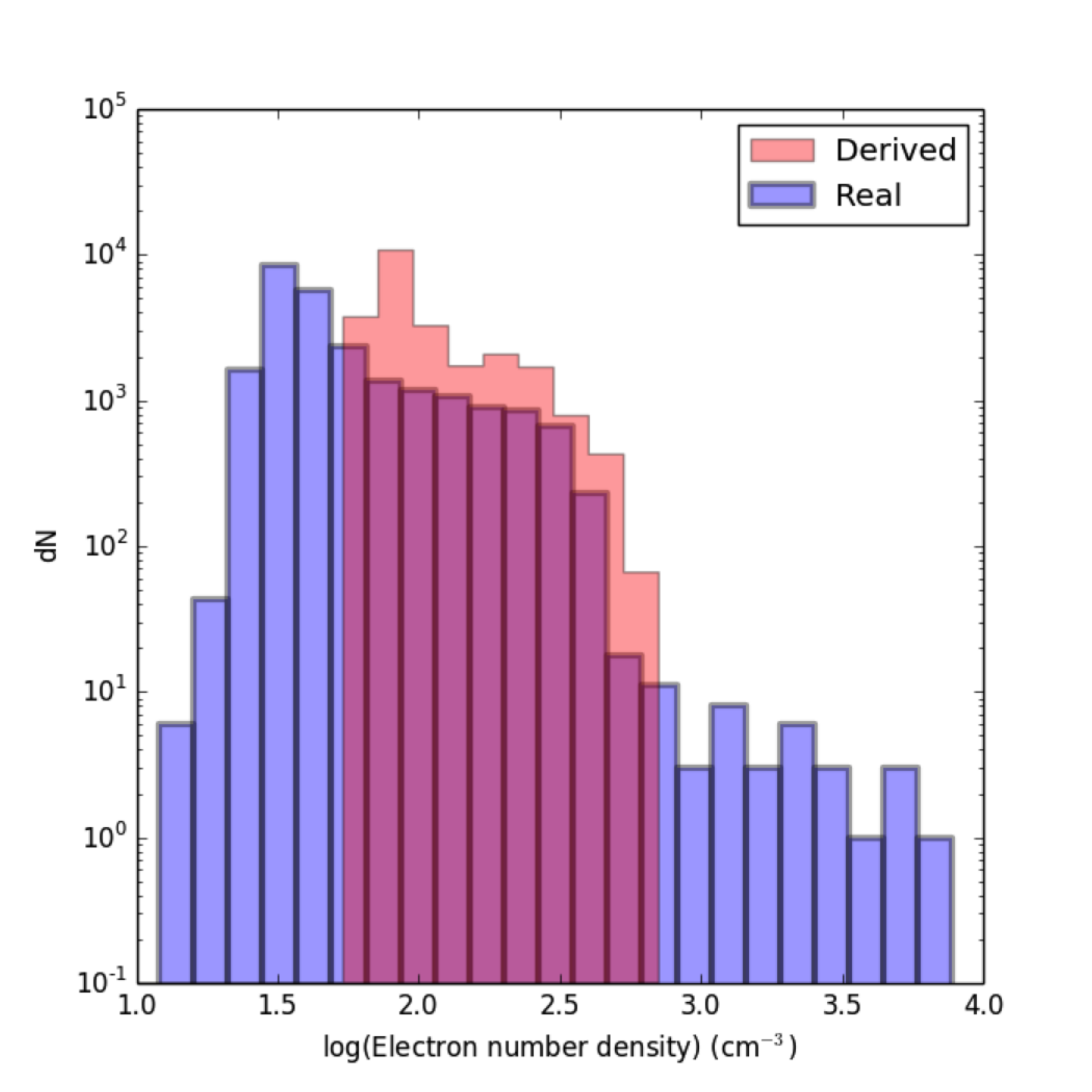}}}
  \caption{Histogram of the electron density of the observed data (panel a) and the simulated pillars (panel b). See text Section 3.3 for discussion.}
  \label{hist_ne}
\end{figure*}

The tips of the pillars show the highest densities with values $>$ 2000 cm$^{-3}$ and we find decreasing density values along the pillar bodies, while the surrounding gas is at densities $<$ 250 cm$^{-3}$ (Fig. \ref{NT}a). The pillar tips having the highest electron densities is a consequence of them also having a much higher column density than the pillar bodies \citep{1999A&A...342..233W}: the dense tips act like caps that shield the pillar bodies from the stellar feedback. Fig. \ref{NT}c shows the electron density map of our simulations, derived via Eq. 2. The density of the HII region is comparable to what we observe, as is the fact that the exposed pillar protrusions that are being ionised display the highest density values. These are lower by a factor of $\sim$ 2, which is consistent with the fact that the radiation field implemented in the simulations is weaker than what is observed in M16. Another factor adding to this is that the simulated pillars, while roughly the same size as the M16 Pillars, are only about 1/10th in the mass. Fig. \ref{hist_ne}a shows that the histogram of the electron density follows a lognormal distribution with a slight tail toward the low-density regime. The lognormal  density distribution is well recovered in our simulations as is shown in Fig. \ref{hist_ne}b, where the electron density as measured from the SPH grid (real) is plotted together with the electron density derived via the [SII] line ratio (derived). In the real electron density histogram it appears that in the simulations the high and low density values are both washed out by the line-ratio calculation, effect arising from smoothing along the line of sight. A lognormal electron density distribution has been observed for the ionised interstellar medium (\citealt{2008ApJ...683..207R}, \citealt{2007ASPC..365..250H}) where the driver for the shape is isothermal turbulence. In the case of the Pillars of Creation we suggest that shocks could produce such a distribution. The presence of shocks could in principle be inferred from emission line ratios, but, as we show in Section 3.5, these shock indicators may be washed out - and therefore not detected - by the dominant influence of photoionisation.

The electron temperature is usually determined via the forbidden line ratio ([OIII]$\lambda$5007+[OIII]$\lambda$4959)/[OIII]$\lambda$4363, but the [OIII]$\lambda$4363 is not covered by MUSE and we therefore derived the electron temperature T$_{e}$ via the [NII] doublet according to equation 5.5 in \cite{2006agna.book.....O} (assuming that the term in the denominator is negligible) 

\begin{equation}
T_{e} = \frac{2.5\times10^{4}}{ln(0.164R)}
\label{te}
\end{equation}

where $R=[NII]\lambda6548,8584/[NII]\lambda5755$. From the electron temperature map shown in Fig. \ref{NT} it is clear that the high density regions at the tip of the pillars and the arc-like structures in P1 display lower temperatures with respect to the surrounding ambient gas, although the map presents a certain level of noise coming from the weak [NII]5755 line. 

\subsection{Abundance tracers and ionic abundances}
\begin{figure*}
\mbox{
\subfloat[]{\includegraphics[scale=0.5]{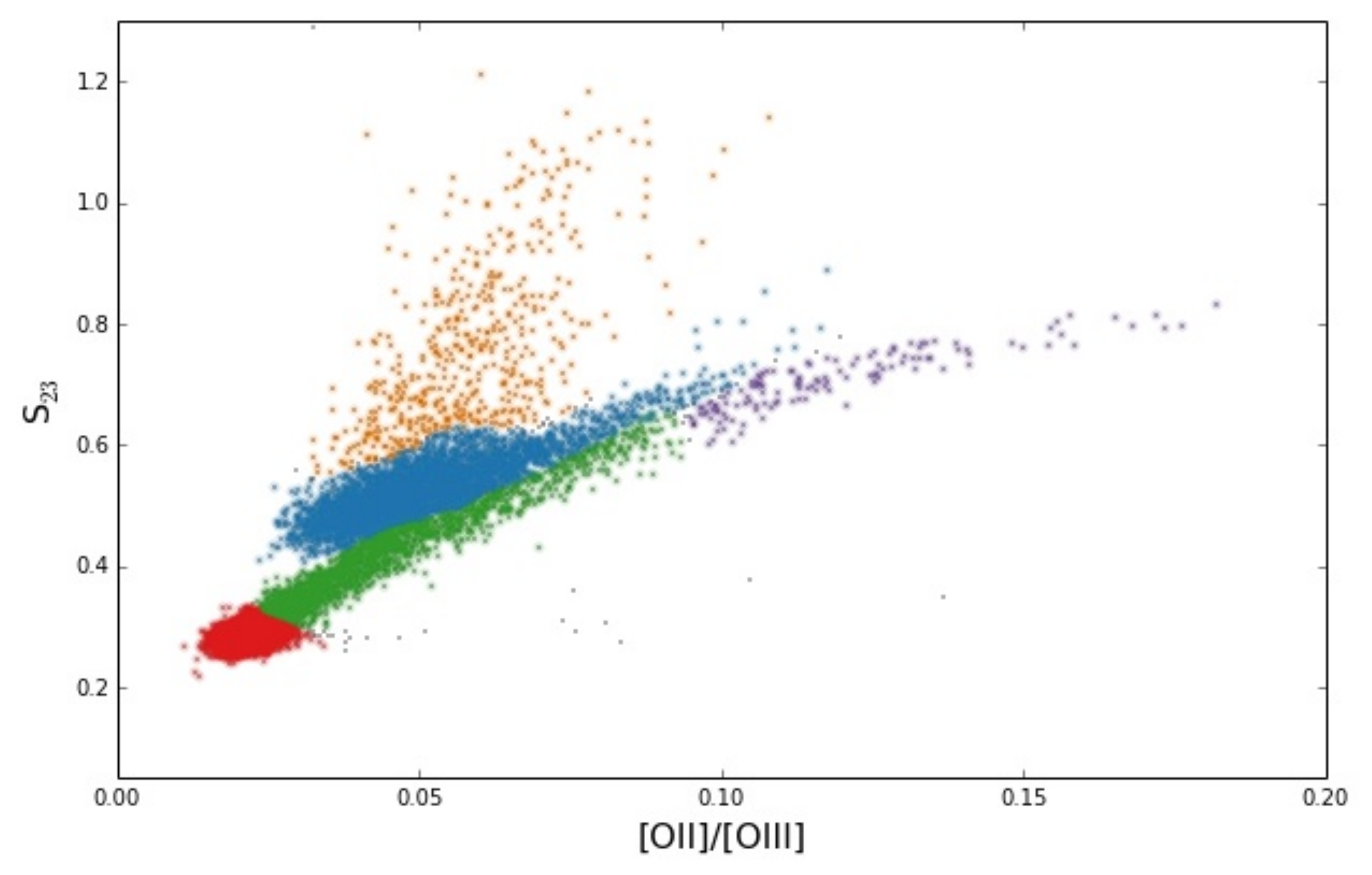}}}
\mbox{
\subfloat[]{\includegraphics[scale=0.5]{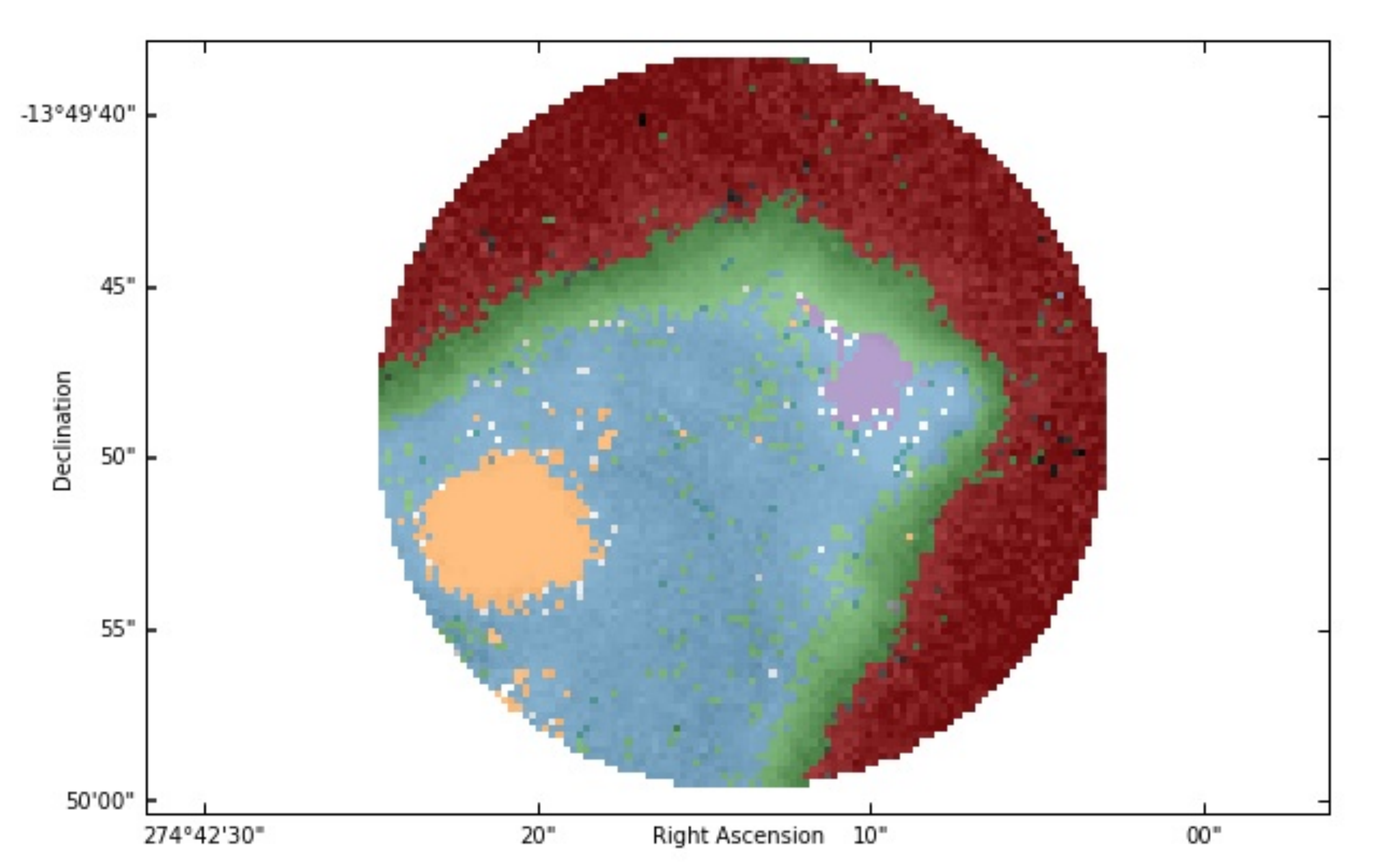}}}
  \caption{O$_{23}$ = [OII]/[OIII] vs. S$_{23}$ for the tip of P2 (panel a) and the S$_{23}$ parameter map of the same region (panel b). The colours in panel a highlight different populations in the scatter plot of panel a which correspond to the same coloured regions in panel (b), as is discussed in Section 3.4.}
  \label{P2_S23}
\end{figure*}

\begin{figure*}
\includegraphics[scale=0.5]{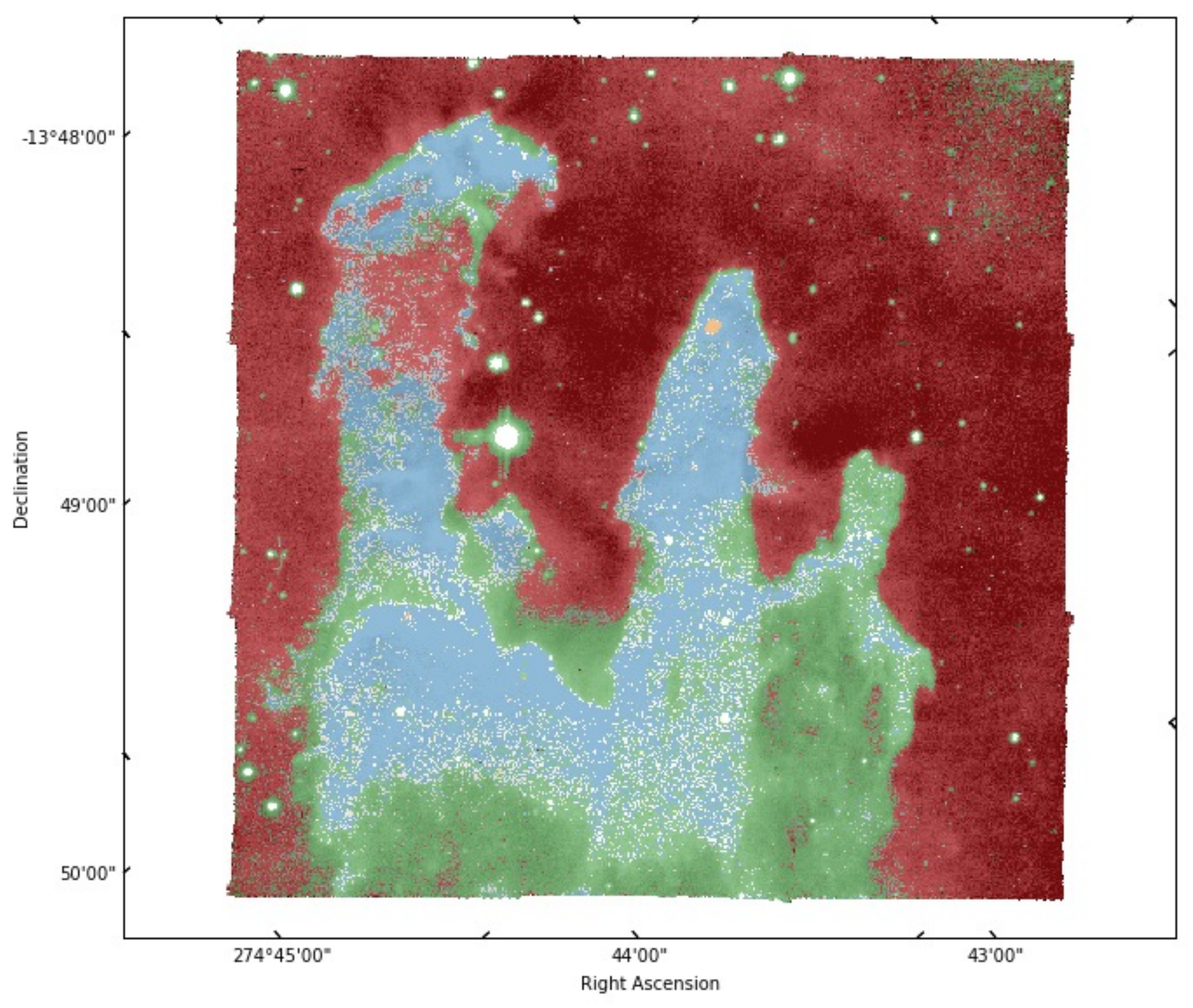}
  \caption{S$_{23}$ parameter map of the entire mosaic, the colour-coding is the same as in Fig. \ref{P2_S23}.}
  \label{brush_all}
\end{figure*}

Among abundance parameters the most used are the oxygen R$_{23}$ and the sulfur S$_{23}$ parameters, which are defined as R$_{23}$ = ([OII]$\lambda$3726,3729+[OIII]$\lambda$4959,5007)/H$\beta$ (\citealt{1979MNRAS.189...95P}) and S$_{23}$ = ([SII]$\lambda$6717,31+[SIII]$\lambda$9068,9532)/H$\beta$ (\citealt{1996MNRAS.280..720V}), and have been extensively used to determine galactic as well as extragalactic chemical abundances, since these differ as a function of position within a galaxy and can therefore be used to study star formation histories and evolutionary scenarios. Other authors have used these parameters, together with the excitation measured by the ratio of [OII]/[OIII], to study the ionisation structure of HII regions (\citealt{2010MNRAS.408.2234G}, \citealt{2011MNRAS.413.2242M}). Here, we exploit these parameters to distinguish between the various regions in our data (pillars, surrounding HII region, stars, young stellar objects, and protostellar outflows). This is demonstrated in Fig. \ref{P2_S23}, where S$_{23}$ (defined as S$_{23}$ = ([SII]$\lambda$6717,31+[SIII]$\lambda$9068)/H$\beta$ because of the lacking coverage of the [SIII]$\lambda$9532 line) is plotted versus [OII]/[OIII] (panel a) for a sub-region that highlights the tip of P2 (panel b): the clearly recognisable different populations in the scatter plot correspond to ambient matter (red), the pillar-ambient interface (green), the pillar matter (blue), a known T Tauri star (magenta, see Section 4.1) and a sulfur-abundant region (orange) of panel b. The brushed S$_{23}$ map of the entire mosaic is shown in Fig. \ref{brush_all}, the stars have been intentionally left white. Fitting in the picture where P1a lies behind NGC 6611 and the other two pillars along the line of sight (see Section 4.2), we see the pillar body of P1a (marked in blue) contaminated by the HII region (marked in red).
The sulfur-abundant region is clearly visible in the S$_{23}$ parameter map (Appendix A) and is visible as a slight over-density of the order of $\sim$ 2000 cm$^{-3}$ in the electron density map. As will be discussed in Section 4.1, we suggest that this region corresponds to where the blue-shifted lobe of a bipolar jet originating from a deeply embedded protostar is pushing its way out through the pillar material. The fact that it is only visible in the integrated intensity maps of the [SII] lines and in [OI]$\lambda$6300 indicates that the outflow is not energetic enough to ionise the detected atoms with E$_{ion}\geq$ E$_{ion, [SII]}$ = 10.36 eV.

Furthermore, we derived ionic and total abundances of the HII region and the pillar tips via the \texttt{IRAF} task \texttt{ionic}. For this, we produced co-added spectra of a region representative of the HII region matter and a region representative of the pillar tips (see Fig. \ref{lineid}), which were fitted with a gaussian routine to determine emission line intensities. The electron density was computed via equation 2, and this was then used to derive ionic abundances and the electron temperature with  \texttt{IRAF}. The derived ionic and total abundances are listed in Table \ref{abun}. The oxygen ionic abundance ratios O$^{+}$/H$^{+}$ and O$^{2+}$/H$^{+}$ were obtained with the [OII]$\lambda$7320,30 and [OIII]$\lambda$4959,5007 lines respectively and assuming T([OII]) $\approx$ T([NII]) and T([OIII]) $\approx$ T([SIII]), as we do not have the wavelength coverage for the [OII]$\lambda$3727,29 and the [OIII]$\lambda$4363 lines to determine T([OIII]) and T([OII]). The total S/H abundance was determined by taking into account the unobservable ionisation stages and therefore using the ionisation correction factor (ICF) for S$^{+}$+S$^{2+}$ as in \cite{2008MNRAS.383..209H}. As is expected from the fact that molecular material is being ionised at the pillar tips, these present the higher ionic abundance values when compared with the HII region.

\begin{table*}
\begin{center}
\caption{Derived ionic and total abundances of O and S (see Section 3.4).}
\begin{tabular}{lcccccc}
\hline
\hline
 Region & 12+log(O$^{+}$/H$^{+}$) & 12+log(O$^{2+}$/H$^{+}$) & 12+log(O/H) & 12+log(S$^{+}$/H$^{+}$) & 12+log(S$^{2+}$/H$^{+}$) & 12+log(S/H) \\
\hline
HII  & 8.00 & 8.16 & 8.39 & 6.05 & 7.53 & 7.59 \\
\hline
Pillar tip & 8.76 & 8.22 & 8.87 & 6.79 & 7.65 & 7.77  \\
\hline
\label{abun}
\end{tabular}
\end{center}
\end{table*}

\subsection{Line ratio maps and BPT diagrams}
BPT (\citealt{1981PASP...93....5B}) diagrams compare collisionally excited lines like [OIII]$\lambda$5007, [NII]$\lambda$6584 and [SII]$\lambda$6717, 31 to hydrogen recombination lines (H$\alpha$, H$\beta$) and use the forbidden/recombination line ratios as indicators of the number of ionisations per unit volume. The position of an object on a BPT diagram is used for classification purposes, e.g. to distinguish star forming galaxies from active galactic nuclei (AGN), as gas excited by photoionisation rather than shock ionisation occupies different regions in such a diagram. \cite{2001ApJ...556..121K} combined photoionisation and stellar population synthesis models to put an upper limit  on the location of star-forming sources in the BPT diagram. This upper limit is given by a maximum starburst line (labelled as Ke01 in Fig. \ref{BPT}, from \citealt{2001ApJ...556..121K}) of pure stellar photoionisation models, and galaxies lying above this line are considered AGN dominated because for them to be found here an additional excitation mechanism other than stellar photoionisation is required. \cite{2003MNRAS.346.1055K} modified the Ke01 plot to account for galaxies that have contribution from both star formation and AGN (labelled as Ka03 in Fig. \ref{BPT}, from \citealt{2003MNRAS.346.1055K}).

\begin{figure*}
\mbox{
  \subfloat[]{\includegraphics[scale=0.39]{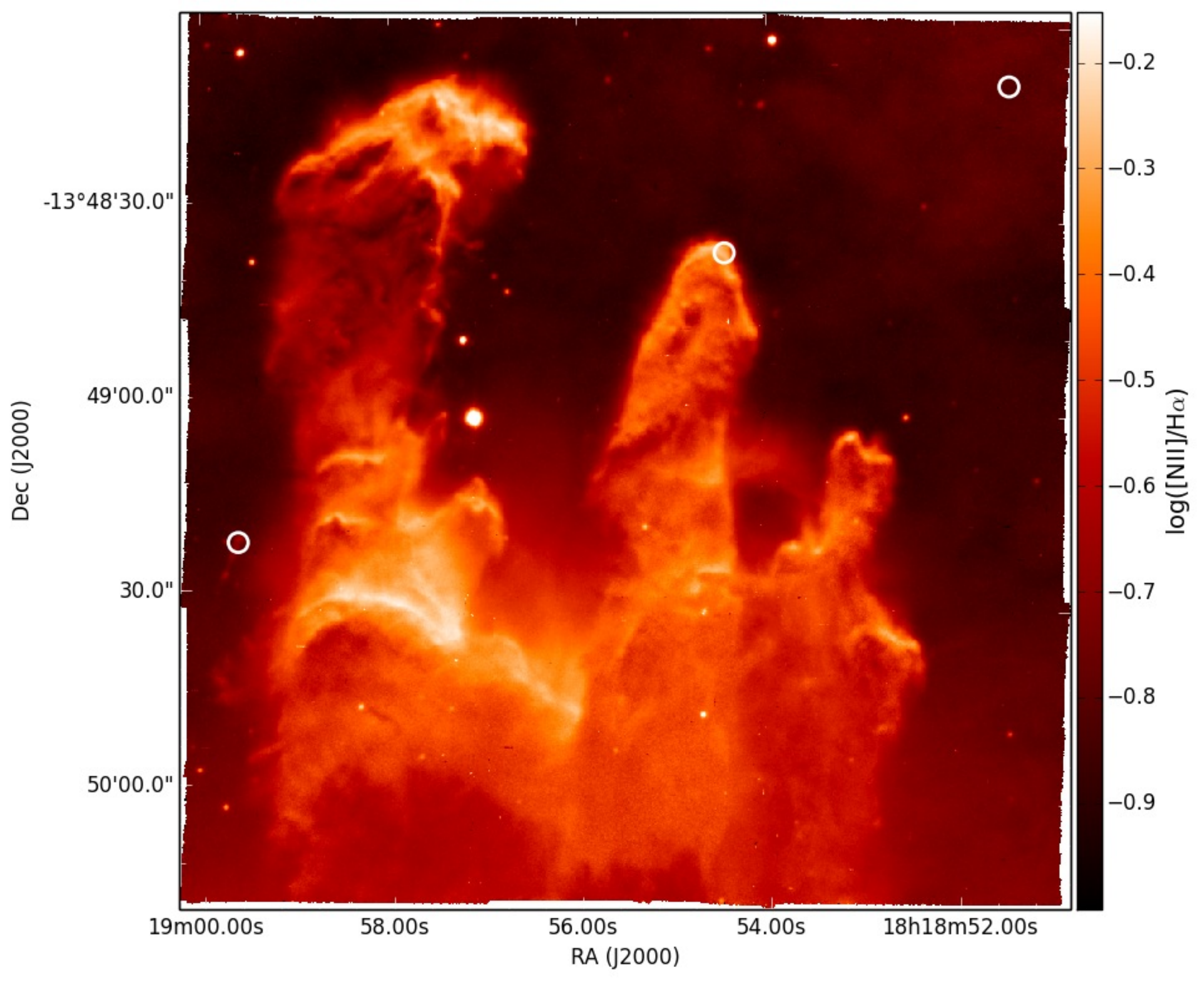}}
  \subfloat[]{\includegraphics[scale=0.39]{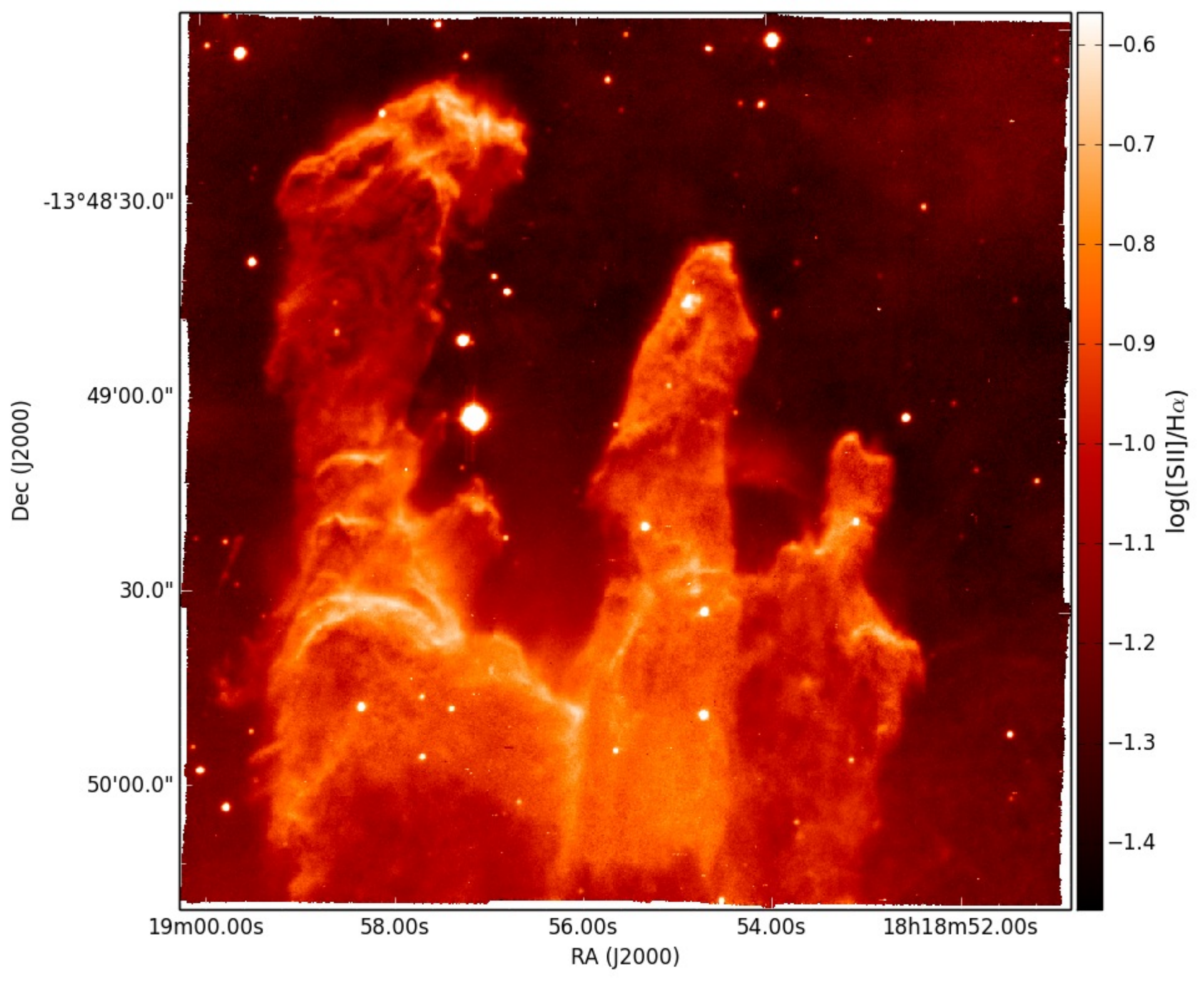}}}
  \mbox{
  \subfloat[]{\includegraphics[scale=0.39]{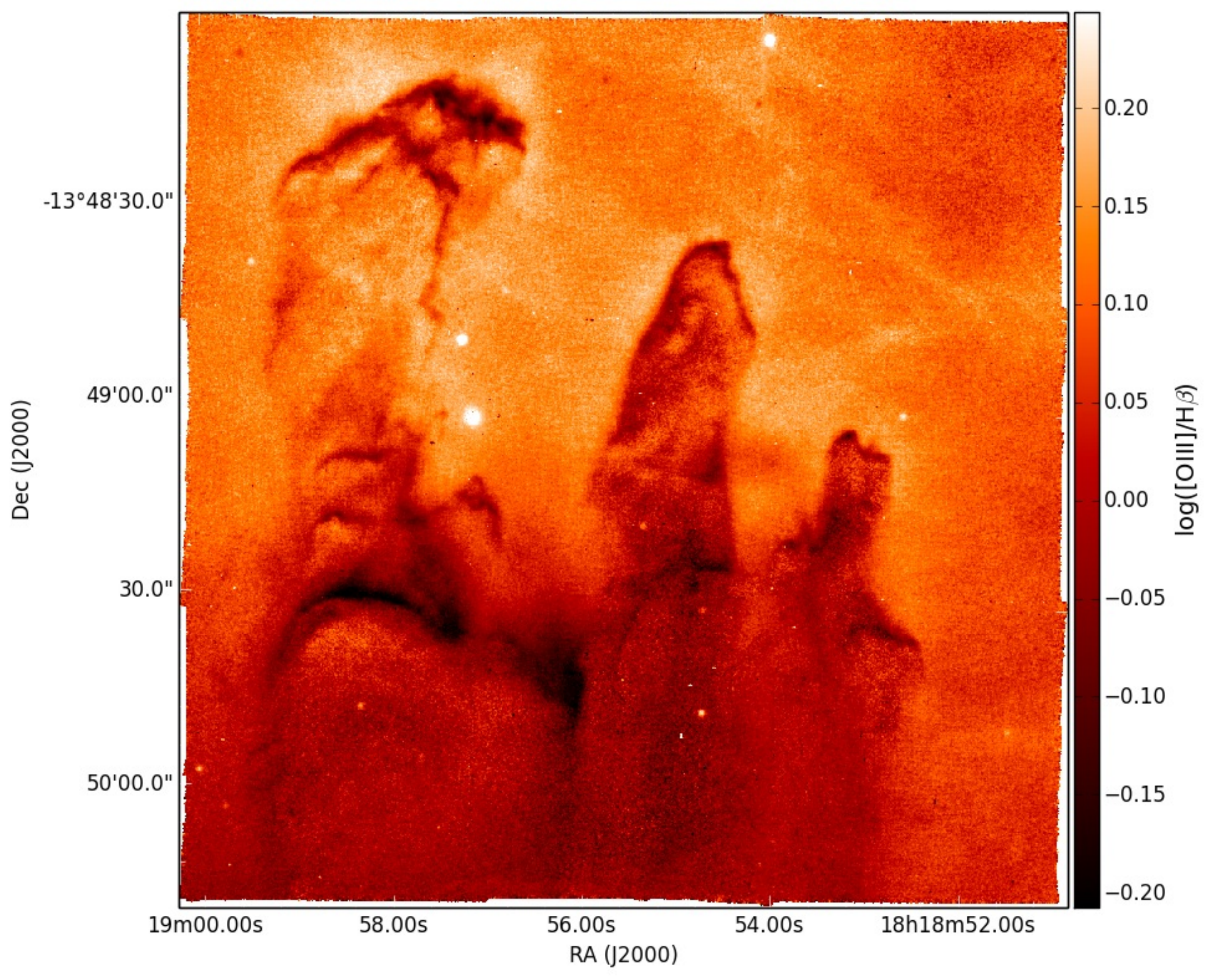}}}
  \caption{Line ratio maps linearly scaled to maximum/minimum: [NII]$\lambda$6584/H$\alpha$ (a), [SII]$\lambda\lambda$6717,6731/H$\alpha$ (b) and [OIII]$\lambda$5007/H$\beta$ (c). The white circles in panel a indicate the regions where the co-added spectra were extracted. The leftmost circle also corresponds to the location of the EGG discussed in Section 1.}
  \label{ratiomaps}
\end{figure*}

\begin{figure*}
\mbox{
  \subfloat[]{\includegraphics[scale=0.22]{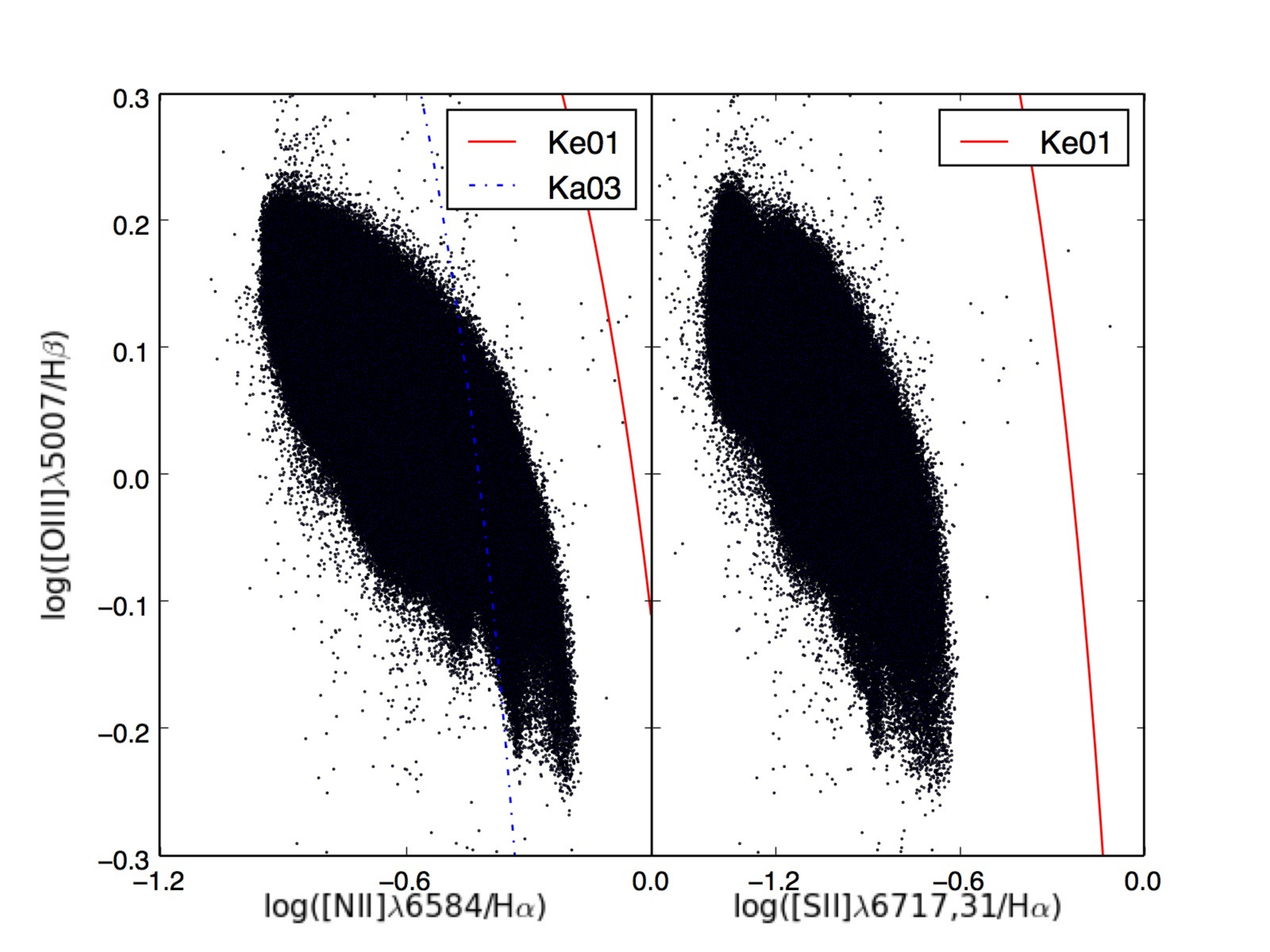}}
  \subfloat[]{\includegraphics[scale=0.22]{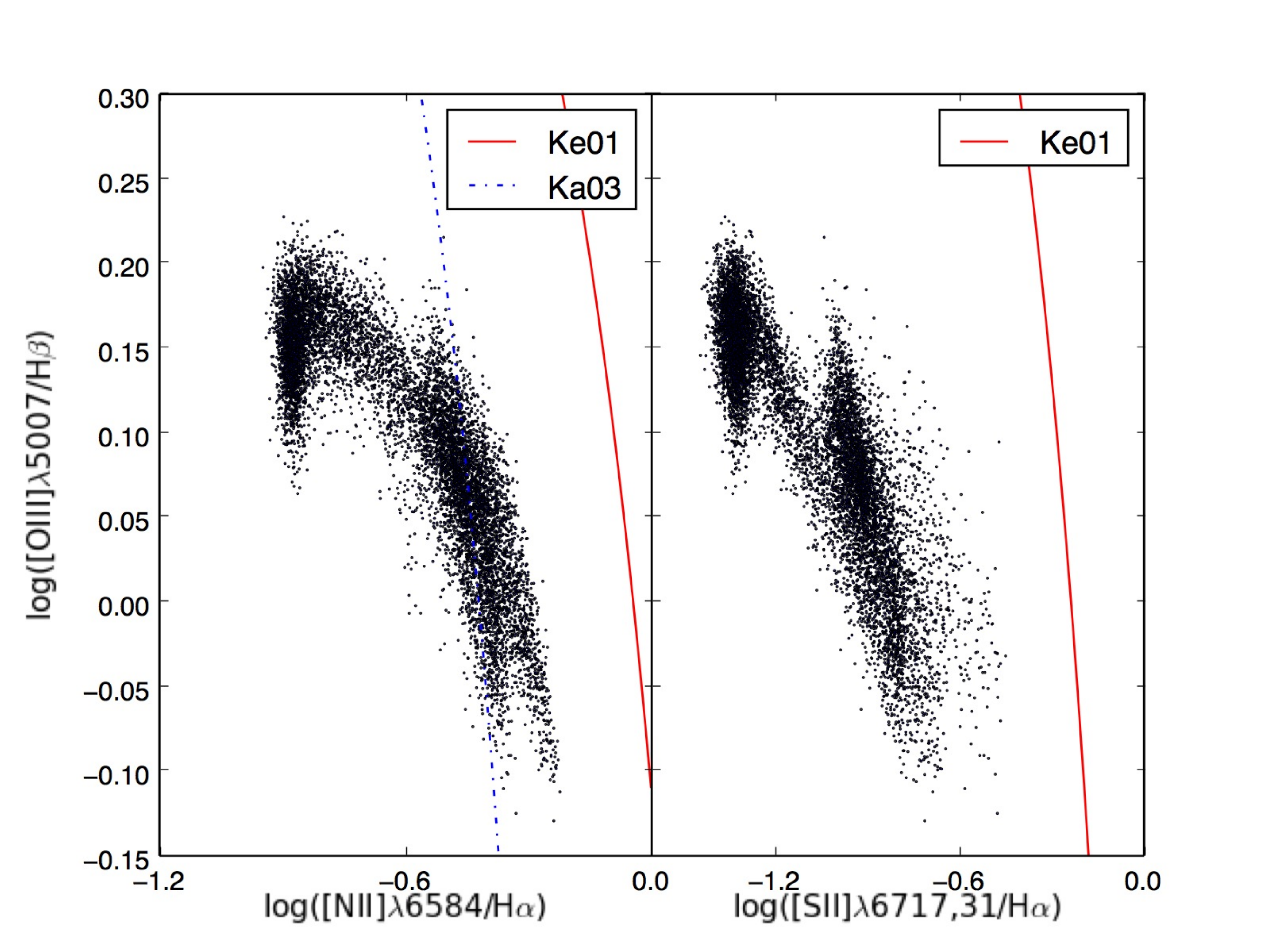}}}
  \mbox{
  \subfloat[]{\includegraphics[scale=0.3]{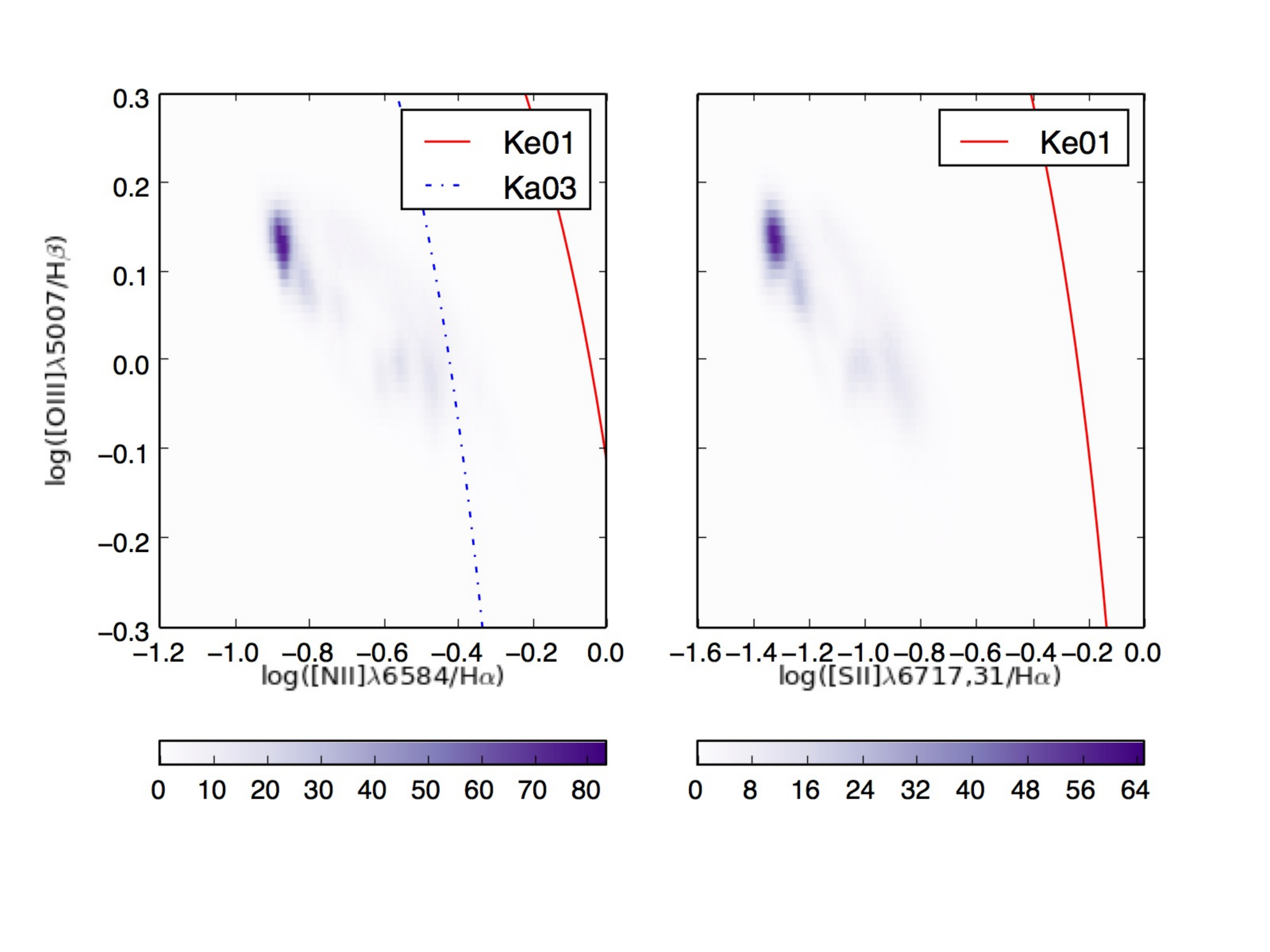}}}
  \caption{Diagnostic BPT diagrams for the entire observed area (panel a) and the tip of P2 only (panel b). Panel c shows a normalised 2D histogram of panel a. See text Section 3.5 for discussion. The Ke01 and Ka03 lines separate regions where photoionisation (below the lines) and AGN feedback (above the lines) dominate. Ka03 accounts for a mixed feedback of both AGN and photoionisation.}
  \label{BPT}
\end{figure*}

The position on the BPT diagram as a diagnostic tool is widely used for spatially unresolved observations of galaxies or nebulae where the integrated emission of the whole source is used to compute the emission line ratio. BPT diagrams have been used for IFU observations of HII regions (\citealt{2011MNRAS.413.2242M}, \citealt{2010MNRAS.408.2234G}) as well, but since these are spatially resolved observations caution is advised: one is comparing a 3D structure with a 1D photoionisation model, and in a 2D projection of a 3D structure the line ratio will show variations from pixel to pixel because of the local gas conditions along the line of sight on each pixel (\citealt{2012MNRAS.420..141E}). Maps of the [NII]$\lambda$6584/H$\alpha$, [SII]$\lambda\lambda$6717,6731/H$\alpha$ and [OIII]$\lambda$5007/H$\beta$ line ratios are shown in Fig. \ref{ratiomaps}.

Here we investigate the commonly used line ratios as indicators of the ionisation structure: [OIII]$\lambda$5007/H$\beta$ is a tracer of the highly ionised gas and [NII]$\lambda$6584/H$\alpha$ as well as [SII]$\lambda\lambda$6717,6731/H$\alpha$ are tracers of the less ionised gas. The [NII]/H$\alpha$ and [SII]/H$\alpha$ line ratio maps are quite similar and show significantly higher line ratios at the tip of the pillars as well as on the edges of the wave-like structures in the pillars' main bodies, correlating with the position of the peaks of the localised emission; the  [OIII]/H$\beta$ ratio on the other hand has its lowest values at the positions of the [SII]/H$\alpha$ and [NII]/H$\alpha$ peaks, as one would expect in an ionisation picture where the double ionised elements are found in the less dense material more exposed to the ionising radiation rather than in the dense molecular environment shielded from the stellar influence.

\begin{figure*}
\includegraphics[scale=0.55]{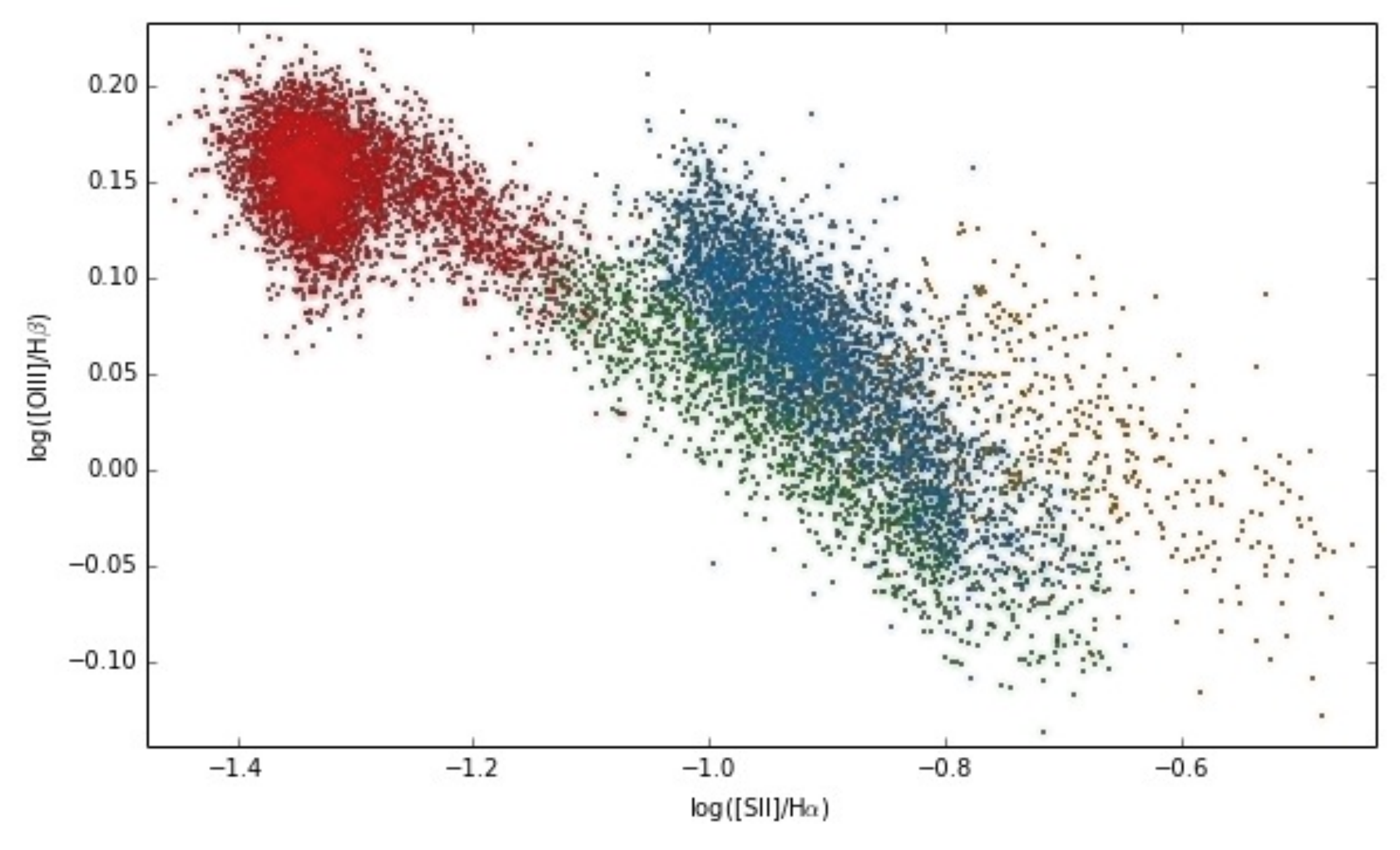}
  \caption{BPT diagram of the tip of P2. The colours in panel (a) highlight different populations in the BPT diagram which correspond to the same coloured regions in panel (b).}
  \label{P2_BPT}
\end{figure*}

Fig. \ref{BPT} shows the [OIII]$\lambda$5007/H$\beta$ versus [NII]$\lambda$6584/H$\alpha$ and [SII]$\lambda\lambda$6717,6731/H$\alpha$ diagnostic diagrams for the entire observed region (panel a) and for a circular region containing the tip of P2 (panel b), the lines correspond to the maximum starburst lines from \cite{2001ApJ...556..121K} and \cite{2003MNRAS.346.1055K} (to better identify structures in panel a, panel c shows a normalised 2D histogram of the panel a). From these diagrams it is clear that the line ratios lie in the the region below Ke01 and therefore in the region where HII regions are expected to be found. Also taking into account the fact that if shocks were strongly contributing to the excitation one would expect [NII]/H$\alpha$ $>$ -0.1 and [SII]/H$\alpha$ $>$ -0.4 \citep{2008ApJS..178...20A} but we do not see values that high, we suggest that the dominant excitation mechanism at the pillar tips is photoionisation. This agrees very well with the BPT diagrams of our SPH simulations (see Fig. 4 and 5 in \citealt{2012MNRAS.420..141E}), where the majority of data points lie in the photoionisation region. The simulated BPT diagrams (Figures 4 and 5 in \citealt{2012MNRAS.420..141E}) do however show a small percentage of points lying in the shock-dominated region, but as the simulations do not include shock models these line ratios are due to 3D effects. Fig. \ref{P2_BPT} shows the BPT diagram of the tip of P2 color coded according to the regions identified in Fig. \ref{P2_S23}b.
 
\section{Emission line fitting}
\begin{figure*}
\mbox{
  \subfloat[]{\includegraphics[scale=0.48]{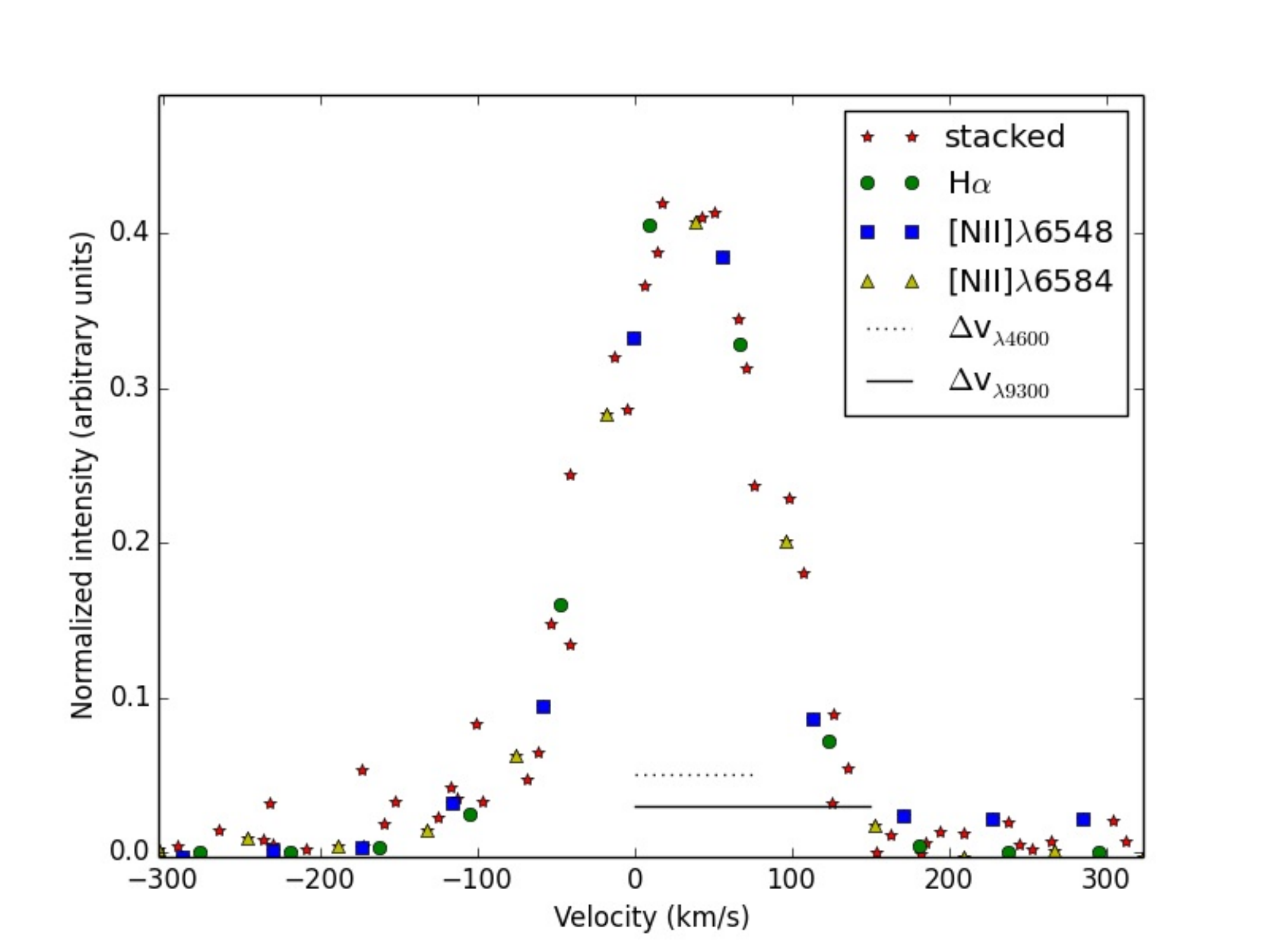}}
  \subfloat[]{\includegraphics[scale=0.48]{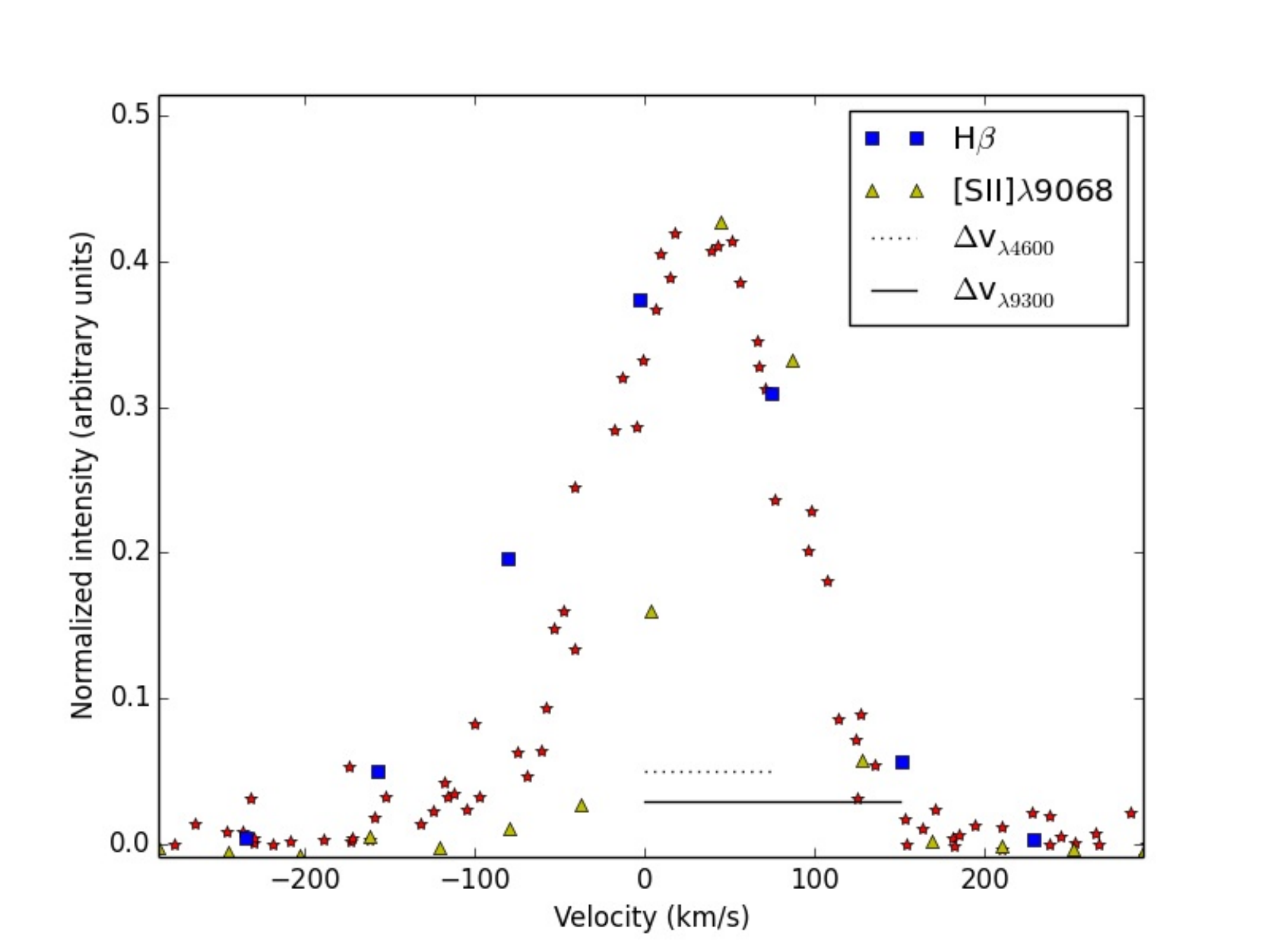}}}
  \caption{Stacked spectrum (red in both panels), shown together with the H$\alpha$ (green circles), [NII]$\lambda$6548 (blue squares) and [NII]$\lambda$6584 (yellow triangles) spectra (panel a), as well as the H$\beta$ (blue squares) and [SIII]$\lambda$9068 (yellow triangles) (panel b). See Section 4 for discussion.}
  \label{fit}
\end{figure*}

MUSE's detectors undersample the line spread function, and we therefore proceeded by first normalising and stacking several emission lines in each single pixel (thus assuming that all the lines in one pixel originate from the same emitting gas) and then by fitting the resulting velocity spectrum in each pixel with a gaussian function to produce a velocity map\footnote{The \textsc{python} packages \textsc{spectral$\_$cube} (spectral-cube.readthedocs.org) and \textsc{pyspeckit} \citep{2011ascl.soft09001G} were used for the fitting routine.}. The lines used for the stacked spectrum are H$\alpha$, the two [NII] lines, the two [SII] lines, [OI]$\lambda$6300,6363 and HeI$\lambda$6678. The undersampling of single lines is shown in Fig. \ref{fit}, where the stacked spectrum and spectra of single lines (H$\alpha$, [NII]$\lambda$6548 and [NII]$\lambda$6584 in panel a, H$\beta$ and [SIII]$\lambda$9068 in panel b) are shown, together with the resolution at 4600 \AA\ ($\Delta$v = 150 km s$^{-1}$) and at 9300 \AA\ ($\Delta$v = 75 km s$^{-1}$). The early version of the data reduction pipeline we used did not include a radial velocity correction, and the derived velocities are therefore topocentric. Furthermore, during the observations the sky lines at $\lambda$5577 \AA\ and $\lambda$6300 \AA\ are calibrated with a set of arc lamps, calibration that yields an offset between the arc and observed lines; this offset is then propagated to the rest of the spectrum, but in a linear way such that the wavelength dependency of the correction is lost. This leads to a velocity offset of the red (e.g. [SIII]) and blue (e.g. [OIII], H$\beta$) lines with respect to the lines close to the calibration wavelength which can be seen in Fig. \ref{pv_profile} (a) and (b), where the topocentric velocities of single emission line species are plotted along the slits shown in Fig. \ref{museptg}. Furthermore, when computing local standard of rest velocities from the MUSE data with \texttt{IRAF} and comparing them to velocities obtained form CO observations (\citealt{1998ApJ...493L.113P}, v$_{LSR}\sim$ 20 - 20 km s$^{-1}$), we find that the two are offset by about 20 km s$^{-1}$, offset we cannot attribute to a physical origin. Further investigation of the science verification data and the early version of the used data reduction pipeline is needed. Because of these problems we only report relative velocities: for each line emitting species the mean velocity of the HII region was computed, value which was then subtracted to the whole velocity map of that line. The relative velocities reported are therefore velocities relative to the HII region.

\subsection{Velocity structure}
\begin{figure*}
\mbox{
  \subfloat[]{\includegraphics[scale=0.4]{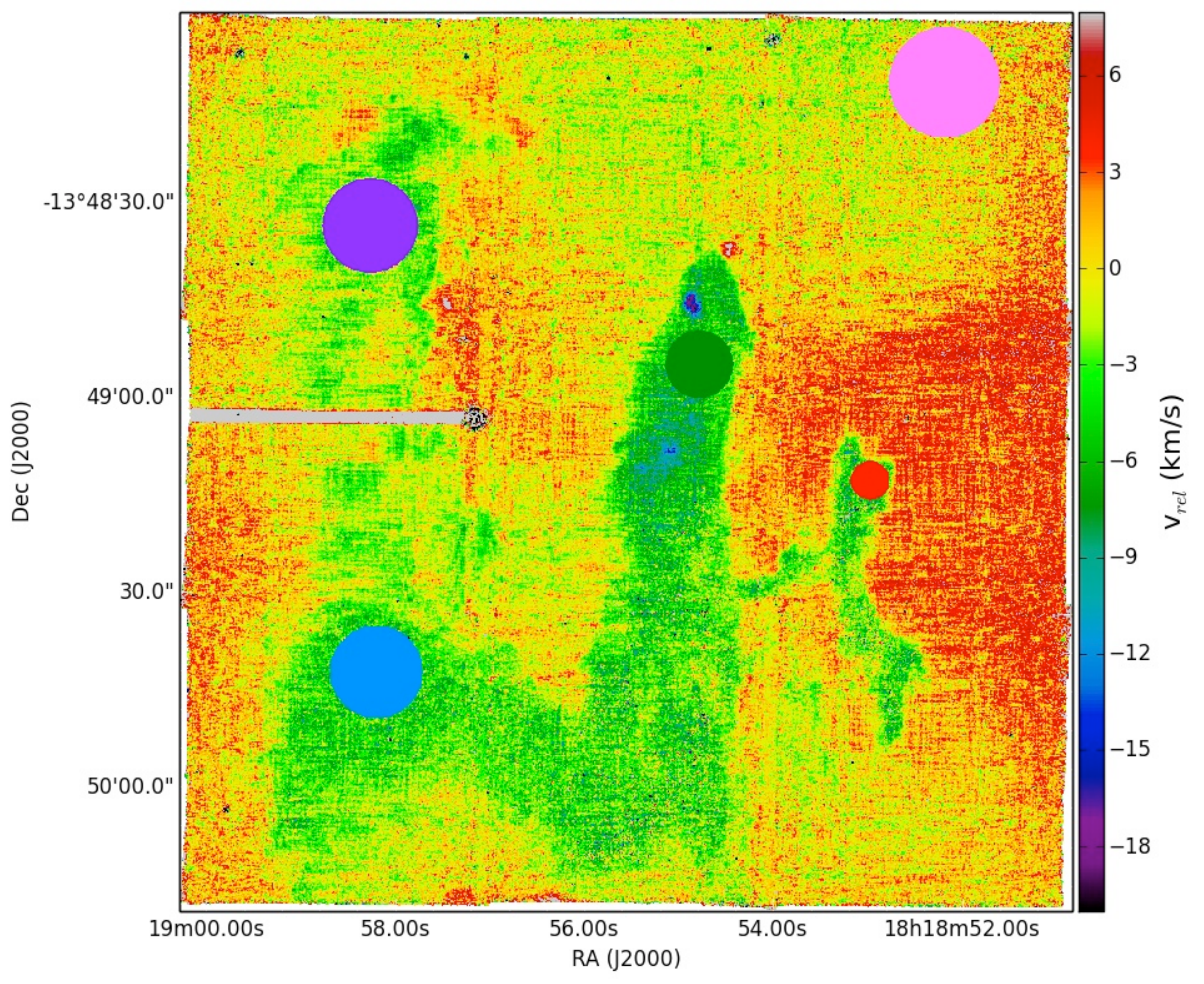}}}
  \mbox{
  \subfloat[]{\includegraphics[scale=0.4]{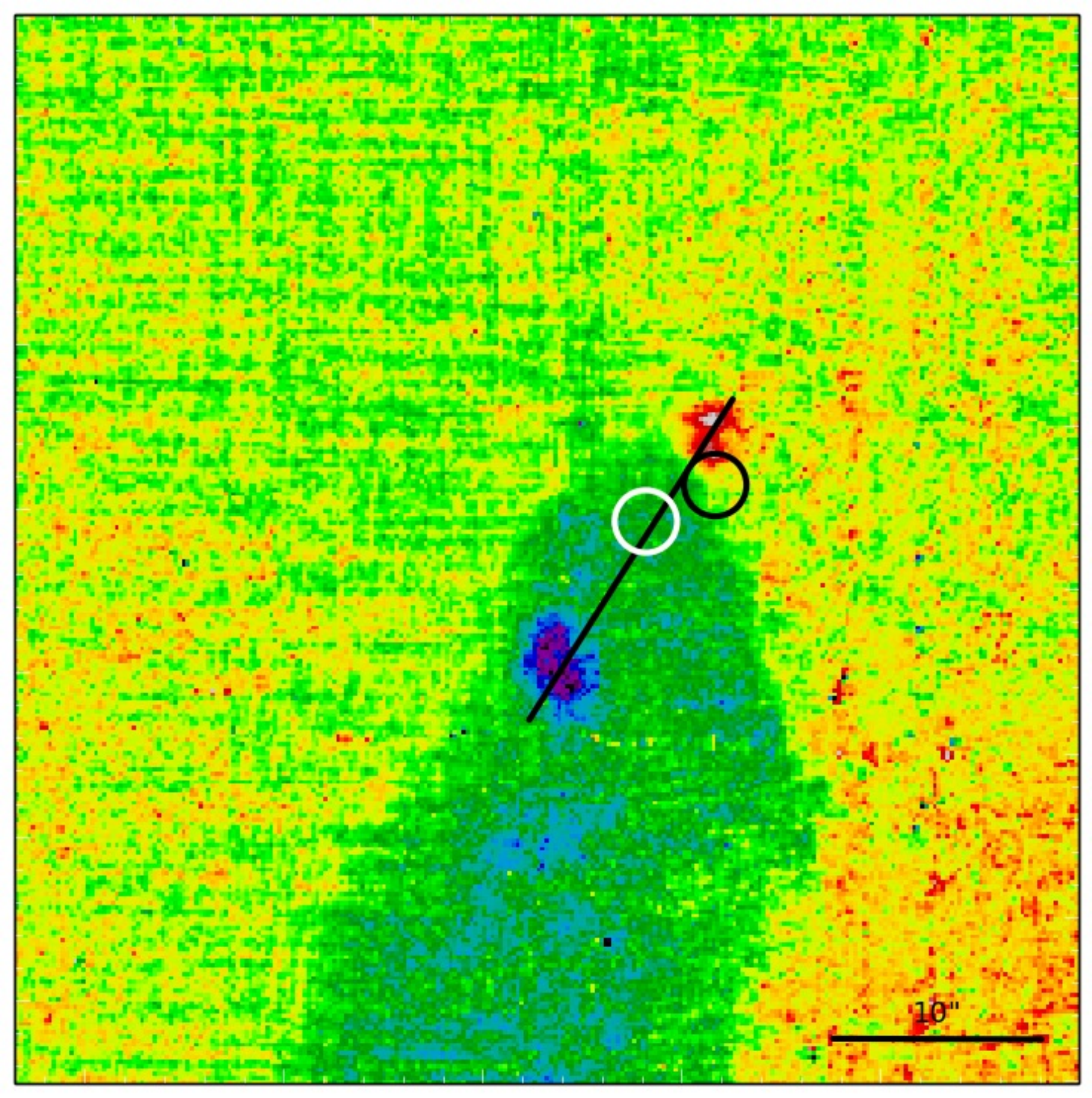}}}
  \caption{Map of relative velocity (to the the HII region) of the entire mosaic (panel a) and a zoom-in on P2 (panels 2). The coloured circles in panel a correspond to the regions used for the histograms in Fig. \ref{histo}. The black line in panel b corresponds to a hypothetical orientation of the bipolar outflow with position angle 107.3 degrees, the black circle is centred on the coordinates of the candidate T Tauri star, while the white circle corresponds to the water maser detected in March 2002 by \citealt{2004ApJ...610..835H} (see text Section 4.1).}
  \label{vel}
\end{figure*}

\begin{figure}
\includegraphics[scale=0.4]{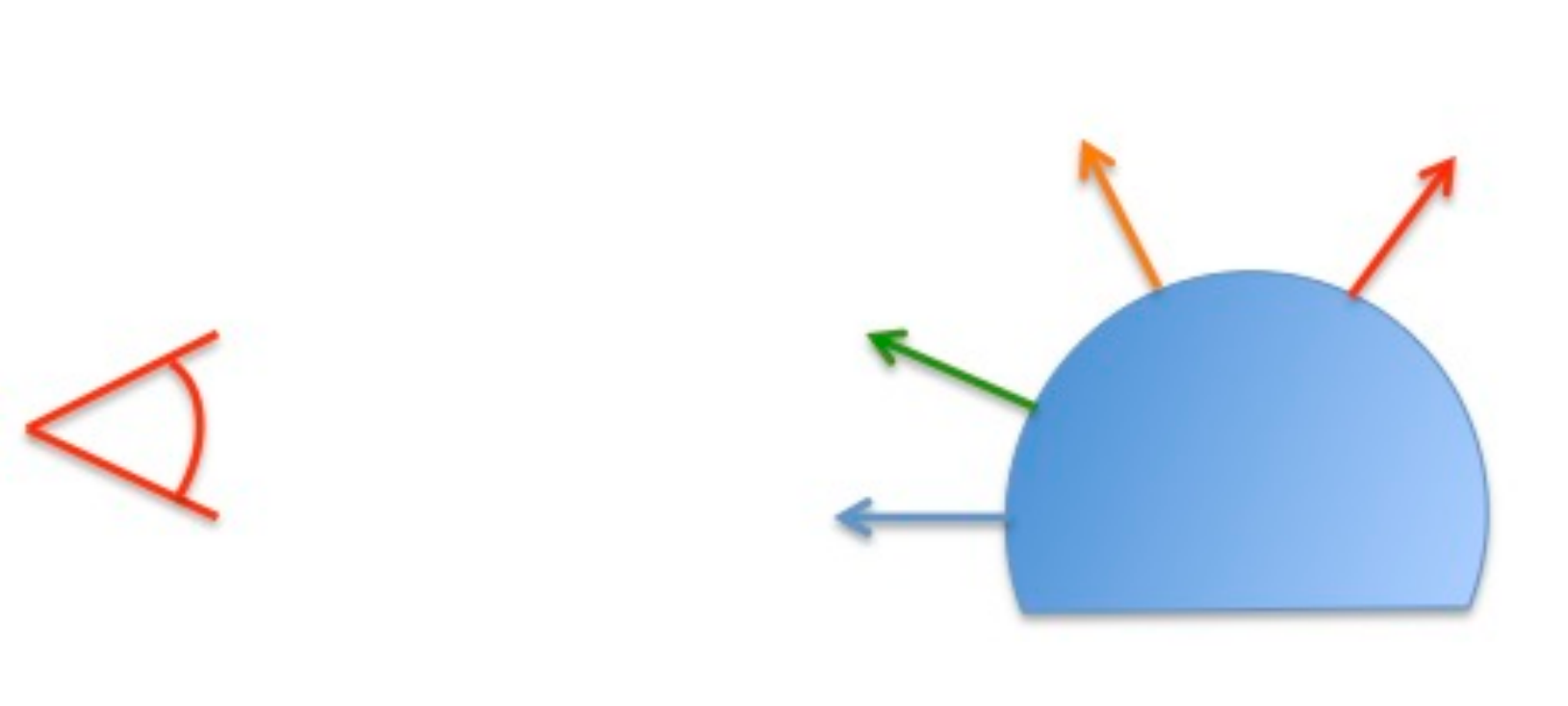}
  \caption{Sketch of the normal velocity vector, corresponding to the photo-evaporative flow, to demonstrate that the pillar tips appear redshifted with respect to the main pillar bodies because of the geometrical effect of the vector component along the line of sight (see text Section 4.1).}
  \label{los_red}
\end{figure}

Based on CO J=3-2 data \cite{1999A&A...342..233W} report complex velocity fields within the pillars and a systematic large scale velocity gradient of about 1.7 km s$^{-1}$ along P2. The resolution of MUSE does not allow the detection of such a small velocity gradient, but Fig. \ref{vel}a shows the velocity map obtained with the above mentioned emission line fitting. The pillars are clearly distinguishable from the ambient gas, and indeed the pillars are blue-shifted with respect to the surrounding gas because of the photo-evaporative flow along the line of sight that is moving radially away from the pillar surfaces. This means that, when moving from the pillar bodies facing our way along the line of sight to the pillar-ambinet interface, we see the pillar border redshifted with respect to the pillar body as a geometrical consequence of the normal vector moving out of the line of sight (this is seen best in P3, and is illustrated in Fig. \ref{los_red}). The velocities of the Pillars and the HII region were estimated first by extracting circular regions for each of them via \textit{brushing}\footnote{The technique of \textit{brushing}, also known as \textit{graphical exploratory data analysis} allows the user to manually select specific data points or subsets from an image or a plot by interactively drawing regions on the latter two. } (Fig. \ref{vel}a), and by then fitting the velocity histograms of these regions with Gaussian distributions (Fig. \ref{histo}). They are listed in Table \ref{vlos}. 
  
The sulfur-abundant region identified at the tip of P2 via the S$_{23}$ parameter is blueshifted (v$_{rel}\approx$ -20 km s$^{-1}$) with respect to the main pillar body (v$_{rel}\approx$ -7 km s$^{-1}$), as can be seen in Fig. \ref{vel}b, and at the very tip of the same pillar a redshifted region is identified (v$_{rel}\approx$ 10 km s$^{-1}$). The red and blue-shifted regions can be connected with a line with a position angle of approximately 107.3 degrees, on which a source identified as a candidate T Tauri star with a disk-like structure (\citealt{2002ApJ...565L..25S}, \citealt{2007PASJ...59..507S}, source M16 ES-2 in \citealt{2002ApJ...570..749T}), as well as water maser \citep{2004ApJ...610..835H} are found. We suggest that the blue- and redshifted regions trace the two counterparts of a bipolar outflow originating from an accreting protostar or young stellar object. As discussed by \cite{2004ApJ...610..835H}, the water maser is too far away from the candidate T Tauri star to be excited by the latter and the two are therefore not associated. With these observations we are unable to say whether the bipolar outflow originates from the T Tauri star or from a deeply embedded object, but considering the fact that the T Tauri star seems to lie somewhat off the hypothetical direction of the jet, we tentatively suggest that the latter is the case. \cite{2004ApJ...610..835H} suggest that the excitation source of the maser is a Class 0 protostar due to the lack of a detected near-IR source, but high angular resolution sub-mm observations are required to identify the maser driving protostar. Optical protostellar jets generally have observed radial velocities from a few tens up to hundreds km s$^{-1}$ \citep{2007prpl.conf..215B}, implying that the one detected here is a low-velocity outflow with v$_{rel, jet}\approx$ 20 km s$^{-1}$. The blue-shifted counterpart is tunnelling its way through the pillar material and is only now starting to emerge, and the redshifted counterpart on the other hand is flowing out from the pillar tip but is immediately dispersed in the strong ionising and unfavourable conditions of the HII region.

We also identify a candidate outflow on the inner side of P1, which is shown in Fig. A10 of the appendix. Also in this case we identify both a blueshifted and a redshifted counterpart in the velocity map that have relative velocities of $\sim$ -10 km s$^{-1}$ and $\sim$ 16 km s$^{-1}$ respectively.  This candidate outflow does not appear to correspond to a sulfur-abundant region and is not seen in the S$_{23}$ map. If this is indeed an outflow, we suggest that the absence of the sulfur-indicator comes from the fact that the blue lobe is only just starting to emerge from the pillar body and did not yet have the time to ionise atoms with E$_{ion}\geq$ E$_{ion, [SII]}$ = 10.36 eV. \cite{2004ApJ...610..835H} do not detect water masers in P1, but \cite{2002ApJ...565L..25S} identify candidate T Tauri star right outside the pillar which corresponds to a continuum source in Fig. \ref{30sRGB}. 
  
The presence of ongoing star formation is also observed in our SPH simulations (\cite{2010ApJ...723..971G} and Fig. A9), where 4 objects have formed at the pillar tips and are still present 0.5 Myr into the simulation. These objects all have angles of 20-50 degrees between the disk (or accretion feature) and the direction of radiation (which in the simulations is roughly perpendicular to the long axis of the pillars). \cite{2010RMxAA..46..179R} simulated outflows from stars forming at the tips of pillars and find that the orientation of the outflows depends on the orientation of the pillars along the line of sight. For pillars that are seen face on, the projected outflow orientations are found to be perpendicular to the main pillar axes. Then again, \cite{2010MNRAS.405.1153S} report that in their HST Herbig-Haro jet sample the outflows are not preferentially found to be perpendicular to the pillars. Here, the projected outflow orientation with respect to the P.A. of $\sim$ 137$^{\circ}$ of P2 is $\sim$ 30$^{\circ}$, which corresponds to an angle of only $\sim$ 7$^{\circ}$ between outflow and the direction of radiation. The disk of the jet-driving source is thus supposedly perpendicular to the incident radiation, which contradicts what is found from the simulations, where the accretion disks of the forming sources are preferentially aligned with the direction of radiation (Gritschneder, in preparation).

\begin{figure*}
\mbox{
\subfloat[]{\includegraphics[scale=0.45]{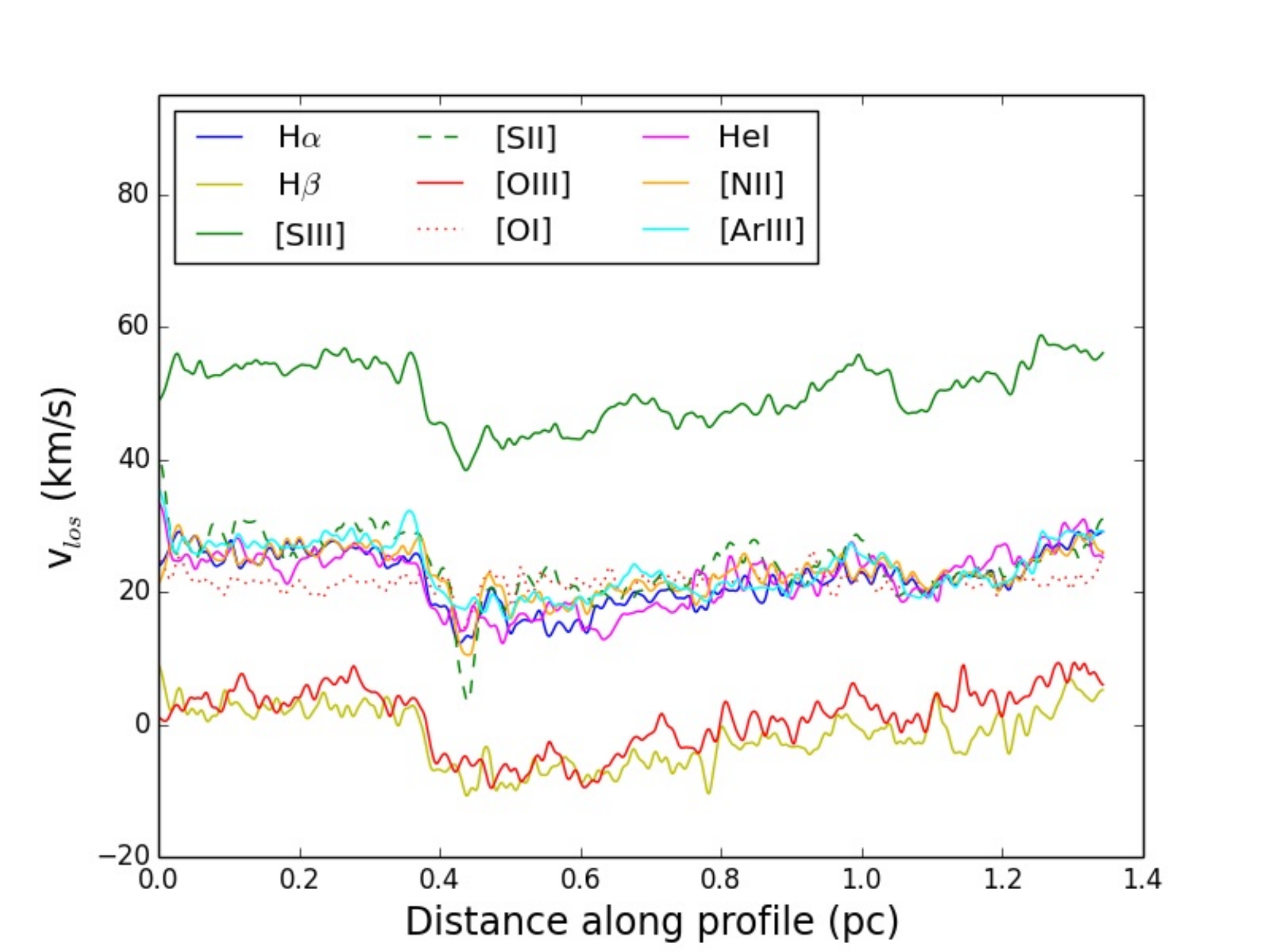}}
\subfloat[]{\includegraphics[scale=0.45]{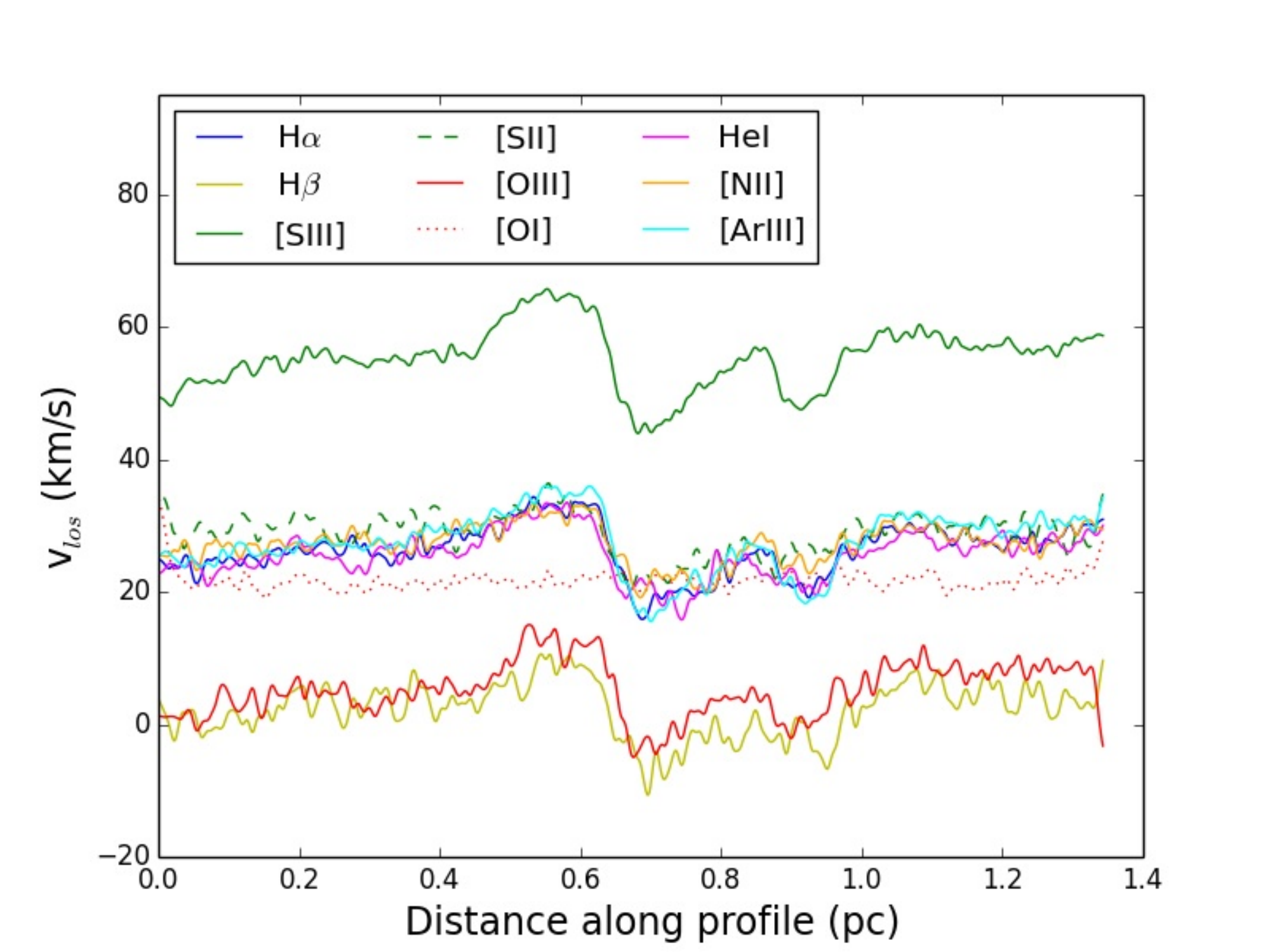}}}
\mbox{
\subfloat[]{\includegraphics[scale=0.45]{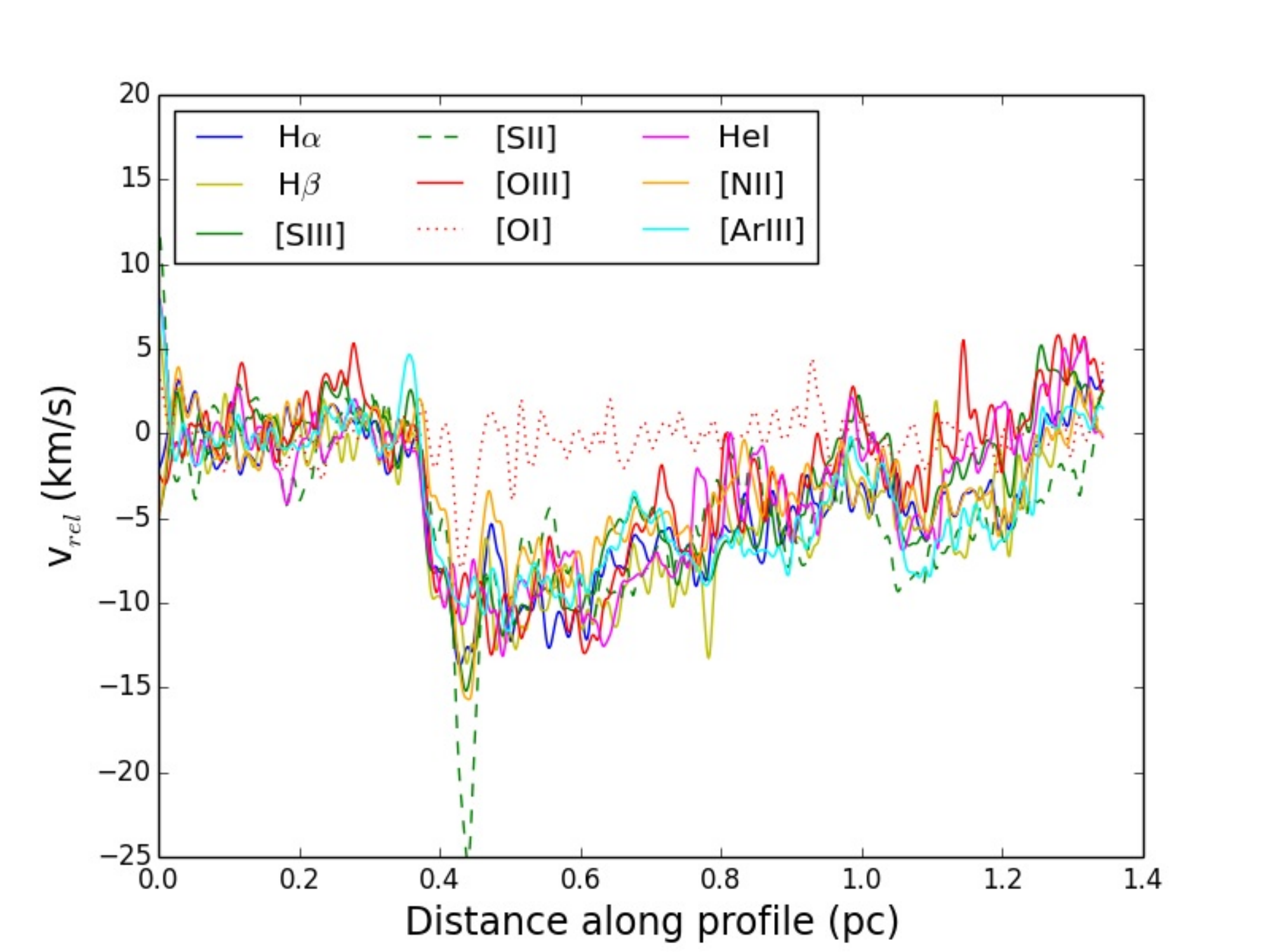}}
\subfloat[]{\includegraphics[scale=0.45]{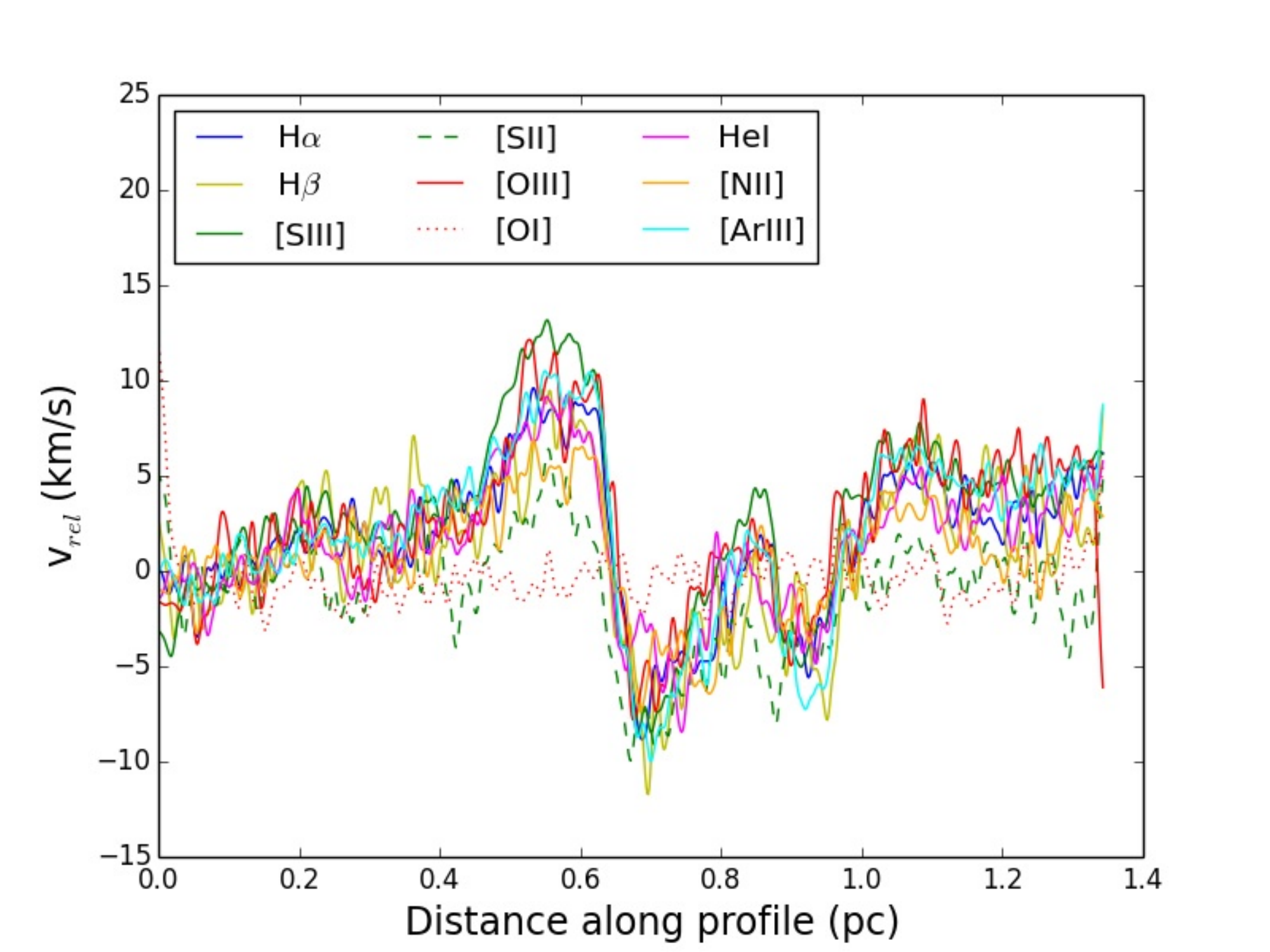}}}
  \caption{Velocity profile of P2 (panels a and c) and P3 (panels b and d) along the same slits as Fig. \ref{intprof}. The lines are: H$\alpha$ (blue), H$\beta$ (yellow), [SIII]9068 (solid green), [SII]6717,31 (dashed green), [OIII]4969,5007 (solid red),  [OI]5577,6300 (dotted red), HeI 5876,6678 (magenta), [ArIII]7135,7751 (cyan), [NII]6548,6584 (orange). Panels (a) and (b) show the observed line of sight velocity (not corrected for radial velocity) with the wavelength-dependent offset between the red ([SIII]), blue ([OIII], H$\beta$) and green (rest) lines originating from the wavelength calibration applied during the observations (see text). Panels (c) and (d) are the same as the other two, but the velocities are now relative to the mean velocity of the region between 0 and 0.2 pc (corresponding to the HII region). See text Section 4.1 for discussion.}
  \label{pv_profile}
\end{figure*}

\begin{figure}
\includegraphics[scale=0.45]{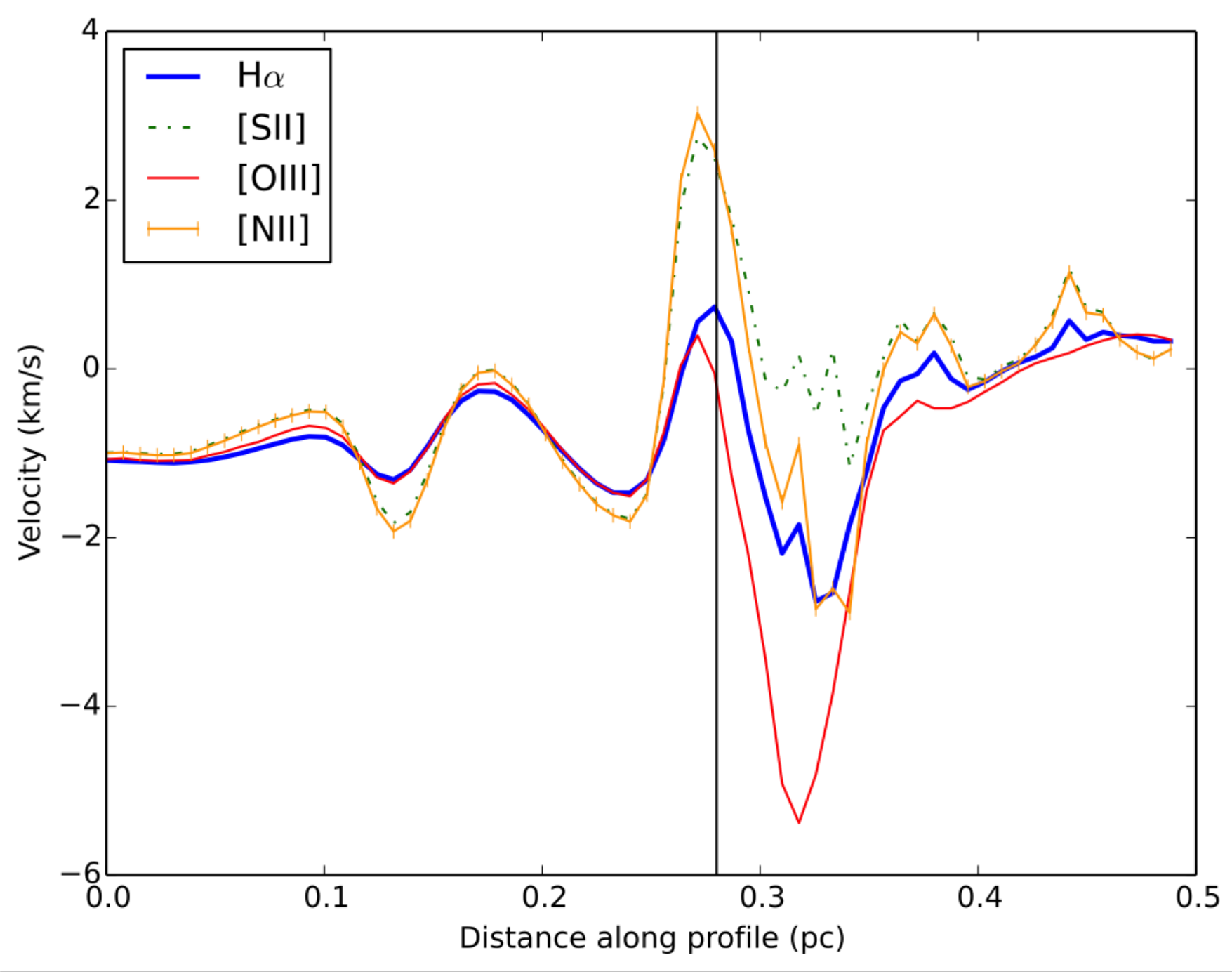}
  \caption{Velocity profile of a simulated pillar along the same slit as Fig. \ref{intprof_sim}b.}
  \label{pv_sim}
\end{figure}

Knowing that the ionisation structure of the photo-evaporative flow is (spatially) stratified, the question whether these emission lines are also offset in position-velocity space occurs naturally. Fig. \ref{pv_profile} (panels c and d) shows the relative velocity\footnote{relative to the HII region} of the single atomic species (obtained by simultaneously fitting all the detected lines for each species) of P2 and P3 along the same slits used for the intensity profiles shown in Fig. \ref{intprof}. The pillar-ambient interface is clearly recognisable from the dip in velocity of $\sim$ 7 and 10 km s$^{-1}$ for P3 and P2 respectively, the dip toward bluer velocities occurring because we are probing the radially outward directed photo-evaporative flow which is thus moving toward us along the line of sight. The neutral [OI] line does not follow the same rise-and-fall profile as the ionisation-tracing lines, while  The [OII] lines are not shown, as their weak intensity to produce a reliable velocity map. Panels (a) and (b) of Fig. \ref{pv_profile} show the uncorrected (topocentric) line of sight velocity: the redshift of the [SIII] line and the blueshift of the H$\beta$ and [OIII] lines are due to the wavelength calibration in the MUSE pipeline, as described at the beginning of this section.
 Fig. \ref{pv_sim} shows the analogous plot for a synthetic pillar: the profile seen in the observations is well recovered by our simulations, but the magnitude of the blueshift is of the order of about 3 - 5 km s$^{-1}$. As already discussed, we don't expect for the simulations and the observations to agree quantitatively because of the different physical conditions.

\subsection{Geometry}
\begin{figure}
\includegraphics[scale=0.45]{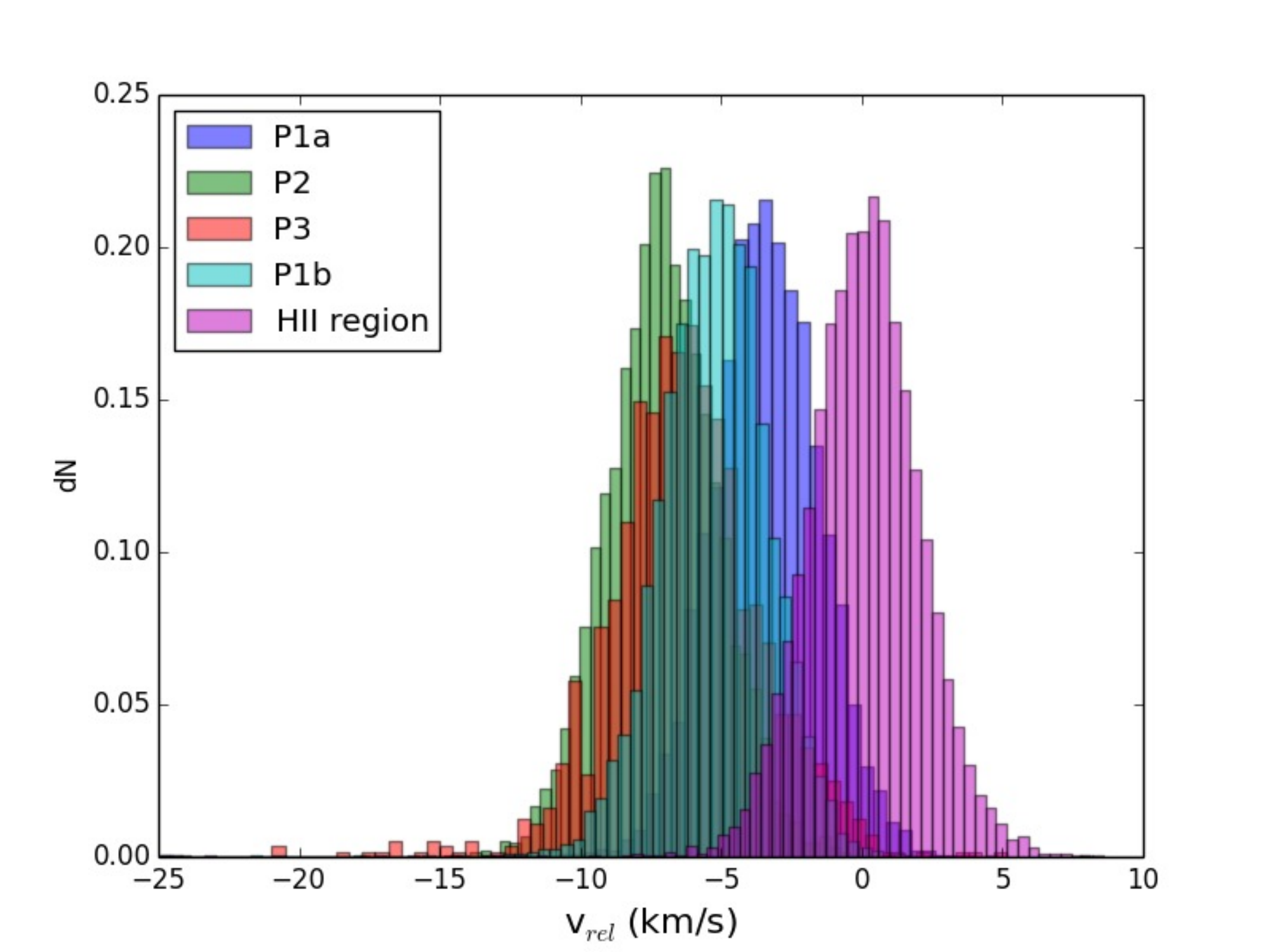}
  \caption{The histograms of the velocity (relative to the HII region, see Section 4.1) of the pillars and the HII region obtained from circular regions extracted from the velocity map.}
  \label{histo}
\end{figure}

Another interesting open question is the 3D geometry of the Pillars and their spatial orientation with respect to NGC 6611, and because for the first time we have the velocity information combined with an extinction map and integrated line intensities for the entire structures, this can now be addressed. In our attempt to understand the geometry of the Pillars, we consider the following:

\begin{itemize}
\item from the integrated intensity and electron density maps it is clear that ionisation mainly occurs at the tip of P1 and P2, that P3 seems to be less exposed to the ionising radiation on the side we can see, and that the exposed protrusions along P1 are also being (though somewhat less than the tip) ionised 
\item the extinction map (Fig. \ref{hahb}b) shows a gradient from P1 (most extinction) to P3 (least extinction)
\item the tips of P1 and P2 are seen as a reflection nebulae in the continuum 3-color composite (Fig. \ref{30sRGB}), while P3 is more seen as a silhouette illuminated from behind
\item the three pillars have different velocities, as is shown in Fig. \ref{histo} and Table \ref{vlos}
\item the velocity of the lower part of P1 is consistent with the velocity of P2 and P3, and seems inconsistent with the upper part of P1
\end{itemize}

\begin{figure*}
\includegraphics[scale=0.4]{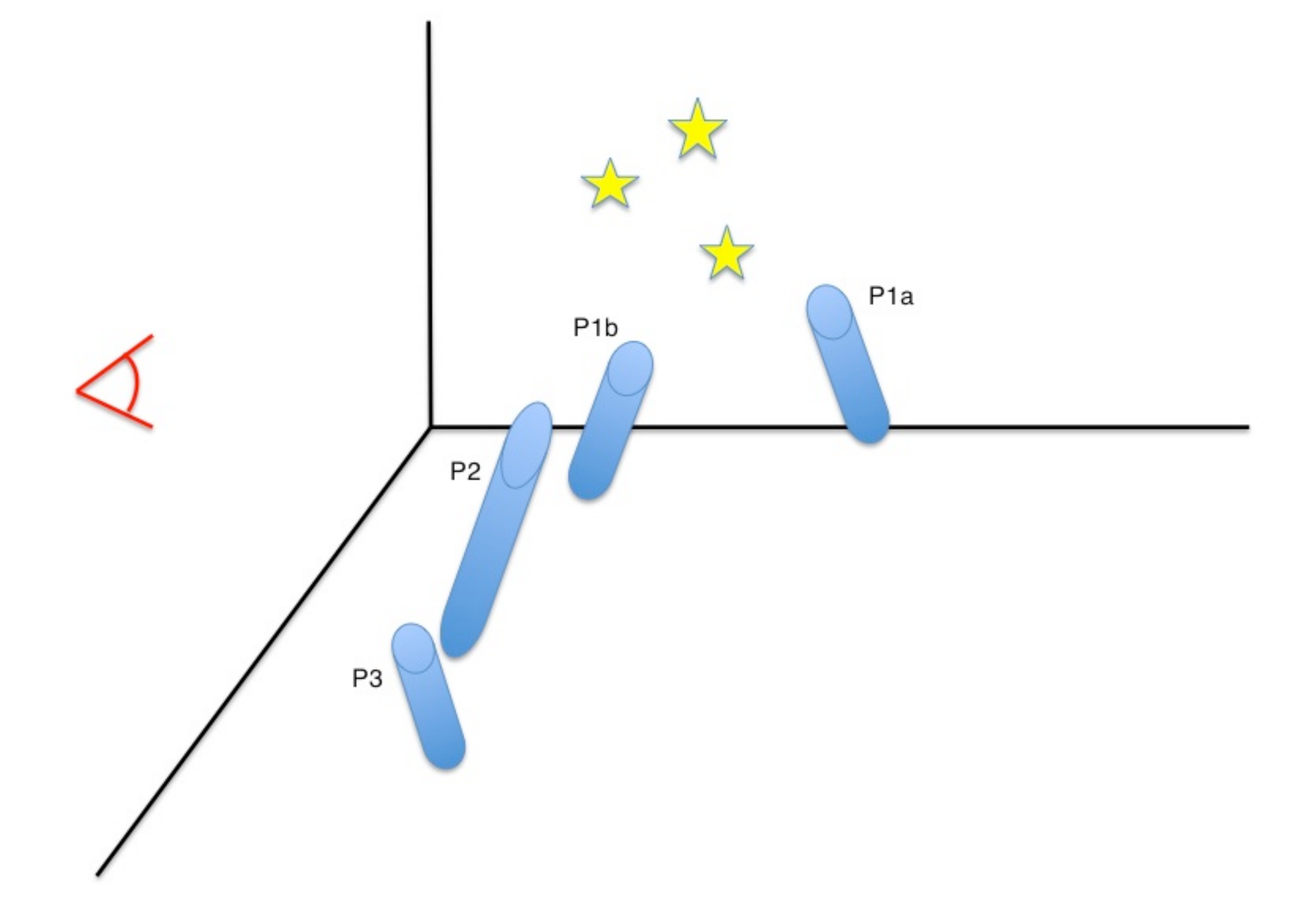}
  \caption{Sketch of the 3D geometry of the Pillars of Creation with respect to the ionising stars in NGC 6611 (not to scale). See text Section 4.2.}
  \label{sketchi}
\end{figure*}

With molecular line data, \cite{1998ApJ...493L.113P} find that P1 is in reality composed of two separate parts, the upper part (P1a) being behind the ionising stars of NGC 6611 with a tail inclined away from us and the lower part (P1b) being in front of them with its tail is inclined toward us. P2 is a single structure with its tail pointing toward us, while for P3 their data is inconsistent and inconclusive. From our observations it is clear that P1a is the least blue-shifted and the most extincted of the three pillars, and its tip shows the highest degree of ionisation. Together with the positive velocity gradient reported in \cite{1998ApJ...493L.113P} and the fact that the protrusions along its body show weak signs of ionisation, we confirm that P1a is behind the ionising O stars and its tip is pointing toward us along the line of sight, while P1b is in front of the stars and inclined with its tip pointing away from us. Because of the lack of ionisation along their bodies, we suggest that both P2 and P3 are also in front of the ionising sources and that both are inclined with their tips pointing away and their tails pointing toward us, that P3 is closer to the observer but because it is slightly less blue-shifted and its tip shows less ionisation than P2, it is inclined with its tip pointing toward us. We sketch this in Fig. \ref{sketchi} to make it easier for the reader to visualise. 

\begin{table}
\begin{center}
\caption{Line of sight velocities relative to the HII region derived from fitting gaussian distributions to the histograms in Fig. \ref{histo} (see Section 4.1).}
\begin{tabular}{lcccc}
\hline
\hline
& P1a & P1b & P2 & P3  \\
 \hline
 v$_{rel}$ (km s$^{-1}$) & -3.43 & -5.34 & -7.14 & -6.62 \\
\hline
\label{vlos}
\end{tabular}
\end{center}
\end{table}

\subsection{Mass loss rate and lifetime}

Because forbidden lines such as [OI]$\lambda$6300 or [SII]$\lambda$6731 are optically thin they can be used to estimate the mass of the line emitting matter, since the line luminosity is proportional to the number of line emitting atoms along the line of sight. For the pillar-ambient interface region (T$_{e}\approx$ 11870 K, N$_{e}\approx$ 2800 cm$^{-3}$, Fig. \ref{NT}), the ionisation fraction is f$_{ion}\sim$ 0.01 and the critical density is N$_{c, [SII]\lambda6731}\cong$ 3.9$\times$10$^{3}$ cm$^{-3}$, and therefore N$ > $N$_{c}$. Because the signal-to-noise ratio is much higher for the [SII] line than for the [OI] line, we use the former to derive the mass loss rate of the pillars. This line originates from the transition from the $^{2}$D$_{3/2}$ (level 1) to the $^{4}$S$_{3/2}$ (level 2) transition which occurs with a rate A$_{21}$ = 8.8$\times$10$^{-4}$ s$^{-1}$ \citep{2006agna.book.....O}. Following the method described in \cite{1995ApJ...452..736H}, we compute the mass of the line emitting matter with expressions for the luminosity L$_{6731}$ of the [SII]$\lambda$6731 line and the total number of sulfur atoms in the lower level, $\eta_{1}$

$$L_{6731}=\frac{g_{2}}{g_{1}}\eta_{1}\quad \exp\left(-\frac{h\nu_{12}}{kT}\right)\left(1+\frac{N_{c}}{N_{e}}\right)^{-1}A_{21}h\nu_{21}$$

and

$$\eta_{1}=\left[\frac{\eta_{1}}{\eta(S)}\right]\left[\frac{\eta(S)}{\eta(H)}\right]\left[\frac{\eta(H)}{\eta_{tot}}\right]\frac{M_{tot}}{\mu m_{H}}$$

which correspond to equations A2 and A5 in \cite{1995ApJ...452..736H} and where g$_{1}$ = g$_{2}$ = 4 are the statistical weights of the two levels, $h\nu_{12}$= 2.95 $\times$10$^{-12}$ erg is the energy of the transition, $\eta$(S)/$\eta$(H) = 5.88$\times$10$^{-5}$ (Table \ref{abun}), $\eta_{1}$/$\eta(S)\approx$ 0.9 (computed with the \texttt{IRAF} task \texttt{ionic}), and assuming $\eta$(H)/$\eta_{tot}\approx$ 0.921 and $\mu$ = 1.24 \citep{1973asqu.book.....A}. With an electron density of 2800 cm$^{-3}$ and a temperature of about 10$^{4}$ K characteristic for the pillar tips, the mass of the line emitting matter then becomes

$$M \simeq 2.69\times10^{-4}\left(1+\frac{N_{c}}{N_{e}}\right)\left(\frac{L_{6731}}{L_{\odot}}\right)M_{\odot}$$

The mass loss rate can then be computed via $\dot{M}$ = Mv/$l$, where v is the velocity of the photo evaporative outflow and $l$ the average size of the line emitting region. In order to estimate the total luminosity of the [SII]$\lambda6731$ line we subtracted the stellar continuum in the [SII]$\lambda6731$ integrated intensity map and summed the line intensity of all pixels to obtain L$_{6731}\cong$ 130 L$_{\odot}$. With v $\approx$ 8 km s$^{-1}$ from the velocity map in Fig. \ref{vel} and $l$ $\cong$ 3 $\times$10$^{16}$ cm (measured from the SII intensity map), $\dot{M}\approx$ 70 M$_{\odot}$ Myr$^{-1}$. With a total mass of the pillars of M$_{tot}\sim$ 200 M$_{\odot}$ \citep{1999A&A...342..233W}, we estimate the lifetime of the pillars to be approximately 3 Myr. This however is an upper limit, as the [SII] line may not be tracing the bulk flow and we therefore might be underestimating the mass loss rate. \cite{1998ApJ...493L.113P} estimated the expected lifetime of the Pillars to be about 20 Myr, but as suggested by \citealt{2013MNRAS.435...30W} (who find a lifetime of about 2 Myr for pillars near NGC 3603) we believe that the used value of the curvature radius was too small, so they overestimated the lifetime.     

We suggest that the identified outflow will not contribute significantly to the mass loss rate. Indeed, we compute the mass of the outflow to be $\dot{M}\cong$ 0.29 M$_{\odot}$ Myr$^{-1}$ (with N$_{e}\sim$ 1400 cm$^{-3}$, L$_{6731}\simeq$ 0.18 L$_{\odot}$), which is only $\sim$ 0.4 \% of the total mass loss rate.

\section{Conclusions}
For the first time the Pillars of Creation were imaged using optical integral field spectroscopy. This technique allowed us to map the entire structures and obtain medium resolution spectra for each pixel in the 3x3 arcmin$^{2}$ mosaic, which were fitted with Gaussian profiles to produce integrated line and line of sight velocity maps in many spectral lines. Also, we computed physical parameters such as the electron density, the electron temperature and the extinction for the entire mosaic. 

In agreement with \cite{1998ApJ...493L.113P}, we find that the three Pillars are actually 4 distinct structures that lie at different distances along the line of sight and have different inclination angles, some pointing away and others pointing toward the observer. The consequence of their position and inclination is a different degree of ionisation not only of the pillar tips, but also of the pillar bodies. We therefore see the strongest emission from ionised species from the tip of P1b, which is pointing toward us but lies behind the ionising O stars of NGC 6611 and we therefore see weak emission coming from the exposed structures along its body as well. This we do not see in P1b, P2 and P3, as these all lie in front of the ionising stars, and furthermore for P1b and P3 because they are inclined with their tips facing the observer and their tails pointing away from us. The effect of their different inclination angles is that the 4 structures show different velocities: all of them are blue-shifted with respect to the ambient HII region because we are tracing the photo-evaporative flow, but because this flow is normal to their surfaces, we will see them at different velocities.

The kinematic analysis also allowed us to reveal a possible protostellar outflow at the tip of P2. Of this outflow, we can identify both lobes as a blue and a redshifted counterpart. The jet seems to be almost parallel to the incident ionising radiation, in fact it is inclined by only 7 degrees with respect to the pillar body and the radiation. This does not seem to agree with what we find in our smoothed particle hydrodynamic simulations, in which we also find stars forming at the pillar tips, but their accretion disks are preferentially aligned with the direction of radiation. We explain this by arguing that the angle between the outflow and the pillar bodies is the consequence of projection effects, as is discussed in \cite{2010RMxAA..46..179R}. The outflow is clearly traced by the S$_{23}$ parameter, which, if plotted against an indicator of the degree of ionisation such as [OII]/[OIII], can be used in future studies with IFU data of star forming regions to detect outflows in dense molecular material. In our data this is the only outflow we can detect in both the velocity map and via the S$_{23}$ parameter. From the kinematics we also identify a candidate outflow on the inner side of P1, which however does not show a sulfur abundance. Using IFU data it would be interesting to apply this method to other star forming regions that have an embedded young stellar population to look for outflows that have not yet fully emerged from the molecular material and would therefore have been missed in previous studies. Furthermore, the outflow and its driving source be an interesting target for the Atacama Large Millimeter/submillimeter Array, which has the required high sensitivity and the frequency coverage to analyse the velocity field and the outflow in great detail.

The Pillars show a classical stratified ionisation structure in which line emitting species peak in a sequential manner throughout the ionisation interface of the molecular material, in  agreement with previous authors. Typical electron densities at the pillar tips are $>$ 2000 cm$^{-3}$, whereas the electron temperature in this region is in the 8000 K - 10 000 K range. The comparison with our SPH+radiative transfer computations is encouraging, as we are able to reproduce the ionisation structure, the kinematics and physical parameters. Finally, by computing the mass of the [SII] line-emitting gas, we estimate the mass loss rate to be of about 70 M$_{\odot}$ Myr$^{-1}$, which, assuming the total mass of the Pillars to be $\sim$ 200 M$_{\odot}$, yields an expected lifetime of about 3 Myr, timescale which does not depend of the presence of an outflow at the tip of the middle pillar.

\section*{Acknowledgments}
This article is based on data obtained with ESO telescopes at the Paranal Observatory under programme ID 60.A-9309(A). We would like to thank Bernd Husemann for the valuable help during data reduction and Eric Feigelson for the insight on statistics and classification. The data reduction and analysis make use of \textsc{aplpy} (http:$//$aplpy.github.io), \textsc{spectral$\_$cube} (spectral-cube.readthedocs.org), \textsc{pyspeckit} (pyspeckit.bitbucket.org, \cite{2011ascl.soft09001G}) and \textsc{glue} (glueviz.org). This research was supported by the DFG cluster of excellence \textit{Origin and Structures of the Universe} (JED, BE).

\bibliography{musebib}

\begin{thebibliography}{}

\bibitem[\protect\citeauthoryear{{Allen}}{{Allen}}{1973}]{1973asqu.book.....A}
{Allen} C.~W.,  1973, {Astrophysical quantities}.
{University of London, Athlone Press, 3rd ed.}

\bibitem[\protect\citeauthoryear{{Allen}, {Groves}, {Dopita}, {Sutherland} \&
  {Kewley}}{{Allen} et~al.}{2008}]{2008ApJS..178...20A}
{Allen} M.~G.,  {Groves} B.~A.,  {Dopita} M.~A.,  {Sutherland} R.~S.,
  {Kewley} L.~J.,  2008, \apjs, 178, 20

\bibitem[\protect\citeauthoryear{{Bacon} et~al.,}{{Bacon}
  et~al.}{2010}]{2010SPIE.7735E..08B}
{Bacon} R.,  et~al., 2010, in Society of Photo-Optical Instrumentation
  Engineers (SPIE) Conference Series Vol.~7735 of Society of Photo-Optical
  Instrumentation Engineers (SPIE) Conference Series, {The MUSE
  second-generation VLT instrument}.
p.~8

\bibitem[\protect\citeauthoryear{{Baldwin}, {Phillips} \&
  {Terlevich}}{{Baldwin} et~al.}{1981}]{1981PASP...93....5B}
{Baldwin} J.~A.,  {Phillips} M.~M.,    {Terlevich} R.,  1981, \pasp, 93, 5

\bibitem[\protect\citeauthoryear{{Bally}, {Reipurth} \& {Davis}}{{Bally}
  et~al.}{2007}]{2007prpl.conf..215B}
{Bally} J.,  {Reipurth} B.,    {Davis} C.~J.,  2007, Protostars and Planets V,
  pp 215--230

\bibitem[\protect\citeauthoryear{{Bisbas}, {W{\"u}nsch}, {Whitworth}, {Hubber}
  \& {Walch}}{{Bisbas} et~al.}{2011}]{2011ApJ...736..142B}
{Bisbas} T.~G.,  {W{\"u}nsch} R.,  {Whitworth} A.~P.,  {Hubber} D.~A.,
  {Walch} S.,  2011, \apj, 736, 142

\bibitem[\protect\citeauthoryear{{Cardelli}, {Clayton} \& {Mathis}}{{Cardelli}
  et~al.}{1989}]{1989ApJ...345..245C}
{Cardelli} J.~A.,  {Clayton} G.~C.,    {Mathis} J.~S.,  1989, \apj, 345, 245

\bibitem[\protect\citeauthoryear{{Chini} \& {Wargau}}{{Chini} \&
  {Wargau}}{1990}]{1990A&A...227..213C}
{Chini} R.,  {Wargau} W.~F.,  1990, \aap, 227, 213

\bibitem[\protect\citeauthoryear{{Dale}, {Ercolano} \& {Bonnell}}{{Dale}
  et~al.}{2012}]{2012MNRAS.427.2852D}
{Dale} J.~E.,  {Ercolano} B.,    {Bonnell} I.~A.,  2012, \mnras, 427, 2852

\bibitem[\protect\citeauthoryear{{Dale}, {Ercolano} \& {Bonnell}}{{Dale}
  et~al.}{2013a}]{2013MNRAS.431.1062D}
{Dale} J.~E.,  {Ercolano} B.,    {Bonnell} I.~A.,  2013a, \mnras, 431, 1062

\bibitem[\protect\citeauthoryear{{Dale}, {Ercolano} \& {Bonnell}}{{Dale}
  et~al.}{2013b}]{2013MNRAS.430..234D}
{Dale} J.~E.,  {Ercolano} B.,    {Bonnell} I.~A.,  2013b, \mnras, 430, 234

\bibitem[\protect\citeauthoryear{{Ercolano}, {Barlow} \& {Storey}}{{Ercolano}
  et~al.}{2005}]{2005MNRAS.362.1038E}
{Ercolano} B.,  {Barlow} M.~J.,    {Storey} P.~J.,  2005, \mnras, 362, 1038

\bibitem[\protect\citeauthoryear{{Ercolano}, {Barlow}, {Storey} \&
  {Liu}}{{Ercolano} et~al.}{2003}]{2003MNRAS.340.1136E}
{Ercolano} B.,  {Barlow} M.~J.,  {Storey} P.~J.,    {Liu} X.-W.,  2003, \mnras,
  340, 1136

\bibitem[\protect\citeauthoryear{{Ercolano}, {Dale}, {Gritschneder} \&
  {Westmoquette}}{{Ercolano} et~al.}{2012}]{2012MNRAS.420..141E}
{Ercolano} B.,  {Dale} J.~E.,  {Gritschneder} M.,    {Westmoquette} M.,  2012,
  \mnras, 420, 141

\bibitem[\protect\citeauthoryear{{Ercolano} \& {Gritschneder}}{{Ercolano} \&
  {Gritschneder}}{2011}]{2011MNRAS.413..401E}
{Ercolano} B.,  {Gritschneder} M.,  2011, \mnras, 413, 401

\bibitem[\protect\citeauthoryear{{Ercolano}, {Young}, {Drake} \&
  {Raymond}}{{Ercolano} et~al.}{2008}]{2008ApJS..175..534E}
{Ercolano} B.,  {Young} P.~R.,  {Drake} J.~J.,    {Raymond} J.~C.,  2008,
  \apjs, 175, 534

\bibitem[\protect\citeauthoryear{{Ferland}, {Porter}, {van Hoof}, {Williams},
  {Abel}, {Lykins}, {Shaw}, {Henney} \& {Stancil}}{{Ferland}
  et~al.}{2013}]{2013RMxAA..49..137F}
{Ferland} G.~J.,  {Porter} R.~L.,  {van Hoof} P.~A.~M.,  {Williams} R.~J.~R.,
  {Abel} N.~P.,  {Lykins} M.~L.,  {Shaw} G.,  {Henney} W.~J.,    {Stancil}
  P.~C.,  2013, \rmxaa, 49, 137

\bibitem[\protect\citeauthoryear{{Fukuda}, {Hanawa} \& {Sugitani}}{{Fukuda}
  et~al.}{2002}]{2002ApJ...568L.127F}
{Fukuda} N.,  {Hanawa} T.,    {Sugitani} K.,  2002, \apjl, 568, L127

\bibitem[\protect\citeauthoryear{{Garc{\'{\i}}a-Benito}, {D{\'{\i}}az},
  {H{\"a}gele}, {P{\'e}rez-Montero}, {L{\'o}pez}, {V{\'{\i}}lchez},
  {P{\'e}rez}, {Terlevich}, {Terlevich} \&
  {Rosa-Gonz{\'a}lez}}{{Garc{\'{\i}}a-Benito}
  et~al.}{2010}]{2010MNRAS.408.2234G}
{Garc{\'{\i}}a-Benito} R.,  {D{\'{\i}}az} A.,  {H{\"a}gele} G.~F.,
  {P{\'e}rez-Montero} E.,  {L{\'o}pez} J.,  {V{\'{\i}}lchez} J.~M.,
  {P{\'e}rez} E.,  {Terlevich} E.,  {Terlevich} R.,    {Rosa-Gonz{\'a}lez} D.,
  2010, \mnras, 408, 2234

\bibitem[\protect\citeauthoryear{{Ginsburg} \& {Mirocha}}{{Ginsburg} \&
  {Mirocha}}{2011}]{2011ascl.soft09001G}
{Ginsburg} A.,  {Mirocha} J., , 2011, {PySpecKit: Python Spectroscopic
  Toolkit}, Astrophysics Source Code Library

\bibitem[\protect\citeauthoryear{{Gritschneder}, {Burkert}, {Naab} \&
  {Walch}}{{Gritschneder} et~al.}{2010}]{2010ApJ...723..971G}
{Gritschneder} M.,  {Burkert} A.,  {Naab} T.,    {Walch} S.,  2010, \apj, 723,
  971

\bibitem[\protect\citeauthoryear{{Gritschneder}, {Naab}, {Burkert}, {Walch},
  {Heitsch} \& {Wetzstein}}{{Gritschneder} et~al.}{2009}]{2009MNRAS.393...21G}
{Gritschneder} M.,  {Naab} T.,  {Burkert} A.,  {Walch} S.,  {Heitsch} F.,
  {Wetzstein} M.,  2009, \mnras, 393, 21

\bibitem[\protect\citeauthoryear{{Guarcello}, {Micela}, {Peres}, {Prisinzano}
  \& {Sciortino}}{{Guarcello} et~al.}{2010}]{2010A&A...521A..61G}
{Guarcello} M.~G.,  {Micela} G.,  {Peres} G.,  {Prisinzano} L.,    {Sciortino}
  S.,  2010, \aap, 521, A61

\bibitem[\protect\citeauthoryear{{H{\"a}gele}, {D{\'{\i}}az}, {Terlevich},
  {Terlevich}, {P{\'e}rez-Montero} \& {Cardaci}}{{H{\"a}gele}
  et~al.}{2008}]{2008MNRAS.383..209H}
{H{\"a}gele} G.~F.,  {D{\'{\i}}az} {\'A}.~I.,  {Terlevich} E.,  {Terlevich} R.,
   {P{\'e}rez-Montero} E.,    {Cardaci} M.~V.,  2008, \mnras, 383, 209

\bibitem[\protect\citeauthoryear{{Hartigan}, {Edwards} \&
  {Ghandour}}{{Hartigan} et~al.}{1995}]{1995ApJ...452..736H}
{Hartigan} P.,  {Edwards} S.,    {Ghandour} L.,  1995, \apj, 452, 736

\bibitem[\protect\citeauthoryear{{Healy}, {Hester} \& {Claussen}}{{Healy}
  et~al.}{2004}]{2004ApJ...610..835H}
{Healy} K.~R.,  {Hester} J.~J.,    {Claussen} M.~J.,  2004, \apj, 610, 835

\bibitem[\protect\citeauthoryear{{Hester}}{{Hester}}{1991}]{1991PASP..103..853H}
{Hester} J.~J.,  1991, \pasp, 103, 853

\bibitem[\protect\citeauthoryear{{Hester} et~al.,}{{Hester}
  et~al.}{1996}]{1996AJ....111.2349H}
{Hester} J.~J.,  et~al., 1996, \aj, 111, 2349

\bibitem[\protect\citeauthoryear{{Hill}, {Reynolds}, {Benjamin} \&
  {Haffner}}{{Hill} et~al.}{2007}]{2007ASPC..365..250H}
{Hill} A.~S.,  {Reynolds} R.~J.,  {Benjamin} R.~A.,    {Haffner} L.~M.,  2007,
  in {Haverkorn} M.,  {Goss} W.~M.,  eds, SINS - Small Ionized and Neutral
  Structures in the Diffuse Interstellar Medium Vol.~365 of Astronomical
  Society of the Pacific Conference Series, {Density Distribution of the Warm
  Ionized Medium}.
p.~250

\bibitem[\protect\citeauthoryear{{Hillenbrand}, {Massey}, {Strom} \&
  {Merrill}}{{Hillenbrand} et~al.}{1993}]{1993AJ....106.1906H}
{Hillenbrand} L.~A.,  {Massey} P.,  {Strom} S.~E.,    {Merrill} K.~M.,  1993,
  \aj, 106, 1906

\bibitem[\protect\citeauthoryear{{Indebetouw}, {Robitaille}, {Whitney},
  {Churchwell}, {Babler}, {Meade}, {Watson} \& {Wolfire}}{{Indebetouw}
  et~al.}{2007}]{2007ApJ...666..321I}
{Indebetouw} R.,  {Robitaille} T.~P.,  {Whitney} B.~A.,  {Churchwell} E.,
  {Babler} B.,  {Meade} M.,  {Watson} C.,    {Wolfire} M.,  2007, \apj, 666,
  321

\bibitem[\protect\citeauthoryear{{Kauffmann}, {Heckman}, {Tremonti},
  {Brinchmann}, {Charlot}, {White}, {Ridgway}, {Brinkmann}, {Fukugita}, {Hall},
  {Ivezi{\'c}}, {Richards} \& {Schneider}}{{Kauffmann}
  et~al.}{2003}]{2003MNRAS.346.1055K}
{Kauffmann} G.,  {Heckman} T.~M.,  {Tremonti} C.,  {Brinchmann} J.,  {Charlot}
  S.,  {White} S.~D.~M.,  {Ridgway} S.~E.,  {Brinkmann} J.,  {Fukugita} M.,
  {Hall} P.~B.,  {Ivezi{\'c}} {\v Z}.,  {Richards} G.~T.,    {Schneider} D.~P.,
   2003, \mnras, 346, 1055

\bibitem[\protect\citeauthoryear{{Kewley}, {Dopita}, {Sutherland}, {Heisler} \&
  {Trevena}}{{Kewley} et~al.}{2001}]{2001ApJ...556..121K}
{Kewley} L.~J.,  {Dopita} M.~A.,  {Sutherland} R.~S.,  {Heisler} C.~A.,
  {Trevena} J.,  2001, \apj, 556, 121

\bibitem[\protect\citeauthoryear{{Linsky}, {Gagn{\'e}}, {Mytyk}, {McCaughrean}
  \& {Andersen}}{{Linsky} et~al.}{2007}]{2007ApJ...654..347L}
{Linsky} J.~L.,  {Gagn{\'e}} M.,  {Mytyk} A.,  {McCaughrean} M.,    {Andersen}
  M.,  2007, \apj, 654, 347

\bibitem[\protect\citeauthoryear{{McCall}}{{McCall}}{1984}]{1984MNRAS.208..253M}
{McCall} M.~L.,  1984, \mnras, 208, 253

\bibitem[\protect\citeauthoryear{{McCaughrean} \& {Andersen}}{{McCaughrean} \&
  {Andersen}}{2002}]{2002A&A...389..513M}
{McCaughrean} M.~J.,  {Andersen} M.,  2002, \aap, 389, 513

\bibitem[\protect\citeauthoryear{{Miao}, {White}, {Nelson}, {Thompson} \&
  {Morgan}}{{Miao} et~al.}{2006}]{2006MNRAS.369..143M}
{Miao} J.,  {White} G.~J.,  {Nelson} R.,  {Thompson} M.,    {Morgan} L.,  2006,
  \mnras, 369, 143

\bibitem[\protect\citeauthoryear{{Mizuta}, {Kane}, {Pound}, {Remington},
  {Ryutov} \& {Takabe}}{{Mizuta} et~al.}{2006}]{2006ApJ...647.1151M}
{Mizuta} A.,  {Kane} J.~O.,  {Pound} M.~W.,  {Remington} B.~A.,  {Ryutov}
  D.~D.,    {Takabe} H.,  2006, \apj, 647, 1151

\bibitem[\protect\citeauthoryear{{Monreal-Ibero}, {Rela{\~n}o}, {Kehrig},
  {P{\'e}rez-Montero}, {V{\'{\i}}lchez}, {Kelz}, {Roth} \&
  {Streicher}}{{Monreal-Ibero} et~al.}{2011}]{2011MNRAS.413.2242M}
{Monreal-Ibero} A.,  {Rela{\~n}o} M.,  {Kehrig} C.,  {P{\'e}rez-Montero} E.,
  {V{\'{\i}}lchez} J.~M.,  {Kelz} A.,  {Roth} M.~M.,    {Streicher} O.,  2011,
  \mnras, 413, 2242

\bibitem[\protect\citeauthoryear{{Ngoumou}, {Hubber}, {Dale} \&
  {Burkert}}{{Ngoumou} et~al.}{2015}]{2015ApJ...798...32N}
{Ngoumou} J.,  {Hubber} D.,  {Dale} J.~E.,    {Burkert} A.,  2015, \apj, 798,
  32

\bibitem[\protect\citeauthoryear{{Oliveira}}{{Oliveira}}{2008}]{2008hsf2.book..599O}
{Oliveira} J.~M.,  2008, {Star Formation in the Eagle Nebula}.
Handbook of Star Forming Regions, Volume II, p.~599

\bibitem[\protect\citeauthoryear{{Orsatti}, {Vega} \& {Marraco}}{{Orsatti}
  et~al.}{2006}]{2006AJ....132.1783O}
{Orsatti} A.~M.,  {Vega} E.~I.,    {Marraco} H.~G.,  2006, \aj, 132, 1783

\bibitem[\protect\citeauthoryear{{Osterbrock} \& {Ferland}}{{Osterbrock} \&
  {Ferland}}{2006}]{2006agna.book.....O}
{Osterbrock} D.~E.,  {Ferland} G.~J.,  2006, {Astrophysics of gaseous nebulae
  and active galactic nuclei}.
University Science Books, 2nd ed.

\bibitem[\protect\citeauthoryear{{Pagel}, {Edmunds}, {Blackwell}, {Chun} \&
  {Smith}}{{Pagel} et~al.}{1979}]{1979MNRAS.189...95P}
{Pagel} B.~E.~J.,  {Edmunds} M.~G.,  {Blackwell} D.~E.,  {Chun} M.~S.,
  {Smith} G.,  1979, \mnras, 189, 95

\bibitem[\protect\citeauthoryear{{Peters}, {Banerjee}, {Klessen} \& {Mac
  Low}}{{Peters} et~al.}{2011}]{2011ApJ...729...72P}
{Peters} T.,  {Banerjee} R.,  {Klessen} R.~S.,    {Mac Low} M.-M.,  2011, \apj,
  729, 72

\bibitem[\protect\citeauthoryear{{Pound}}{{Pound}}{1998}]{1998ApJ...493L.113P}
{Pound} M.~W.,  1998, \apjl, 493, L113

\bibitem[\protect\citeauthoryear{{Raga}, {Lora} \& {Smith}}{{Raga}
  et~al.}{2010}]{2010RMxAA..46..179R}
{Raga} A.~C.,  {Lora} V.,    {Smith} N.,  2010, \rmxaa, 46, 179

\bibitem[\protect\citeauthoryear{{Redfield} \& {Falcon}}{{Redfield} \&
  {Falcon}}{2008}]{2008ApJ...683..207R}
{Redfield} S.,  {Falcon} R.~E.,  2008, \apj, 683, 207

\bibitem[\protect\citeauthoryear{{Seaton}}{{Seaton}}{1979}]{1979MNRAS.187P..73S}
{Seaton} M.~J.,  1979, \mnras, 187, 73P

\bibitem[\protect\citeauthoryear{{Smith}, {Bally} \& {Walborn}}{{Smith}
  et~al.}{2010}]{2010MNRAS.405.1153S}
{Smith} N.,  {Bally} J.,    {Walborn} N.~R.,  2010, \mnras, 405, 1153

\bibitem[\protect\citeauthoryear{{Sugitani}, {Tamura}, {Nakajima}, {Nagashima},
  {Nagayama}, {Nakaya}, {Pickles}, {Nagata}, {Sato}, {Fukuda} \&
  {Ogura}}{{Sugitani} et~al.}{2002}]{2002ApJ...565L..25S}
{Sugitani} K.,  {Tamura} M.,  {Nakajima} Y.,  {Nagashima} C.,  {Nagayama} T.,
  {Nakaya} H.,  {Pickles} A.~J.,  {Nagata} T.,  {Sato} S.,  {Fukuda} N.,
  {Ogura} K.,  2002, \apjl, 565, L25

\bibitem[\protect\citeauthoryear{{Sugitani}, {Watanabe}, {Tamura}, {Kandori},
  {Hough}, {Nishiyama}, {Nakajima}, {Kusakabe}, {Hashimoto}, {Nagayama},
  {Nagashima}, {Kato} \& {Fukuda}}{{Sugitani}
  et~al.}{2007}]{2007PASJ...59..507S}
{Sugitani} K.,  {Watanabe} M.,  {Tamura} M.,  {Kandori} R.,  {Hough} J.~H.,
  {Nishiyama} S.,  {Nakajima} Y.,  {Kusakabe} N.,  {Hashimoto} J.,  {Nagayama}
  T.,  {Nagashima} C.,  {Kato} D.,    {Fukuda} N.,  2007, \pasj, 59, 507

\bibitem[\protect\citeauthoryear{{Thompson}, {Smith} \& {Hester}}{{Thompson}
  et~al.}{2002}]{2002ApJ...570..749T}
{Thompson} R.~I.,  {Smith} B.~A.,    {Hester} J.~J.,  2002, \apj, 570, 749

\bibitem[\protect\citeauthoryear{{Tremblin}, {Audit}, {Minier}, {Schmidt} \&
  {Schneider}}{{Tremblin} et~al.}{2012}]{2012A&A...546A..33T}
{Tremblin} P.,  {Audit} E.,  {Minier} V.,  {Schmidt} W.,    {Schneider} N.,
  2012, \aap, 546, A33

\bibitem[\protect\citeauthoryear{{Vilchez} \& {Esteban}}{{Vilchez} \&
  {Esteban}}{1996}]{1996MNRAS.280..720V}
{Vilchez} J.~M.,  {Esteban} C.,  1996, \mnras, 280, 720

\bibitem[\protect\citeauthoryear{{Walch}, {Whitworth}, {Bisbas}, {W{\"u}nsch}
  \& {Hubber}}{{Walch} et~al.}{2013}]{2013MNRAS.435..917W}
{Walch} S.,  {Whitworth} A.~P.,  {Bisbas} T.~G.,  {W{\"u}nsch} R.,    {Hubber}
  D.~A.,  2013, \mnras, 435, 917

\bibitem[\protect\citeauthoryear{{Westmoquette}, {Dale}, {Ercolano} \&
  {Smith}}{{Westmoquette} et~al.}{2013}]{2013MNRAS.435...30W}
{Westmoquette} M.~S.,  {Dale} J.~E.,  {Ercolano} B.,    {Smith} L.~J.,  2013,
  \mnras, 435, 30

\bibitem[\protect\citeauthoryear{{White}, {Nelson}, {Holland}, {Robson},
  {Greaves}, {McCaughrean}, {Pilbratt}, {Balser}, {Oka}, {Sakamoto},
  {Hasegawa}, {McCutcheon}, {Matthews}, {Fridlund}, {Tothill}, {Huldtgren} \&
  {Deane}}{{White} et~al.}{1999}]{1999A&A...342..233W}
{White} G.~J.,  {Nelson} R.~P.,  {Holland} W.~S.,  {Robson} E.~I.,  {Greaves}
  J.~S.,  {McCaughrean} M.~J.,  {Pilbratt} G.~L.,  {Balser} D.~S.,  {Oka} T.,
  {Sakamoto} S.,  {Hasegawa} T.,  {McCutcheon} W.~H.,  {Matthews} H.~E.,
  {Fridlund} C.~V.~M.,  {Tothill} N.~F.~H.,  {Huldtgren} M.,    {Deane} J.~R.,
  1999, \aap, 342, 233

\end{thebibliography}
\bibliographystyle{mn2e}

\appendix
\section[]{}
\begin{figure*}
\mbox{
  \subfloat[]{\includegraphics[scale=0.5]{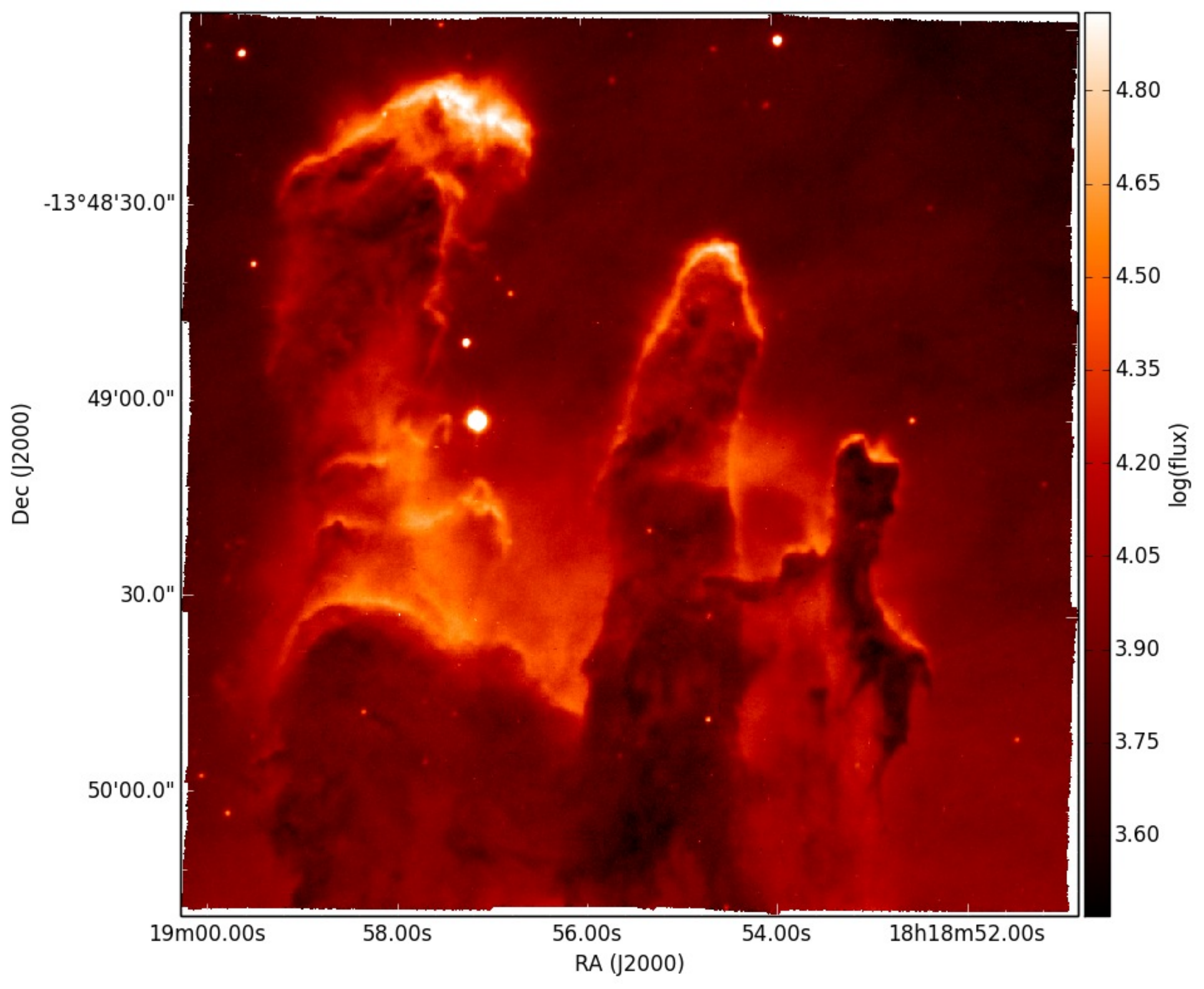}}}
  \mbox{
  \subfloat[]{\includegraphics[scale=0.5]{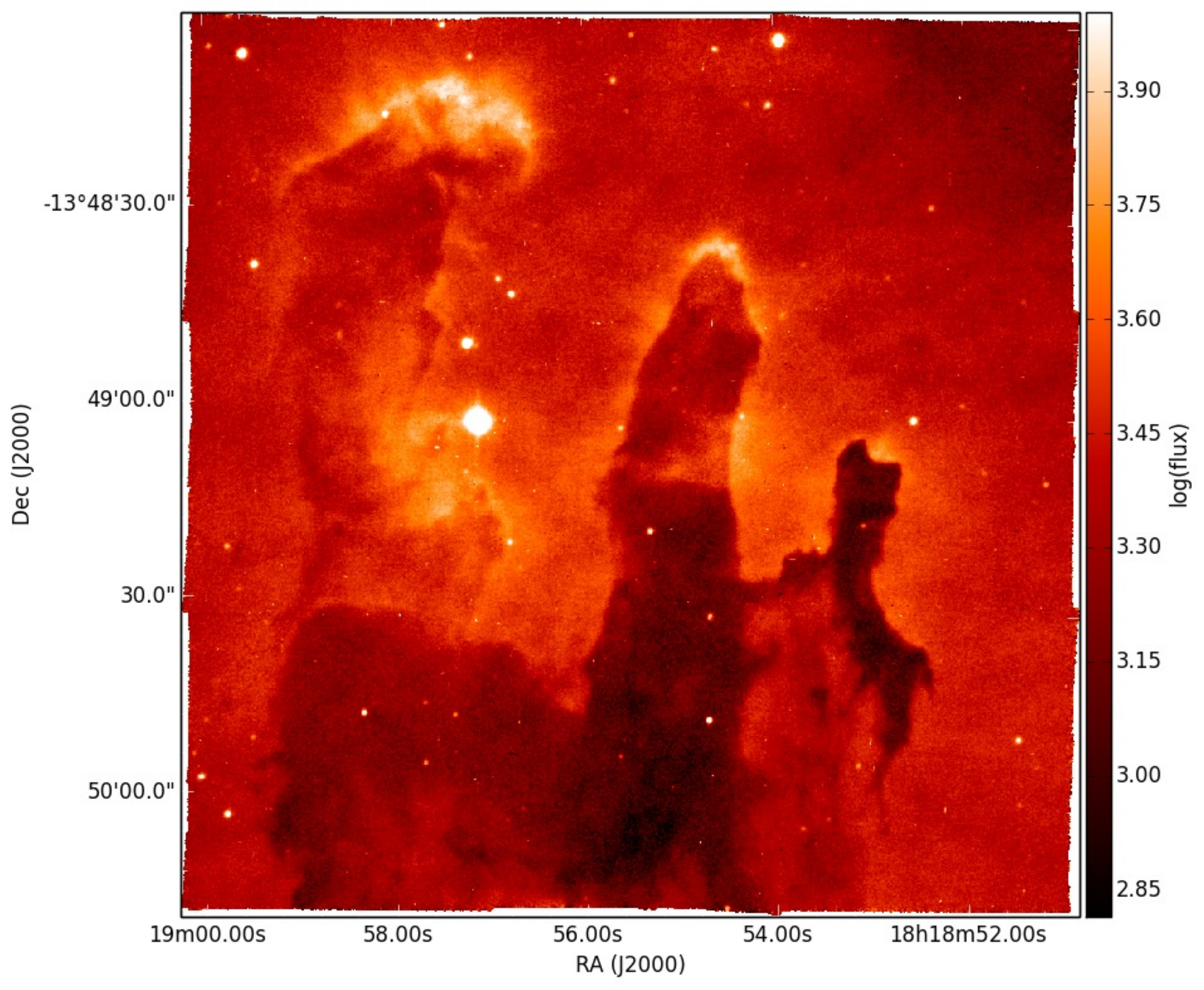}}}
   \caption{NII6584, SIII9068.}
  \label{othermaps1}
\end{figure*}

\begin{figure*}
  \mbox{
  \subfloat[]{\includegraphics[scale=0.5]{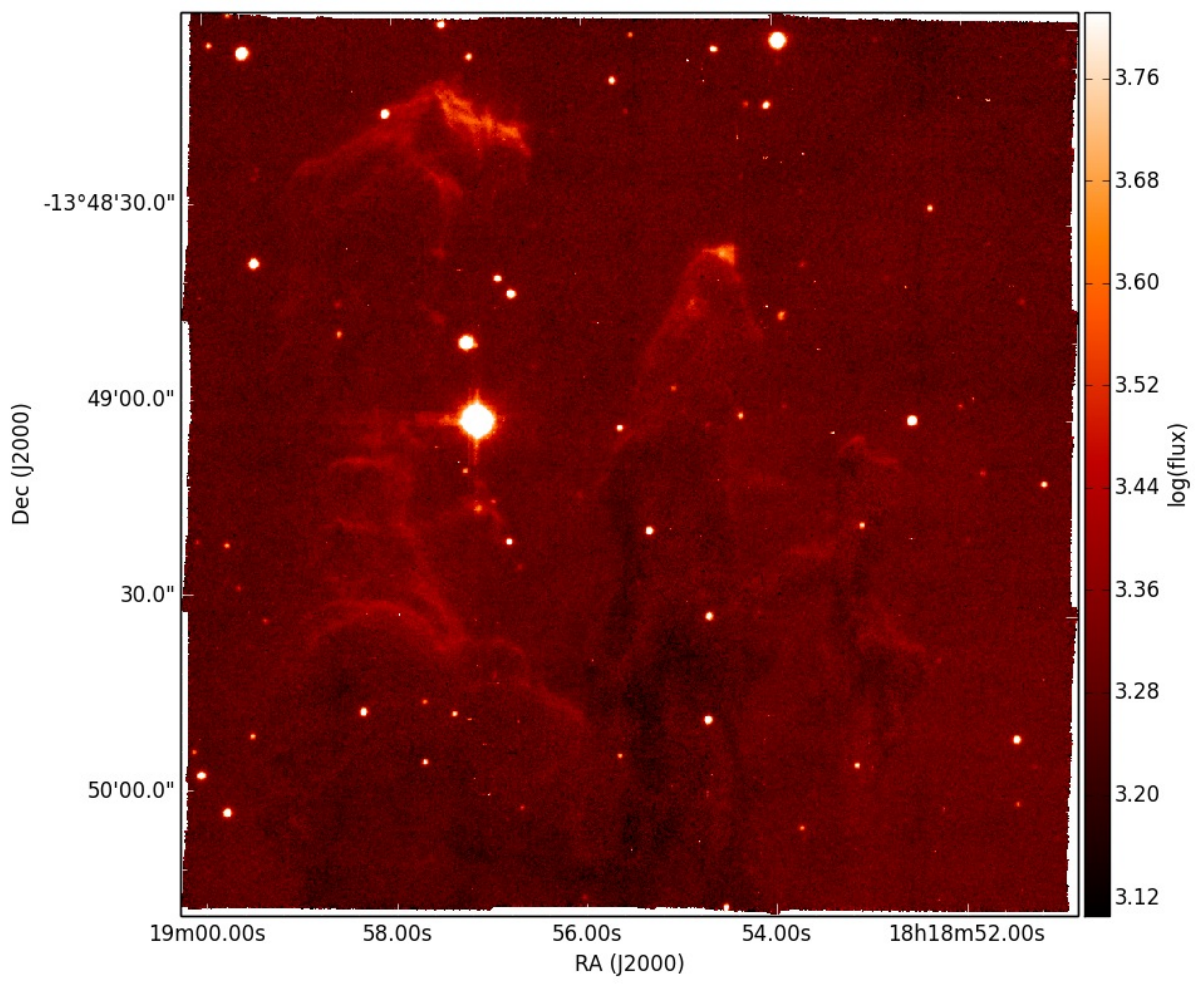}}}
  \mbox{
   \subfloat[]{\includegraphics[scale=0.5]{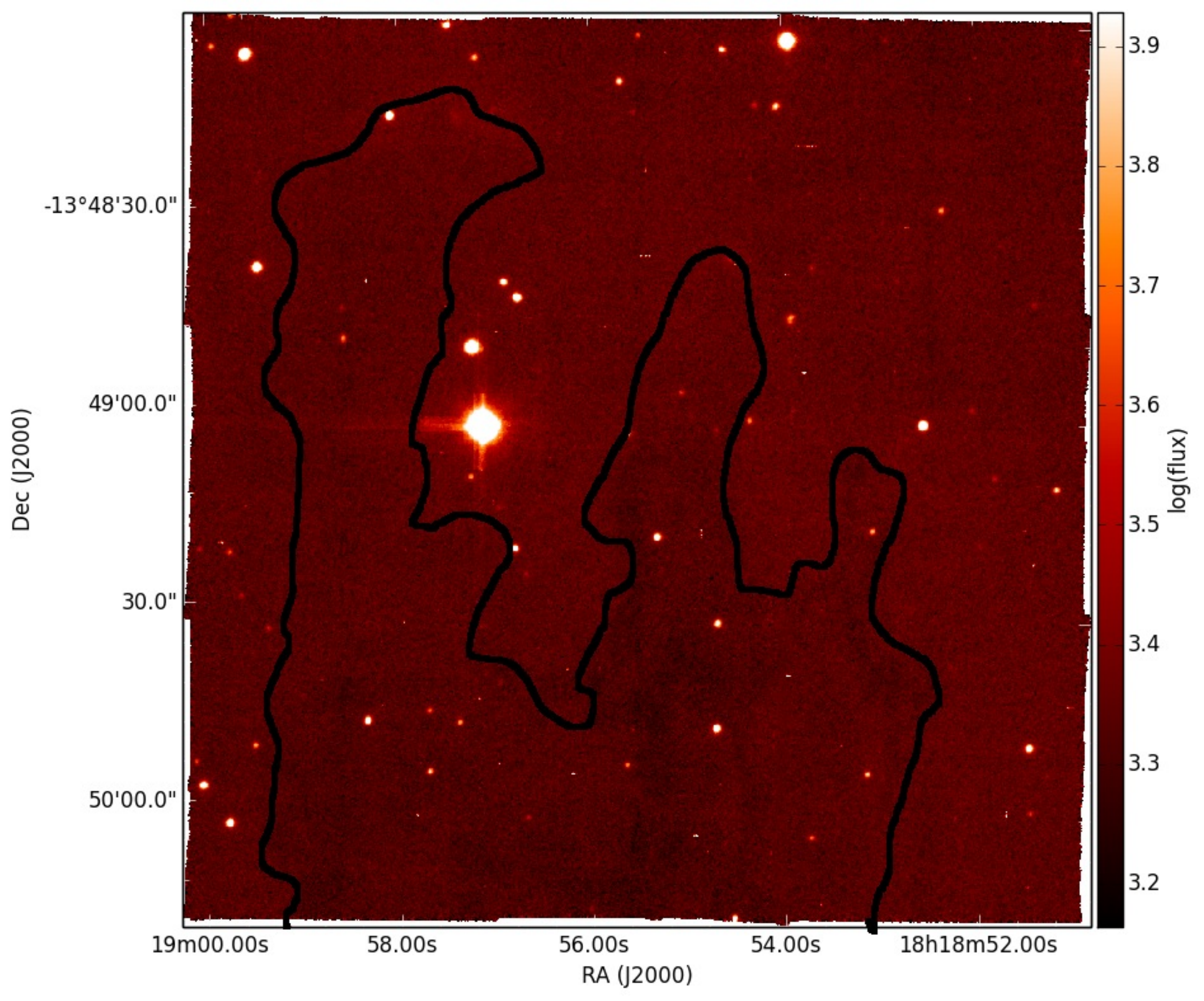}}}
  \caption{OI6300, OI5577. To help the reader identify the location of the Pillars has been marked with a schematic black contour.}
  \label{othermaps2}
\end{figure*}

\begin{figure*}
\mbox{
  \subfloat[]{\includegraphics[scale=0.5]{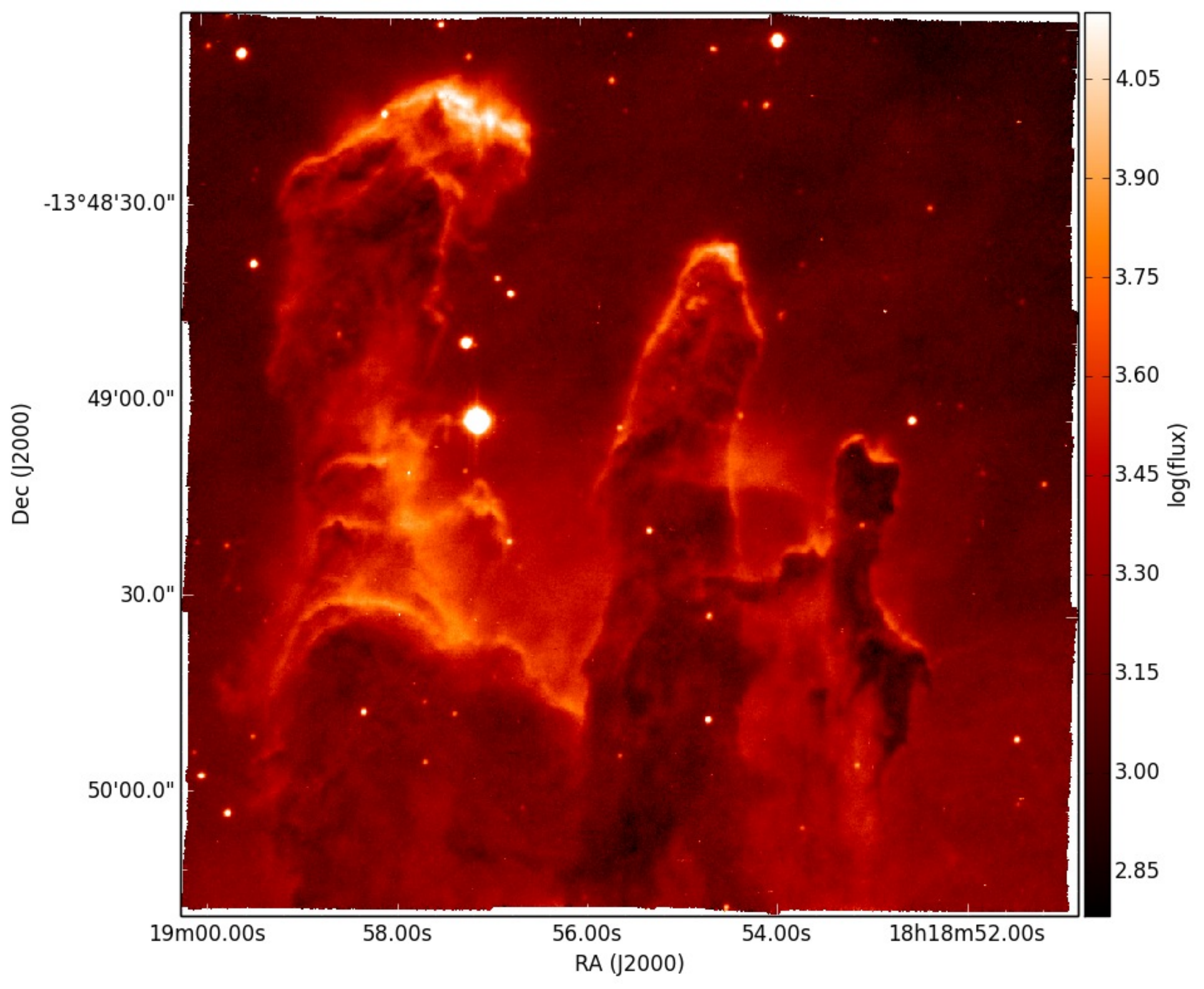}}}
  \mbox{
  \subfloat[]{\includegraphics[scale=0.5]{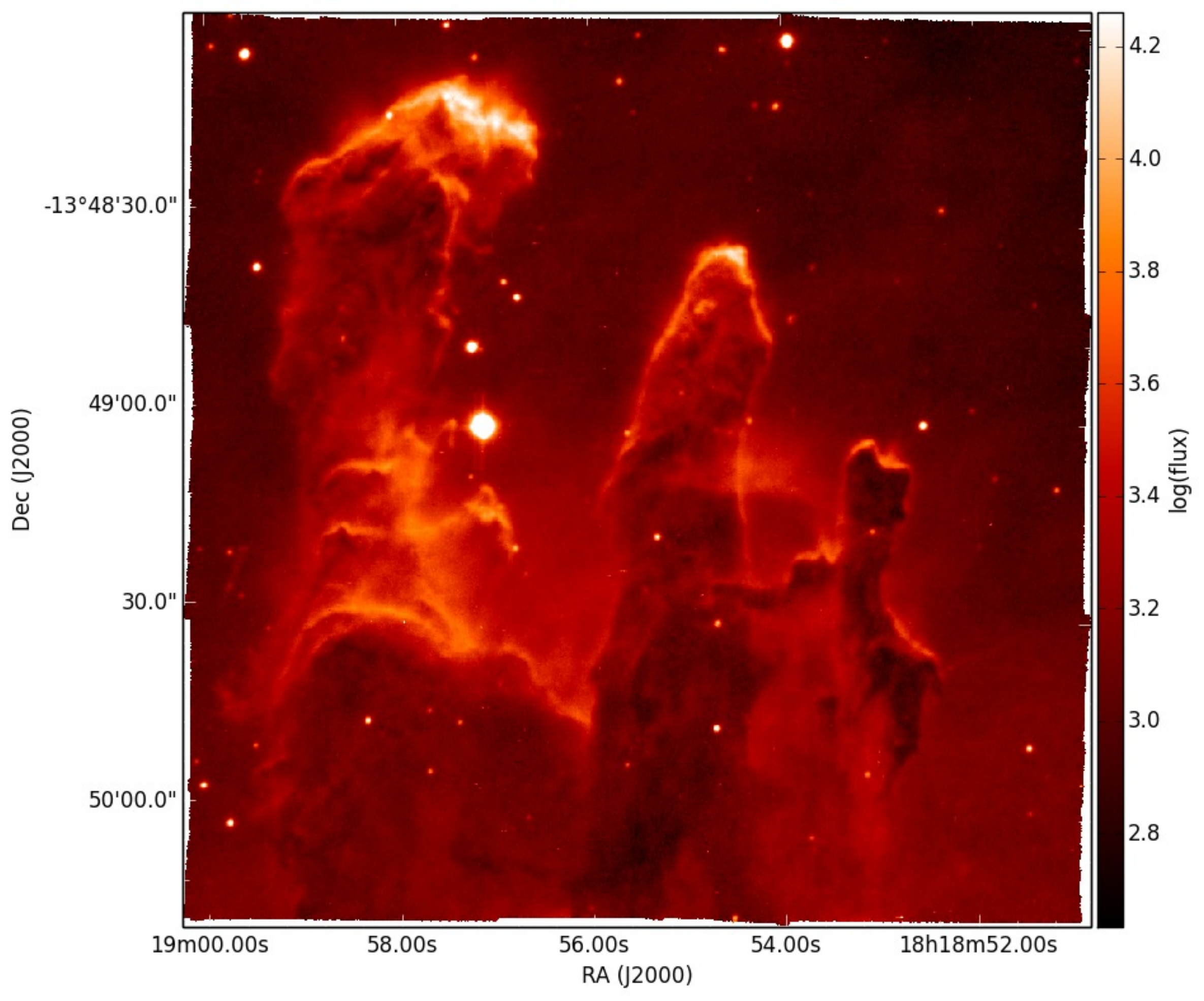}}}
    \caption{SII6717, SII6731.}
  \label{othermaps3}
\end{figure*}

\begin{figure*}
\mbox{
  \subfloat[]{\includegraphics[scale=0.5]{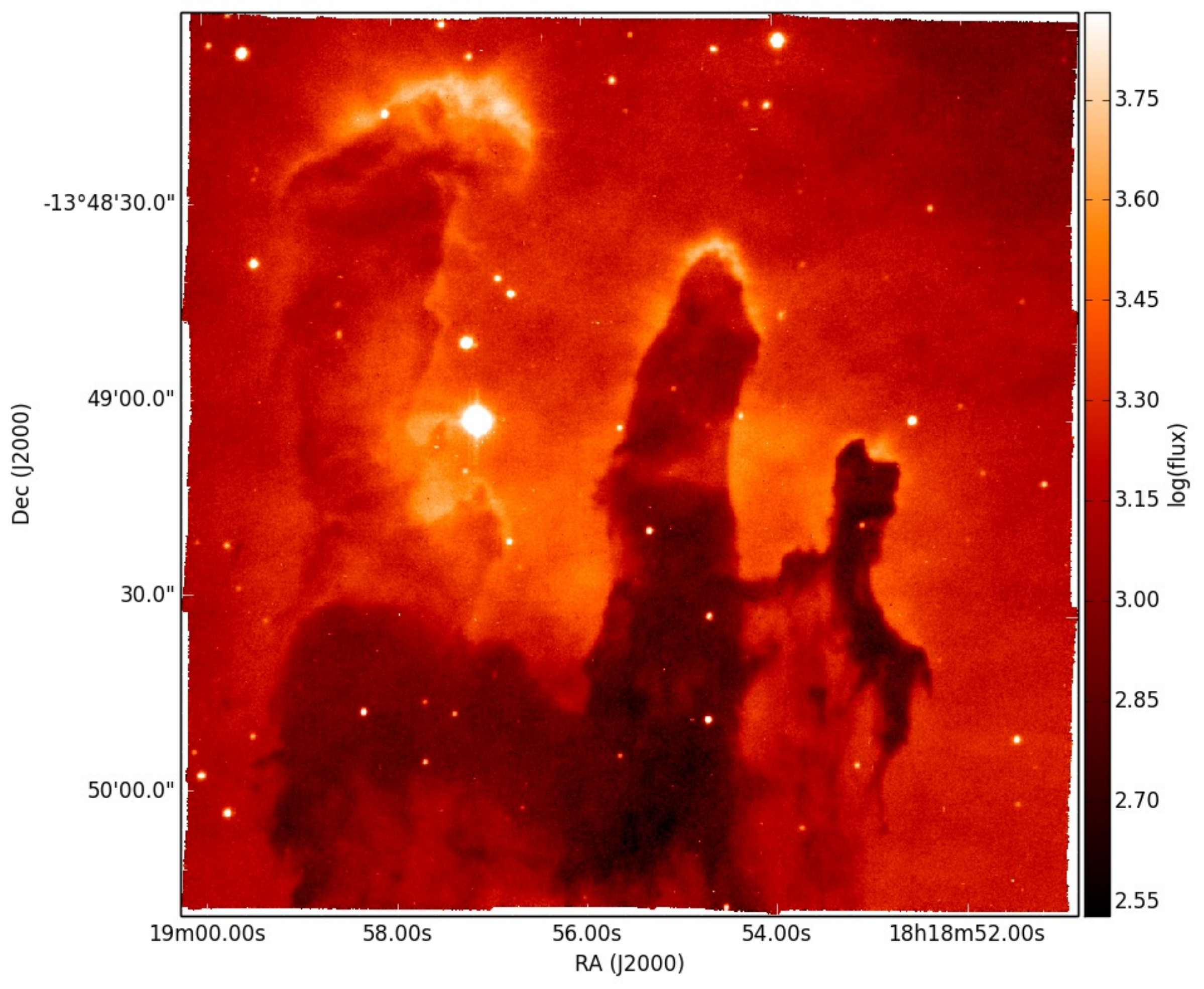}}}
  \mbox{
  \subfloat[]{\includegraphics[scale=0.5]{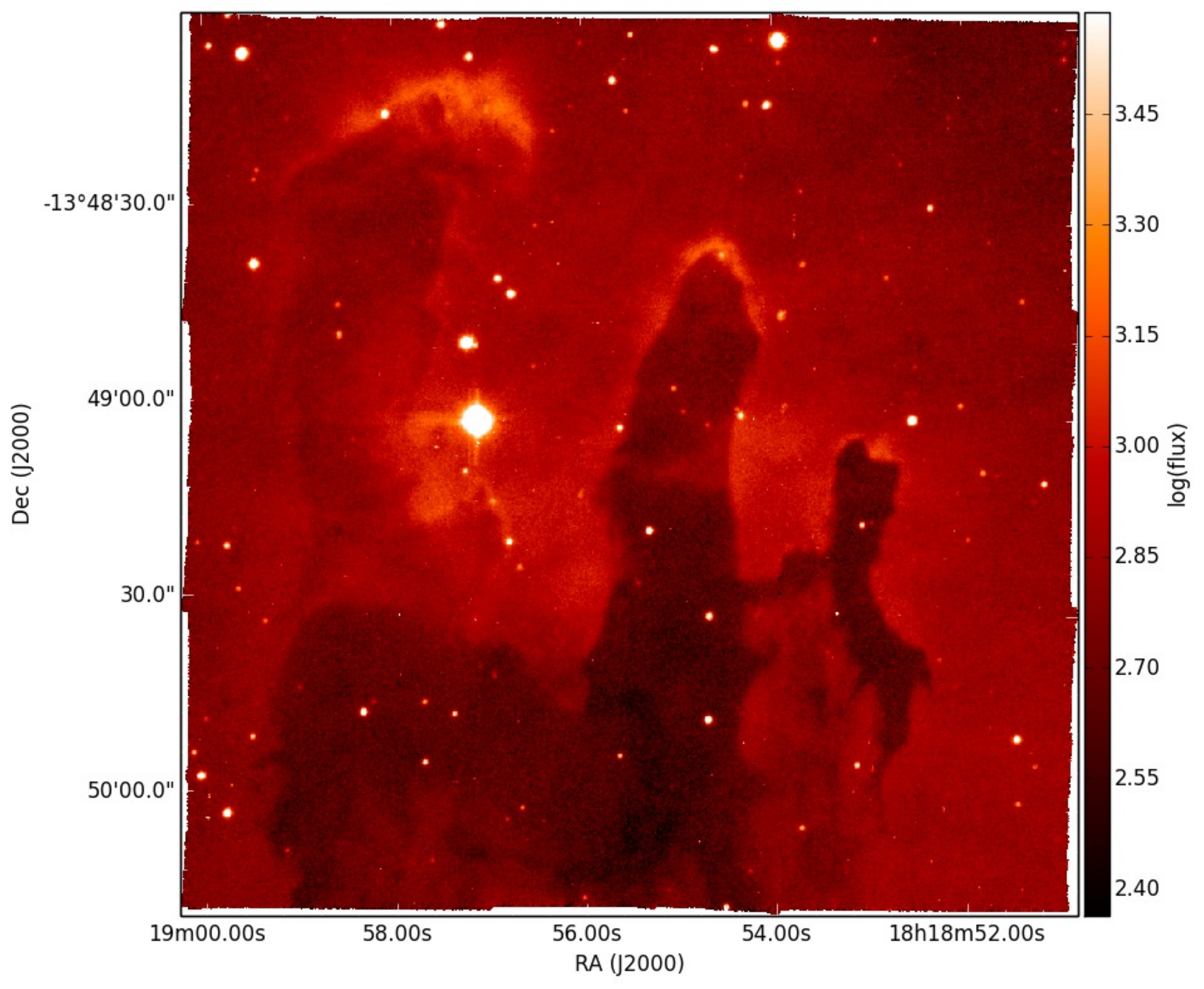}}}
    \caption{ArIII7135, ArIII7751.}
  \label{othermaps5}
\end{figure*}

\begin{figure*}
  \mbox{
  \subfloat[]{\includegraphics[scale=0.5]{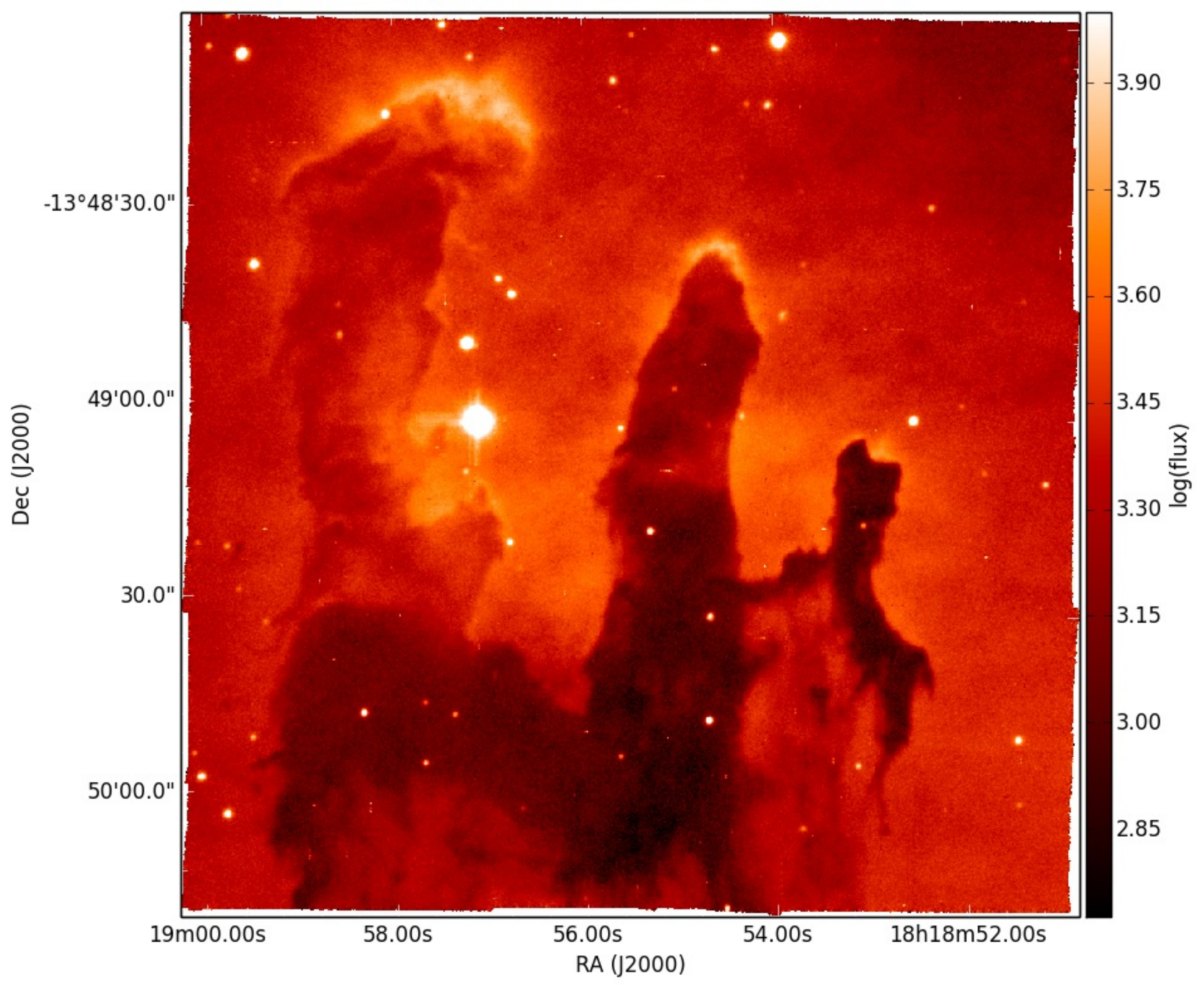}}}
  \mbox{
  \subfloat[]{\includegraphics[scale=0.5]{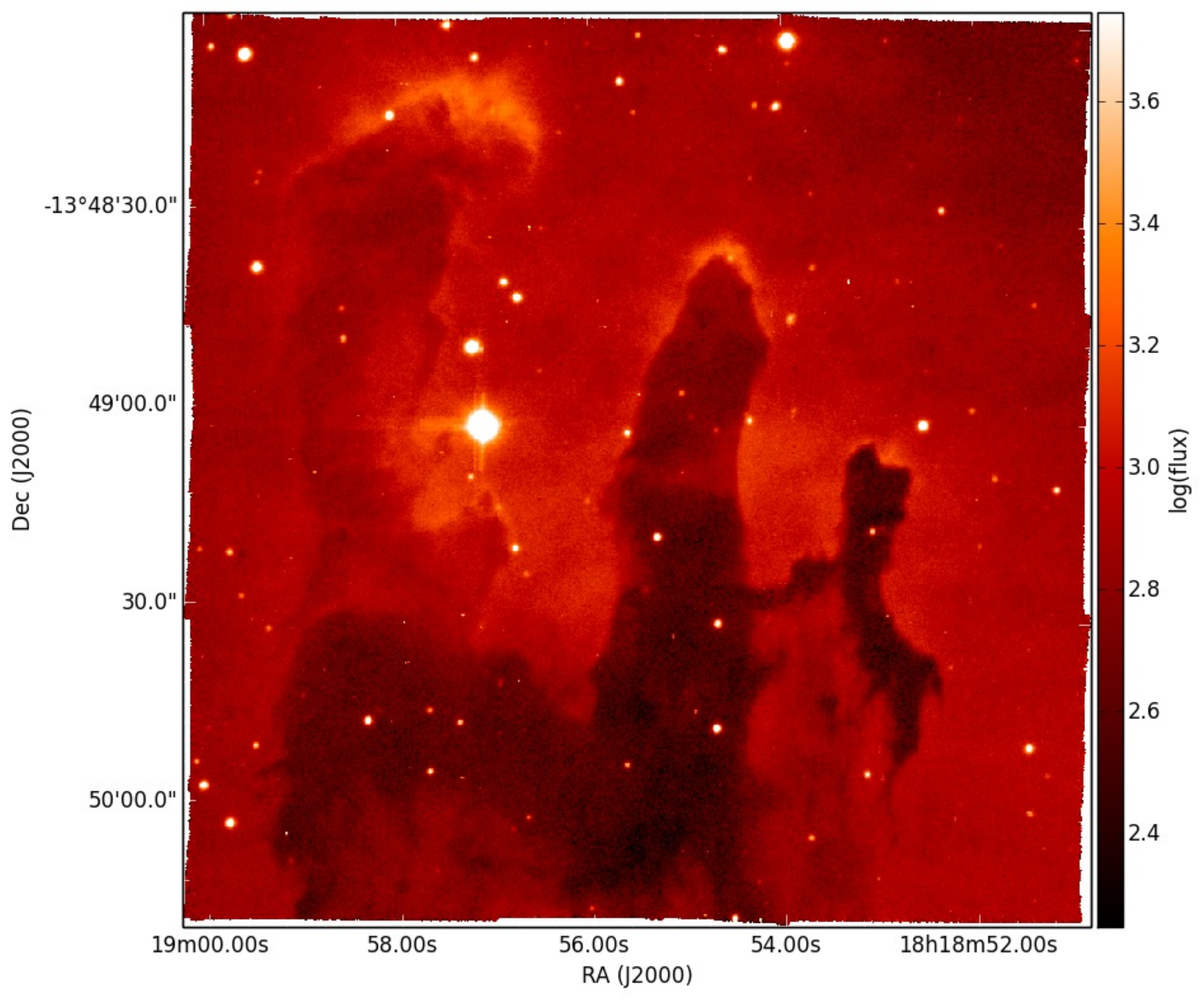}}}
  \caption{HeI5876, HeI6678.}
  \label{othermaps6}
\end{figure*}

\begin{figure*}
  \mbox{
  \subfloat[]{\includegraphics[scale=0.5]{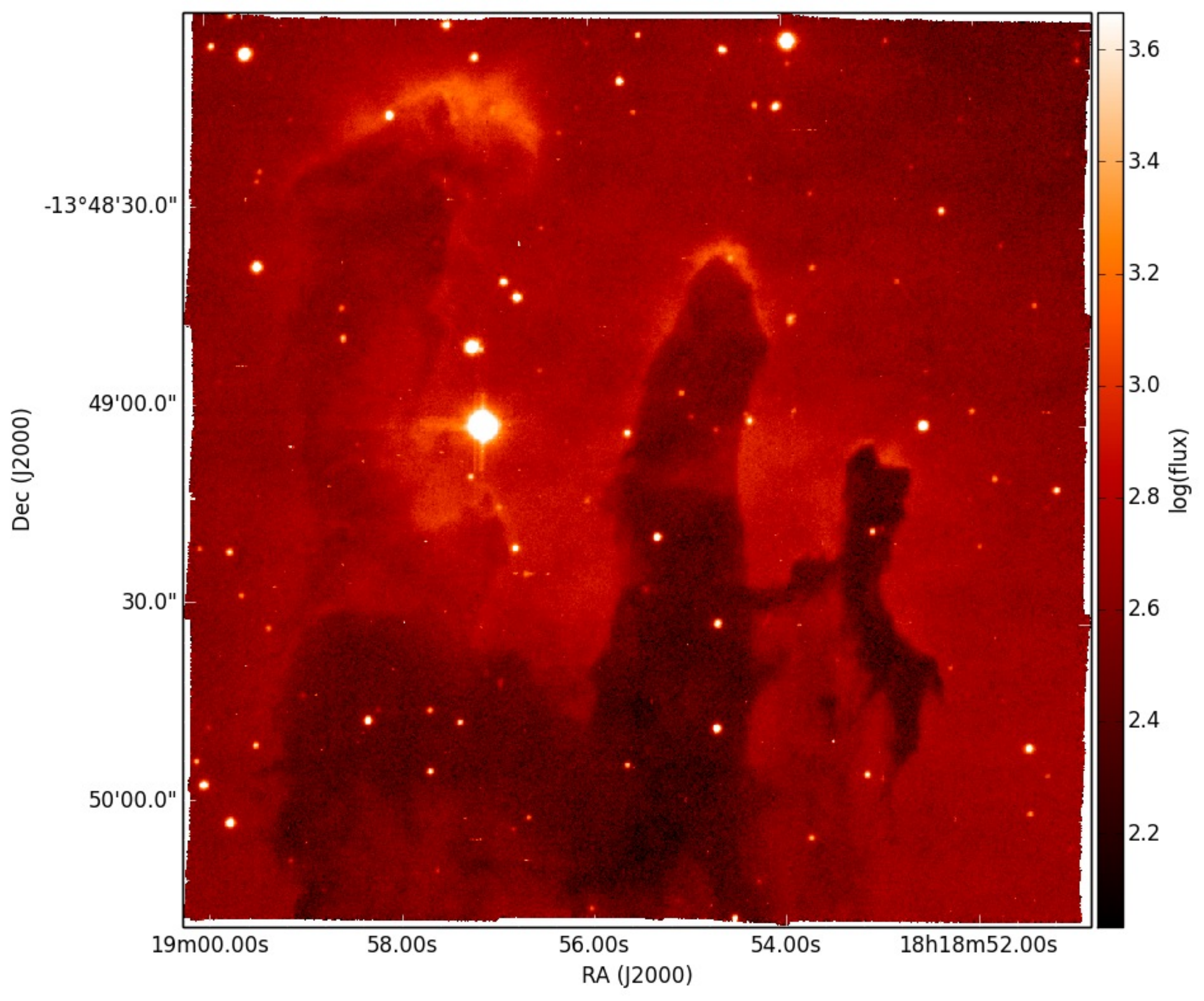}}}
  \mbox{
  \subfloat[]{\includegraphics[scale=0.5]{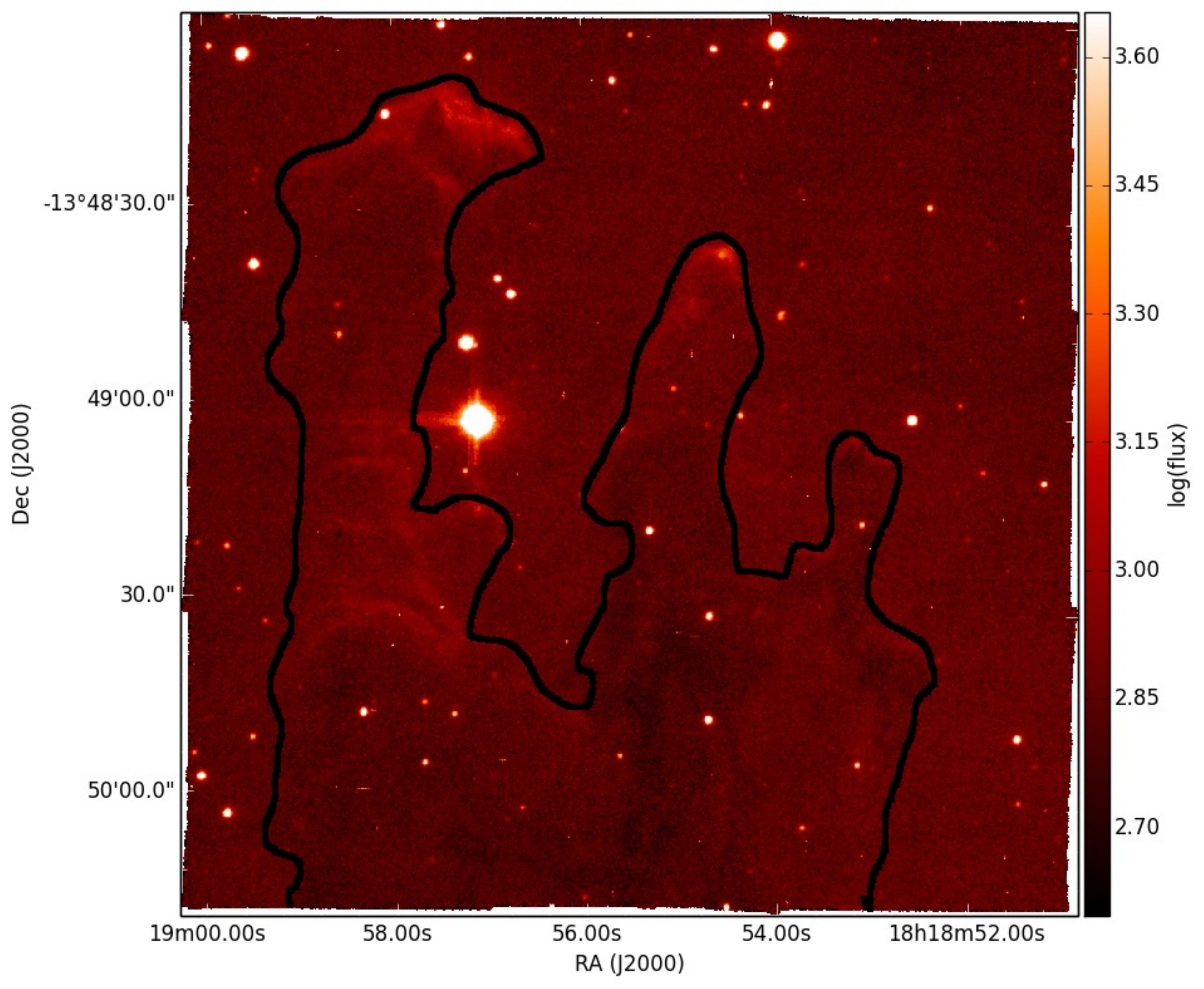}}}
  \caption{HeI7065, OI6363.To help the reader identify the location of the Pillars has been marked with a schematic black contour.}
  \label{othermaps9}
\end{figure*}

\begin{figure*}
  \mbox{
  \subfloat[]{\includegraphics[scale=0.5]{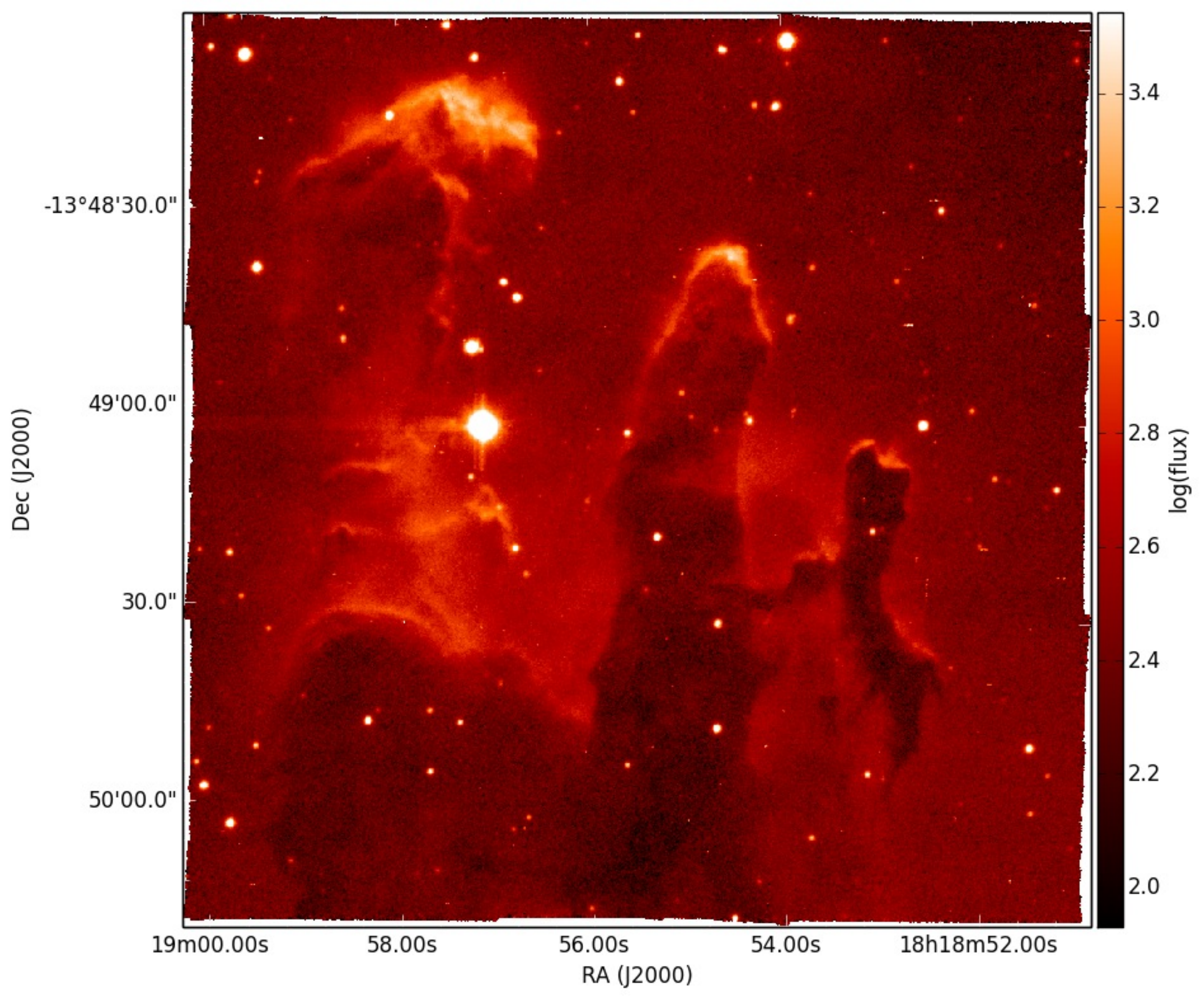}}}
  \mbox{
  \subfloat[]{\includegraphics[scale=0.5]{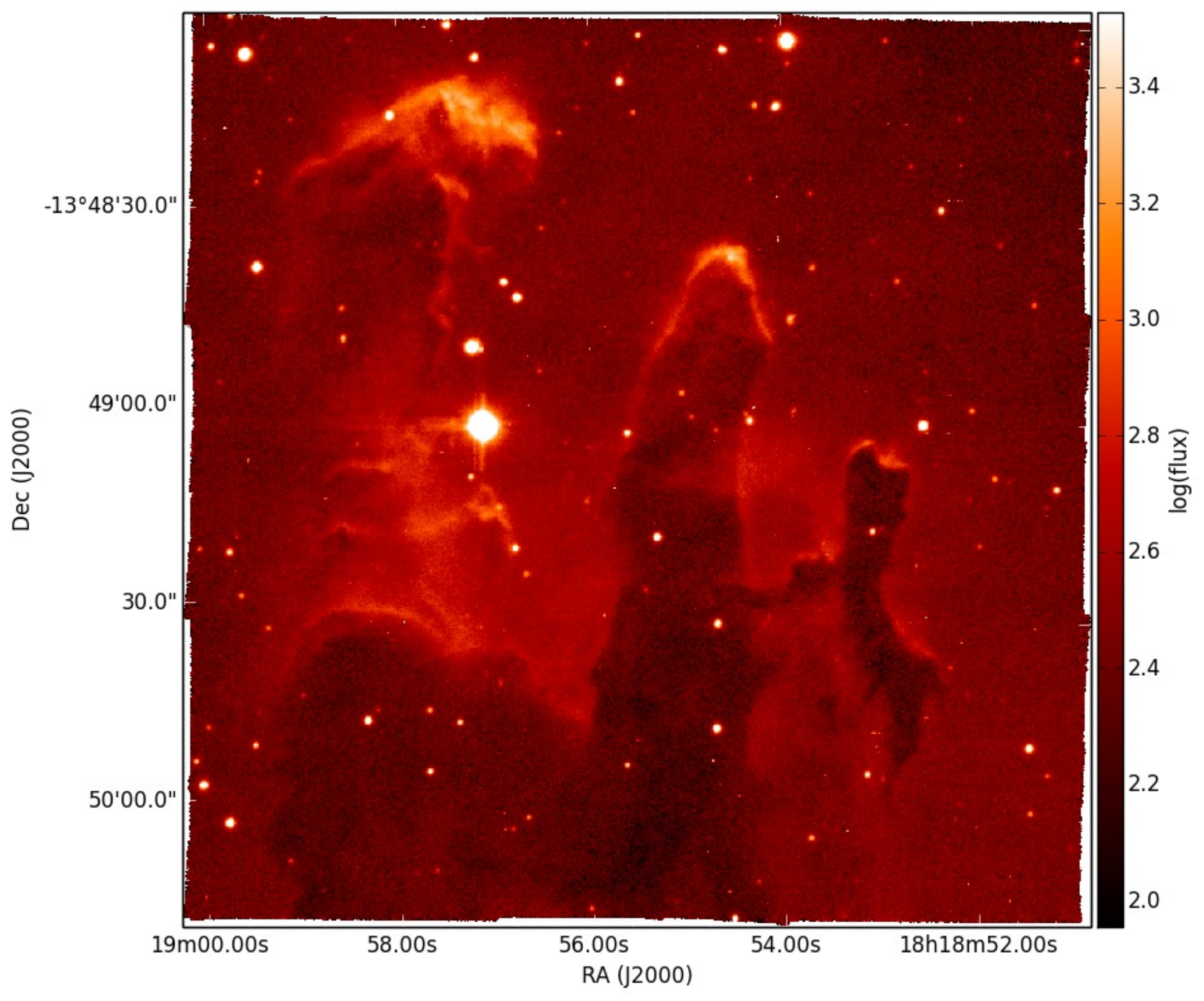}}}
  \caption{OII7320, OII7330.}
  \label{othermaps11}
\end{figure*}

\begin{figure*}
  \mbox{
  \subfloat[]{\includegraphics[scale=0.5]{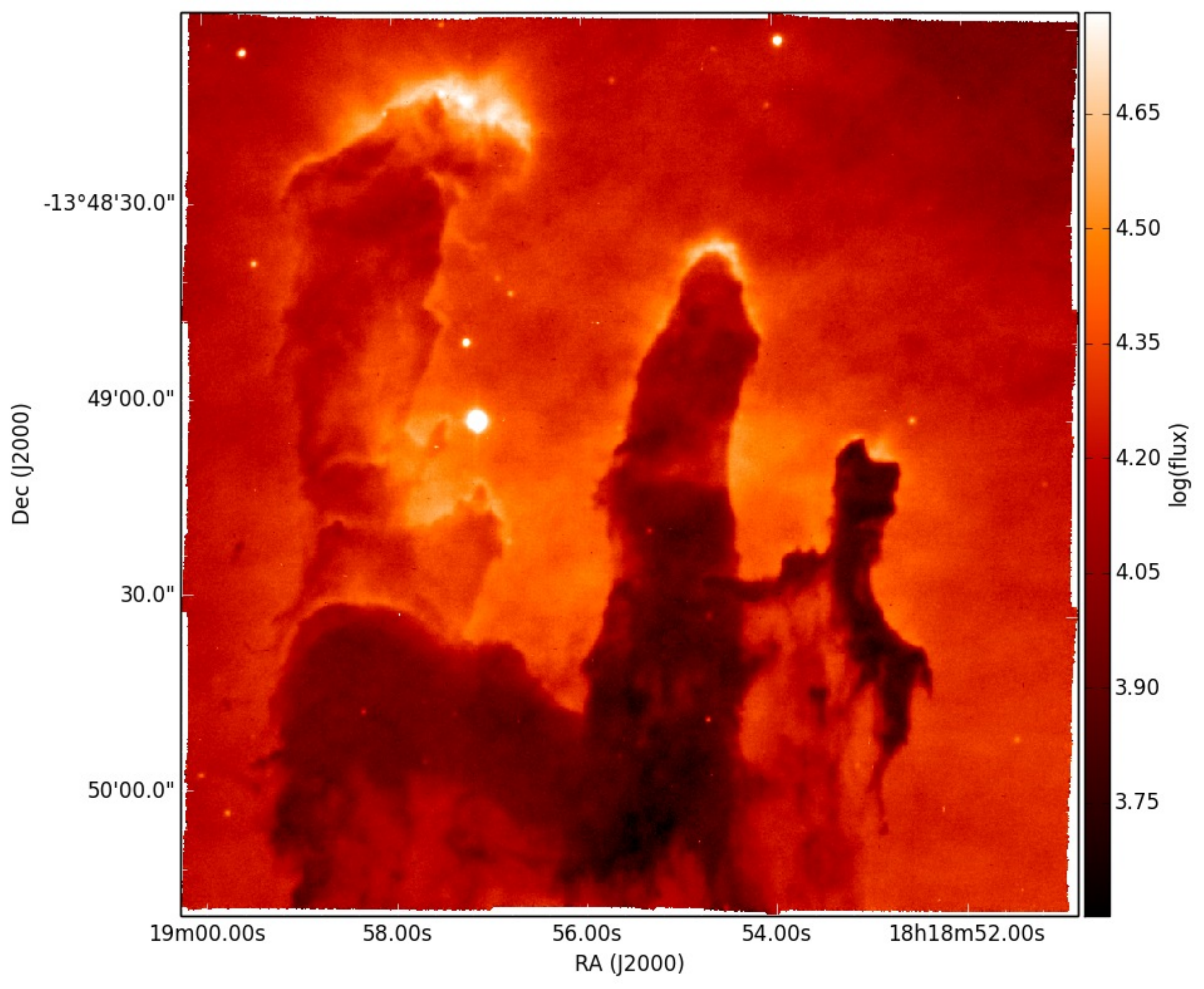}}}
  \mbox{
  \subfloat[]{\includegraphics[scale=0.5]{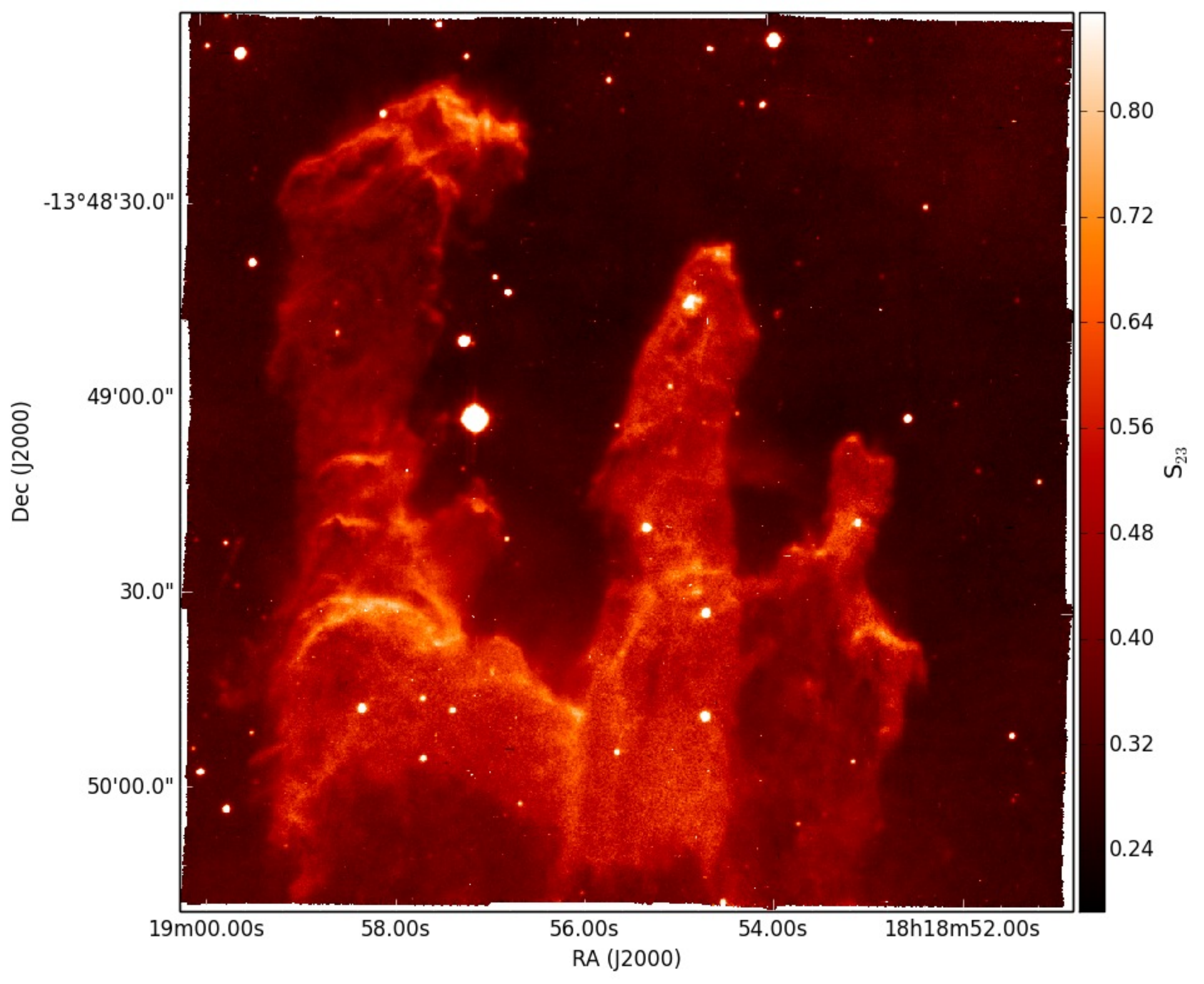}}}
  \caption{H$\beta$, S$_{23}$ parameter (see text for explanation).}
  \label{othermaps7}
\end{figure*}

\begin{figure*}
\includegraphics[scale=0.7,angle=-90]{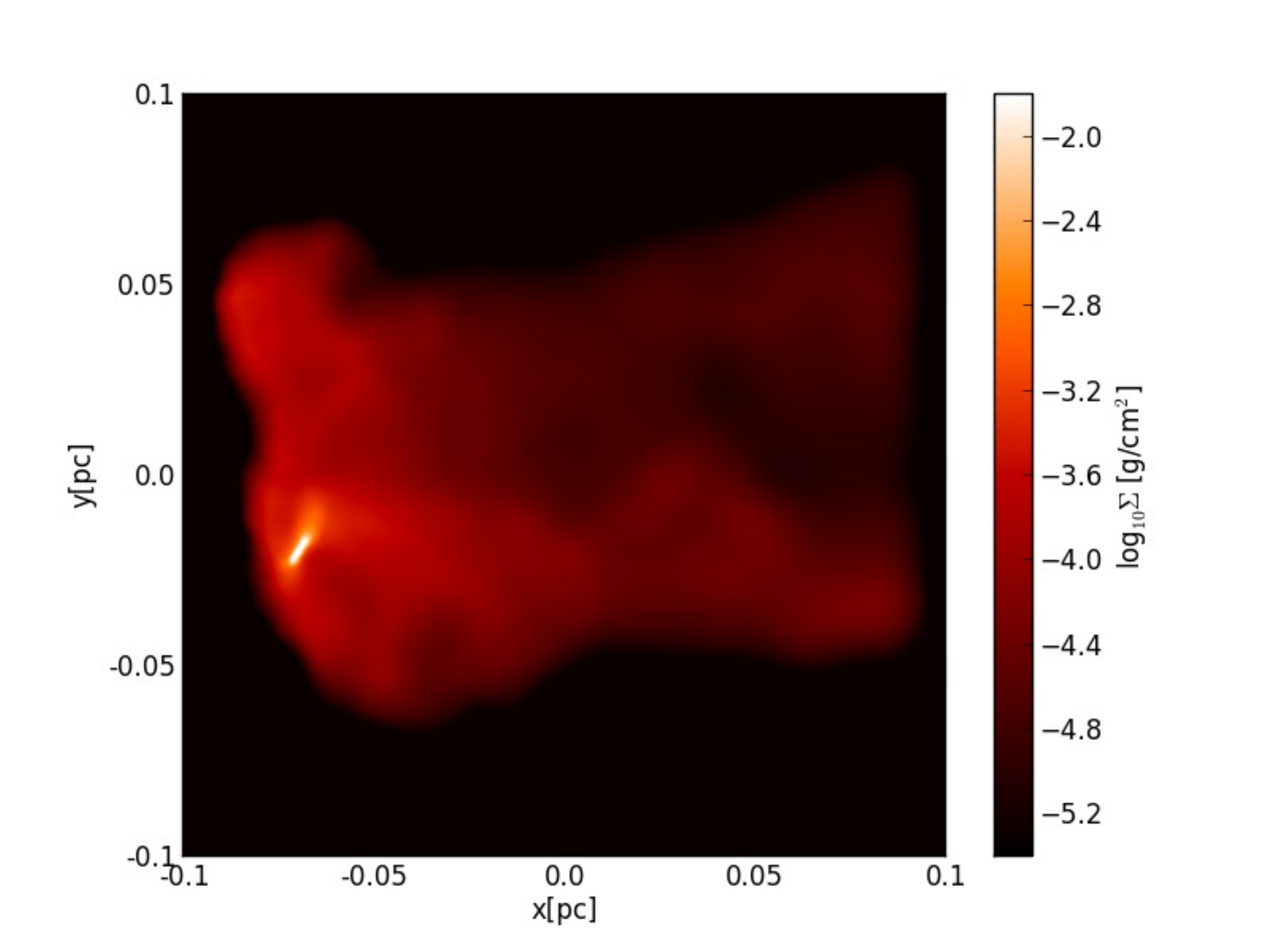}
  \caption{Zoom-in onto one of the simulated pillars: a disk-like structure is seen tracing a forming stellar object at the tip of the pillar (see text Section 4.1).}
  \label{othermaps8}
\end{figure*}

\begin{figure*}
\includegraphics[scale=0.7]{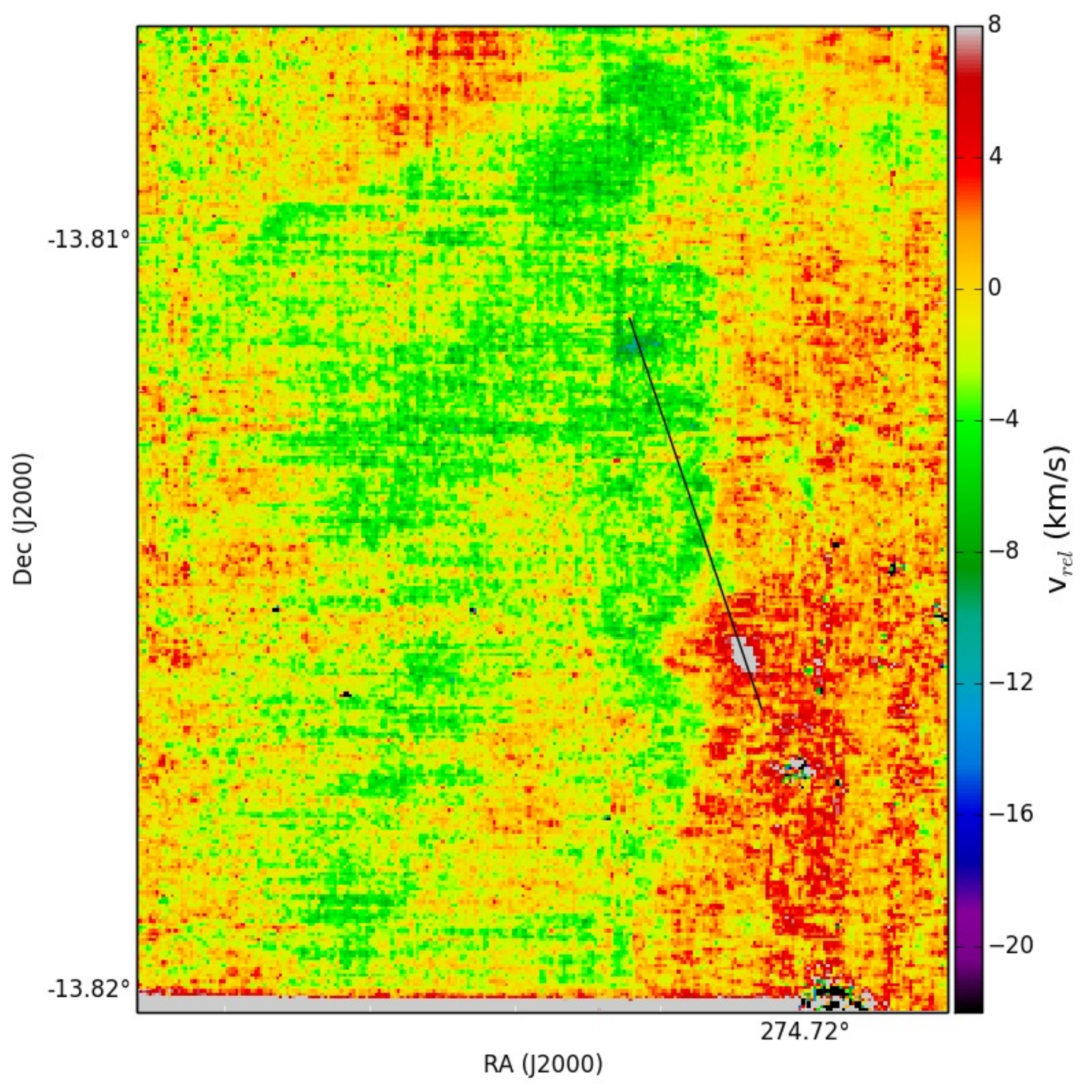}
  \caption{Zoom-in of the velocity map shown in Fig. 16a onto P1. The black line shows the orientation of the outflow candidate discussed in Section 4.1.}
  \label{othermaps10}
\end{figure*}

This appendix is available as supplementary online material. Here we show the extinction corrected integrated line maps of the detected emission lines (Fig. \ref{othermaps1} to Fig. \ref{othermaps7}a, linearly scaled to minimum/maximum, the flux is in units of $10^{-20}$ erg s$^{-1}$ cm$^{-2}$ pixel$^{-1}$), the S$_{23}$ parameter as well as a disk-like structure in one of our simulated pillars (Fig. \ref{othermaps8}) and a zoom-in of the velocity map onto the tip of P1 where the candidate outflow is located (Fig. \ref{othermaps10}).

\bsp

\label{lastpage}

\end{document}